\documentclass[aps,floatfix,superscriptaddress,preprint]
{revtex4}

\usepackage{titlesec}
\usepackage{amssymb,amsmath,amsfonts,latexsym,graphicx,epsfig,bm}
\usepackage{epstopdf}

\titleformat{\section}{\large\bfseries}{\thesection}{1em}{}

\newcommand{\bea}{\begin{eqnarray}}
\newcommand{\ena}{\end{eqnarray}}
\newcommand{\nn}{\nonumber\\}
\newcommand{\be}{\begin{equation}}
\newcommand{\en}{\end{equation}}

\newcommand{\ed}{\end{document}} 

\newcommand{\ds}{\displaystyle}

\newcommand{\slp}{p\kern-5pt/}
\newcommand{\Tr}{\mbox{\rm{tr}}}

\begin{document}

\title{Analyzing new physics in the decays 
\boldmath{$\bar{B}^0 \to D^{(\ast)}\tau^-\bar\nu_{\tau}$} 
with form factors obtained from the covariant quark model}

\author {M. A. Ivanov}
\email{ivanovm@theor.jinr.ru}
\affiliation{
Bogoliubov Laboratory of Theoretical Physics,
Joint Institute for Nuclear Research,
141980 Dubna, Russia}

\author{J. G. K\"{o}rner}
\email{jukoerne@uni-mainz.de}
\affiliation{PRISMA Cluster of Excellence, Institut f\"{u}r Physik, 
Johannes Gutenberg-Universit\"{a}t, 
D-55099 Mainz, Germany}

\author{C. T. Tran}
\email{ctt@theor.jinr.ru,tranchienthang1347@gmail.com}
\affiliation{
Bogoliubov Laboratory of Theoretical Physics,
Joint Institute for Nuclear Research,
141980 Dubna, Russia}
\affiliation{Advanced Center for Physics, Institute of Physics, Vietnam 
Academy of Science and Technology, 100000 Hanoi, Vietnam}
\affiliation{Department of General and Applied Physics, 
Moscow Institute of Physics and Technology, 141700 Dolgoprudny, Russia}
\date{\today}

\begin{abstract}
We study possible new physics (NP) effects in the exclusive decays
$\bar{B}^0 \to D^{(\ast)} \tau^- \bar{\nu}_{\tau}$. We extend the Standard Model
by taking into account right-handed vector (axial), left- and right-handed
(pseudo)scalar, and tensor current contributions. The $\bar{B}^0 \to D^{(\ast)}$ 
transition form factors are calculated in the full kinematic $q^2$ range 
by employing a covariant quark model developed by us. 
We provide constraints on NP operators based on 
measurements of the ratios of branching fractions 
$R(D^{(\ast)}) \equiv \mathcal{B}(\bar{B}^0 \to D^{(\ast)} \tau^- \bar{\nu}_{\tau})/
\mathcal{B}(\bar{B}^0 \to D^{(\ast)} \mu^- \bar{\nu}_{\mu})$ and consider
the effects of these operators on physical observables in different NP
scenarios. We also derive the fourfold angular distribution for the cascade
decay $\bar {B}^0\to D^{\ast\,+}(\to D^0\pi^+)\tau^-\bar\nu_\tau$ which allows one
to analyze the polarization of the $D^{\ast}$ meson in the presence of NP
effects. We discuss several strategies to distinguish between various NP
contributions. 
\end{abstract}

\maketitle

\section{Introduction}
\label{sec:intro} 
In the last few years, the semileptonic decays
$\bar{B}^0 \to D^{(\ast)} \tau^- \bar{\nu}_{\tau}$ have been widely discussed in
the literature as candidates for testing the Standard
Model (SM) and searching for possible
new physics (NP) in charged-current interactions. At $B$ factories, the Belle
and \textit{BABAR} collaborations  have been continuously updating their
measurements with better precision based on electron-positron colliders.
Recently, the LHCb Collaboration has also entered the game with data taken
at the LHC hadron collider. The three groups have reported measurements of the
ratios in Refs.~\cite{Lees:2012xj,Huschle:2015rga,Aaij:2015yra,Sato:2016svk,Abdesselam:2016xqt}. These measurements provide the average ratios
\be
R(D)|_{\rm expt} = 0.397 \pm 0.049 ,
\qquad
R(D^\ast)|_{\rm expt} = 0.308 \pm 0.017 ,
\label{eq:RD-expt}
\en
which exceed the SM  expectations 
\cite{Na:2015kha,Fajfer:2012vx}
\be
R(D)|_{\rm SM} = 0.300 \pm 0.008 ,
\qquad
R(D^\ast)|_{\rm SM} = 0.252 \pm 0.003 ,
\label{eq:RD-SM}
\en
by 1.9~$\sigma$ and 3.3~$\sigma$, respectively.

The excess of $R(D^{(\ast)})$ over SM predictions has attracted a great deal
of attention in the particle physics community and has  led to many theoretical
studies looking for NP explanations. Some studies focus on specific NP models
including two-Higgs-doublet models~\cite{Hou:1992sy,Crivellin:2015hha,Crivellin:2012ye,Celis:2012dk,Nandi:2016wlp}, the minimal
supersymmetric standard model~\cite{Martin:1997ns,Deshpand:2016cpw}, leptoquark
models~\cite{Buchmuller:1986zs,Calibbi:2015kma,Sakaki:2013bfa,Bauer:2015knc,Fajfer:2015ycq,Li:2016vvp}, and other extensions of the SM~\cite{Greljo:2015mma,Boucenna:2016wpr}. Other studies adopt a model-independent approach, in which a
general effective Hamiltonian for the $b \to c \ell \nu$ transition in the
presence of NP is imposed to investigate the impact of various NP operators
on different physical observables~\cite{Fajfer:2012vx,Datta:2012qk,Becirevic:2012jf,Tanaka:2012nw,Biancofiore:2013ki,Duraisamy:2013kcw,Duraisamy:2014sna,Bhattacharya:2015ida,Alok:2016qyh}.
Most of the theoretical studies rely on the heavy quark effective theory
(HQET)~\cite{Neubert:1993mb,grozin2004heavy} to evaluate the hadronic form
factors, which are expressed through a few universal functions in the heavy
quark limit (HQL). In the present analysis, we employ an alternative
approach to calculate the NP-induced hadronic transitions
based on our covariant quark model with embedded infrared
confinement [for short, covariant confined quark model (CCQM)], which has
been developed in some earlier papers by us
(see Ref.~\cite{Branz:2009cd,Ivanov:2011aa} and references therein). 

In a recent paper~\cite{Ivanov:2015tru}, we have provided a thorough study of
the leptonic and semileptonic decays $B^- \to \ell^- \bar\nu_{\ell}$ 
and  $\bar{B}^0 \to D^{(\ast)} \ell^-\bar\nu_{\ell}$ within the SM. We have also
considered the HQL in the heavy-to-heavy transition $\bar{B}^0 \to D (D^\ast)$
and found agreement with the HQET predictions. In this paper we follow the authors of Refs.~\cite{Datta:2012qk,Duraisamy:2013kcw,Duraisamy:2014sna,Tanaka:2012nw,Biancofiore:2013ki,Fajfer:2012vx} to include NP operators in the effective Hamiltonian and investigate their effects on physical observables of the decays $\bar{B}^0 \to D^{(\ast)} \ell^-\bar\nu_{\ell}$. We define a full set of form factors corresponding to SM+NP operators and calculate them by employing the CCQM. In the CCQM the
transition form factors can be determined in the full range of  momentum
transfer, making the calculations straightforward without any extrapolation. This provides an opportunity to investigate NP operators in a self-consistent manner, and independently from the HQET. We first constrain the NP operators using experimental data, then analyze their effects on various observables including the ratios of branching fractions, the forward-backward asymmetries, and a set of polarization observables. We also derive the fourfold angular distribution for the cascade decay $\bar {B}^0\to D^{\ast\,+}(\to D^0\pi^+)\tau^-\bar\nu_\tau$ to analyze the polarization of the $D^{\ast}$ meson in the presence of NP by using the traditional helicity amplitudes. A similar study was done by the authors of Refs.~\cite{Duraisamy:2013kcw,Duraisamy:2014sna}, in which the angular distribution is expressed via the transversality amplitudes. The remaining part of the paper is organized as follows. In Sec.~\ref{sec:FF} we set up our framework by introducing the effective Hamiltonian. In this section we also describe in some detail the calculation technique used in our approach in order to derive the $\bar{B}^0 \to D^{(\ast)}$ transition form factors. In Sec.~\ref{sec:constraint} we use the helicity technique to derive the twofold distribution and provide experimental constraints on the NP operators. In Sec.~\ref{sec:NP} we define various physical observables obtained from the fourfold distribution and analyze their sensitivity to different NP operators. And finally, we provide a brief summary of our main results in Sec.~\ref{sec:summary}.
\section{Effective Hamiltonian and form factors}
\label{sec:FF}

We start with the effective Hamiltonian for the quark-level transition 
$b \to c \tau^- \bar{\nu}_{\tau}$
\bea
{\mathcal H}_{eff} &=&
2\sqrt {2}G_F V_{cb}[(1+V_L)\mathcal{O}_{V_L}+V_R\mathcal{O}_{V_R}
+S_L\mathcal{O}_{S_L}+S_R\mathcal{O}_{S_R} +T_L\mathcal{O}_{T_L}],
\label{eq:Heff}
\ena
where the four-Fermi operators are written as
\bea
\mathcal{O}_{V_L} = 
\left(\bar{c}\gamma^{\mu}P_Lb\right)\left(\bar{\tau}\gamma_{\mu}P_L\nu_{\tau}
\right),
\nn
\mathcal{O}_{V_R} =
\left(\bar{c}\gamma^{\mu}P_Rb\right)
\left(\bar{\tau}\gamma_{\mu}P_L\nu_{\tau}\right),
\nn
\mathcal{O}_{S_L} =\left(\bar{c}P_Lb\right)\left(\bar{\tau}P_L\nu_{\tau}\right),
\nn
\mathcal{O}_{S_R} =\left(\bar{c}P_Rb\right)\left(\bar{\tau}P_L\nu_{\tau}\right),
\nn
\mathcal{O}_{T_L} =\left(\bar{c}\sigma^{\mu\nu}P_Lb\right)
\left(\bar{\tau}\sigma_{\mu\nu}P_L\nu_{\tau}\right).
\label{eq:operators}
\ena
Here, $\sigma_{\mu\nu}=i\left[\gamma_{\mu},\gamma_{\nu}\right]/2$, 
$P_{L,R}=(1\mp\gamma_5)/2$ are the left and right projection operators, and 
$V_{L,R}$, $S_{L,R}$, and $T_L$ are the complex Wilson coefficients governing 
the NP contributions. In the SM one has $V_{L,R}=S_{L,R}=T_L=0$. We assume that 
NP only affects leptons of the third generation.

The invariant matrix element of the semileptonic decays 
$\bar{B}^0\to D^{(\ast)}\tau\bar{\nu}_\tau$ can be written as
\bea
\mathcal{M}&=&
\frac{G_FV_{cb}}{\sqrt{2}}\Big[
(1+V_R+V_L)\langle D^{(\ast)}|\bar{c}\gamma^\mu b|\bar{B}^0\rangle 
\bar{\tau}\gamma_\mu(1-\gamma^5)\nu_\tau\nn
&&
+(V_R-V_L)\langle D^{(\ast)}|\bar{c}\gamma^\mu\gamma^5b|\bar{B}^0\rangle
\bar{\tau}\gamma_\mu(1-\gamma^5)\nu_\tau\nn
&&+(S_R+S_L)\langle D^{(\ast)}|\bar{c}b|\bar{B}^0\rangle
\bar{\tau}(1-\gamma^5)\nu_\tau\nn
&&+(S_R-S_L)\langle D^{(\ast)}|\bar{c}\gamma^5 b|\bar{B}^0\rangle 
\bar{\tau}(1-\gamma^5)\nu_\tau\nn
&&+T_L\langle D^{(\ast)}|\bar{c}\sigma^{\mu\nu}(1-\gamma^5)b|\bar{B}^0\rangle
\bar{\tau}\sigma_{\mu\nu}(1-\gamma^5)\nu_\tau\Big].
\label{eq:amplitude-full}
\ena
Note that the axial and pseudoscalar hadronic currents do not contribute to 
the  $\bar{B}^0\to D$ transition, while the scalar hadronic current does not 
contribute to the $\bar{B}^0\to D^\ast$ transition. Therefore, assuming that NP 
appears in both transitions, the cases of pure $V_R-V_L$ or $S_R\pm S_L$ 
couplings are ruled out, as mentioned in Ref.~\cite{Datta:2012qk}.

The hadronic matrix elements are parametrized by a set of invariant form 
factors as follows:
\bea
T_1^\mu &\equiv &
\langle D(p_2)
|\bar{c} \gamma^\mu b
| \bar{B}^0(p_1) \rangle
= F_+(q^2) P^\mu + F_-(q^2) q^\mu,
\nn
T_2 &\equiv &
\langle D(p_2)
|\bar{c}b
| \bar{B}^0(p_1) \rangle = (m_1+m_2)F^S(q^2),
\label{eq:BD-ff}\\
T_3^{\mu\nu} &\equiv &
\langle D(p_2)|\bar{c}\sigma^{\mu\nu}(1-\gamma^5)b|\bar{B}^0(p_1)\rangle 
=\frac{iF^T(q^2)}{m_1+m_2}\left(P^\mu q^\nu - P^\nu q^\mu 
+i \varepsilon^{\mu\nu Pq}\right),
\nonumber
\ena
for the $\bar{B}^0\to D$ transition, and
\bea
\epsilon^\dagger_{2\alpha} \mathcal{T}_{1L(R)}^{\mu\alpha} &\equiv  &
\langle D^\ast(p_2)
|\bar{c} \gamma^\mu(1\mp\gamma^5)b
| \bar{B}^0(p_1) \rangle
\nn
&=& \frac{\epsilon^{\dagger}_{2\alpha}}{m_1+m_2}
\left( \mp g^{\mu\alpha}PqA_0(q^2) \pm P^{\mu}P^{\alpha}A_+(q^2)
       \pm q^{\mu}P^\alpha A_-(q^2) 
+ i\varepsilon^{\mu\alpha P q}V(q^2)\right),
\nn
\epsilon^\dagger_{2\alpha}\mathcal{T}_2^\alpha &\equiv &
\langle D^\ast(p_2)
|\bar{c}\gamma^5 b
| \bar{B}^0(p_1) \rangle = \epsilon^\dagger_{2\alpha}P^\alpha G^S(q^2),
\nn
\epsilon^\dagger_{2\alpha}\mathcal{T}_3^{\mu\nu\alpha} &\equiv &
\langle D^\ast(p_2)|\bar{c}\sigma^{\mu\nu}(1-\gamma^5)b|\bar{B}^0(p_1)\rangle
\label{eq:BDv-ff}\\
&=&-i\epsilon^\dagger_{2\alpha}\Big[
\left(P^\mu g^{\nu\alpha} - P^\nu g^{\mu\alpha} 
+i \varepsilon^{P\mu\nu\alpha}\right)G_1^T(q^2)
\nn
&&+\left(q^\mu g^{\nu\alpha} - q^\nu g^{\mu\alpha}
+i \varepsilon^{q\mu\nu\alpha}\right)G_2^T(q^2)
\nn
&&+\left(P^\mu q^\nu - P^\nu q^\mu 
+ i\varepsilon^{Pq\mu\nu}\right)P^\alpha\frac{G_0^T(q^2)}{(m_1+m_2)^2}
\Big],
\nonumber
\ena
for the $\bar{B}^0\to D^\ast$ transition.
Here, $P=p_1+p_2$, $q=p_1-p_2$, and $\epsilon_2$ is the polarization vector
of the $D^\ast$ meson so that $\epsilon_2^\dagger\cdot p_2=0$.
The particles are on their mass shells: $p_1^2=m_1^2=m_B^2$ and
 $p_2^2=m_2^2=m_{D^{(\ast)}}^2$.

In Ref.~\cite{Ivanov:2015tru} we have given a detailed description of the CCQM
framework for the calculation of the semileptonic transitions
$\bar{B}^0 \to D^{(\ast)} \tau^- \bar{\nu}_{\tau}$ in the SM. It therefore
suffices to briefly describe the main steps in the corresponding calculation
for a more general set of quark-level transition operators.

The CCQM provides a field-theoretic frame work for
the calculation of particle transitions in the constituent quark
model~(see e.g.,~Refs.~\cite{Efimov:1988yd,Efimov:zg,Gutsche:2013oea,Gutsche:2012ze,Dubnicka:2010ev,Ivanov:2015woa}). In the CCQM, particle transitions are calculated from Feynman diagrams involving quark
loops. For example, the $\bar{B}^0 \to D^{(\ast)}$ transitions are described by a
one-loop diagram requiring a genuine one-loop calculation. The high-energy
behavior of quark loops is tempered by nonlocal Gaussian-type vertex
functions with a Gaussian-type falloff behavior. The particle-quark vertices
have nonlocal interpolating current structure. In the Feynman diagrams one uses
the usual free local quark propagators $(m\, -\!\not\! p)^{-1}$. The
normalization of the
particle-quark vertices is provided by the compositeness condition which
embodies the correct charge normalization of the respective hadron~\cite{Z=0}. The
compositeness condition can be viewed as the field-theoretic equivalent of the
normalization of the wave function of a quantum-mechanical state. A universal
infrared cutoff provides for an effective confinement of quarks. There are
therefore no free quark thresholds in the Feynman diagrams even if they are
allowed by the kinematics of the process. We mention that the authors of Ref.~\cite{Cheung:2014cka} have pursued a related program to calculate heavy meson
transitions covariantly via one-loop integrals.

The loop integrations are done with the help of the Fock-Schwinger 
representation of the quark propagator. 
The use of the Fock-Schwinger representation allows 
one to do tensor loop integrals in a very efficient way since one can
convert loop momenta into derivatives of the exponent function which are
simple to handle. 
We mention that the same idea to treat tensor loop integrals has been used
in the evaluation of loop integrals in local quantum field theory~\cite{Anastasiou:1999bn,Fiorentin:2015vha}.

The model parameters, namely, the hadron size parameter $\Lambda$, the
constituent quark masses $m_{q_i}$, and the universal infrared cutoff parameter $\lambda$, are determined by
fitting calculated quantities 
of a multitude of basic processes to available experimental data or 
lattice simulations (for details, see Ref.~\cite{Branz:2009cd}, 
where a set of weak and electromagnetic decays was used).
In this paper we will use the updated least-squares fit
performed in Refs.~\cite{Gutsche:2015mxa,Ganbold:2014pua,Issadykov:2015iba}.
Those model parameters 
involved in this paper are given in Eq.~(\ref{eq:modelparam}) (all in GeV): 
\be
\begin{tabular}{ c c c  c  c c c c }
\quad $m_{u/d}$ \quad & \quad $m_s$ \quad & \quad $m_c$ \quad &
\quad $m_b$ \quad
& \quad $\lambda$ \quad & \quad $\Lambda_{D^\ast}$ \quad &
\quad $\Lambda_D$ \quad  
& \quad $\Lambda_B$ \quad 
\\\hline
 \quad 0.241 \quad & \quad 0.428 \quad & \quad 1.67 \quad &
\quad 5.04 \quad & \quad 0.181 \quad 
& \quad 1.53 \quad & \quad 1.60 \quad & 
\quad 1.96 \quad  \\
\end{tabular}
.
\label{eq:modelparam}
\en
Once these parameters are fixed, one can employ the CCQM as a frame-independent
tool for hadronic calculation.
Model-independent parameters and other physical constants are taken  from Ref.~\cite{Agashe:2014kda}.

The form factors in our model are represented by three-fold integrals
which are calculated by using \textsc{fortran} codes in the full kinematical
momentum 
transfer region. Our numerical results for the form factors are well represented
by a double-pole parametrization
\be
F(q^2)=\frac{F(0)}{1 - a s + b s^2}, \quad s=\frac{q^2}{m_1^2}. 
\label{eq:DPP}
\en 
The double-pole approximation is quite accurate: the error
relative to the exact results is less than $1\%$ over the entire $q^{2}$ 
range.
For the $\bar{B}^0 \to D$ and $\bar{B}^0 \to D^\ast$ transitions the parameters 
of the  approximation are listed in Eqs.~(\ref{eq:ff_param_BD})
and~(\ref{eq:ff_param_BDv}), respectively:
\be
\begin{array}{c|cccc}
 & \quad F_+ \quad & \quad  F_- \quad & \quad F^S \quad  & \quad F^T \quad\\[1.1ex]
\hline
F(0) &  0.79   & -0.36 &  0.80 & 0.77  
\\[1ex]
F(q^2_{\rm max}) &  1.14   & -0.53 &  0.89 & 1.11  
\\[1ex]
F^{HQET}(q^2_{\rm max}) &  1.14   & -0.54 &  0.88 & 1.14  
\\[1ex]
a    &  0.75   &  0.77 &  0.22 & 0.76  
\\[1ex] 
b    &  0.039  & 0.046 & -0.098 & 0.043 
\\[1.1ex] 
\end{array}
\label{eq:ff_param_BD}
,
\en
\vspace{1.2ex}   
\be
\begin{array}{c|cccccccc}
 & \quad A_0 \quad  & \quad A_+ \quad & 
  \quad A_- \quad & \quad V & \quad G^S \quad  & \quad G^T_0 \quad & 
  \quad G^T_1 \quad & \quad G^T_2 \quad \\[1.1ex]
\hline
F(0) &  1.62 & 0.67  & -0.77 & 0.77 & -0.50 & -0.073 & 0.73 & -0.37
\\[1ex]
F(q^2_{\rm max}) &  1.91 & 0.99  & -1.15 & 1.15 & -0.73 & -0.13 & 1.10 & -0.55
\\[1ex]
F^{HQET}(q^2_{\rm max}) &  1.99 & 1.12  & -1.12 & 1.12 & -0.62 & 0 & 1.12 & -0.50
\\[1ex] 
a    &  0.34 & 0.87  &  0.89 & 0.90 & 0.87 & 1.23 & 0.90 & 0.88
\\[1ex] 
b    & -0.16 & 0.057 & 0.070 & 0.075 & 0.060 & 0.33 & 0.074 & 0.064
\\[1.1ex] 
\end{array}
\label{eq:ff_param_BDv}
.
\en
\vspace{1.2ex}

We have also listed the zero-recoil values of our model form factors (row three)
which, using the relations in Appendix~\ref{app:HQET}, can be compared to the
corresponding HQET results (row four). The agreement between the two sets of
zero-recoil values is quite good $[O(1-10) \%]$ showing that our model
calculation, which includes the full heavy mass dependence, is quite
close to the heavy quark limit. It is quite noteworthy that we obtain a nonzero result for the form factor
$G_0^T$ in our model calculation at zero recoil, which is predicted to vanish in the HQL.

\begin{figure}[htbp]
\begin{tabular}{lr}
\includegraphics[scale=0.45]{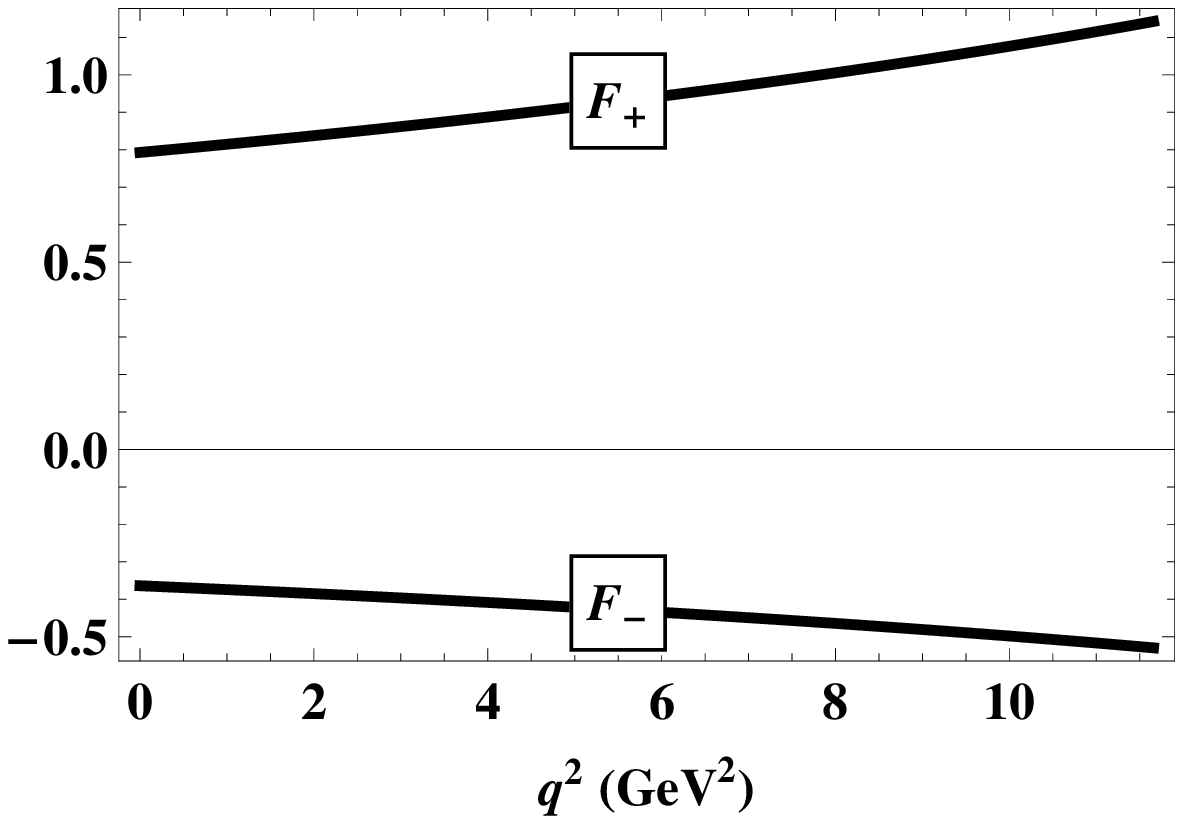}
& 
\includegraphics[scale=0.45]{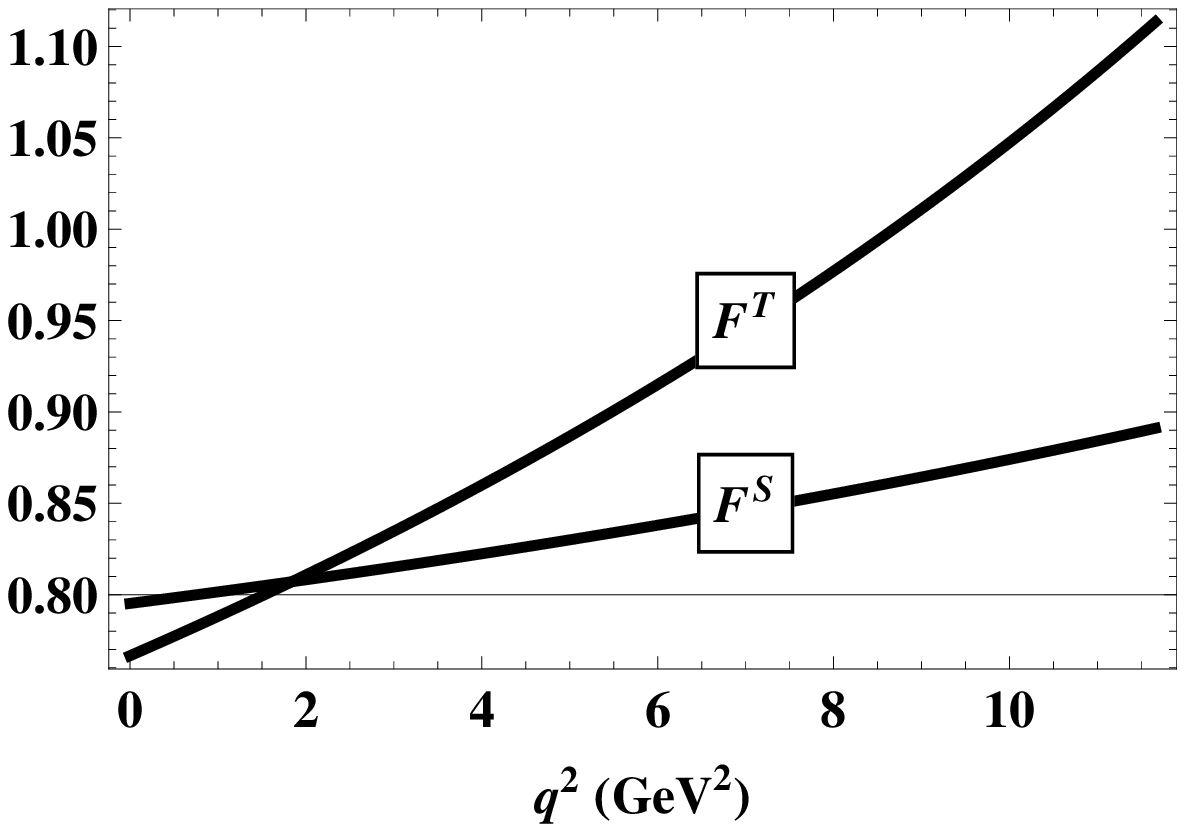}\\
\includegraphics[scale=0.45]{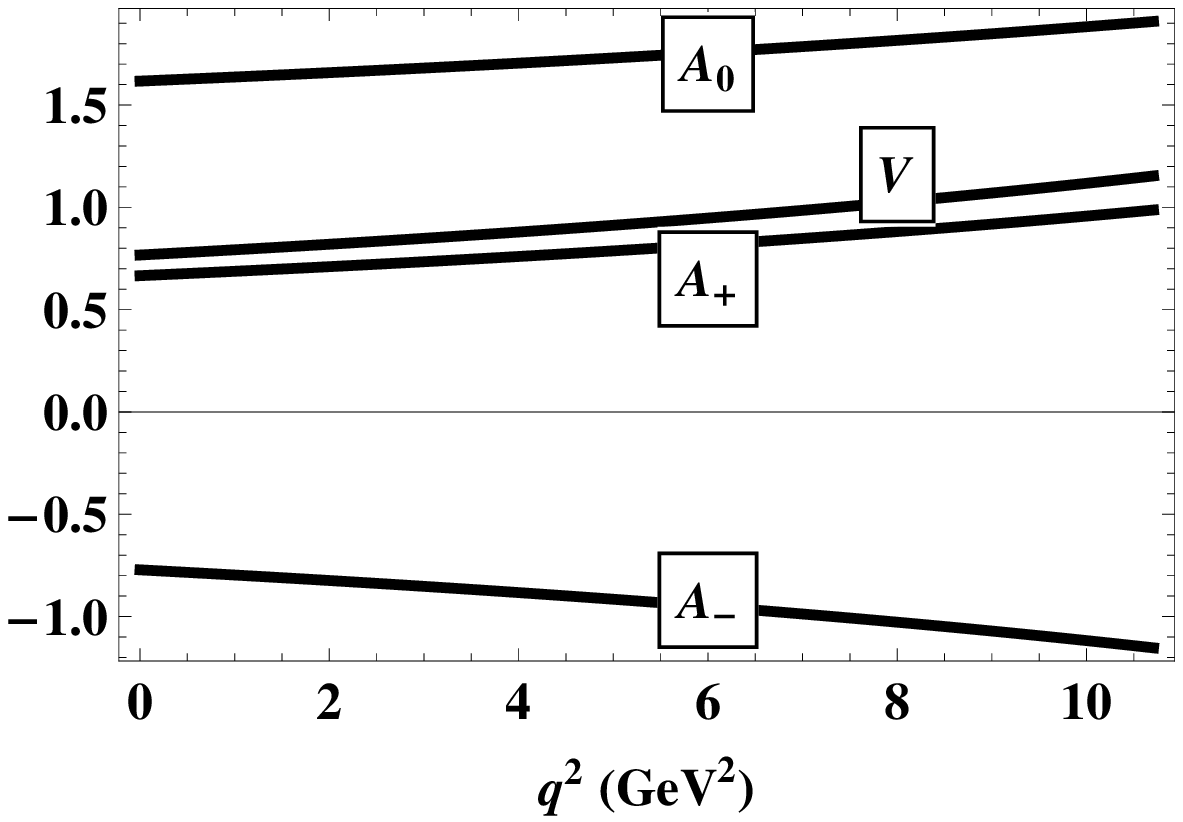}
& 
\includegraphics[scale=0.45]{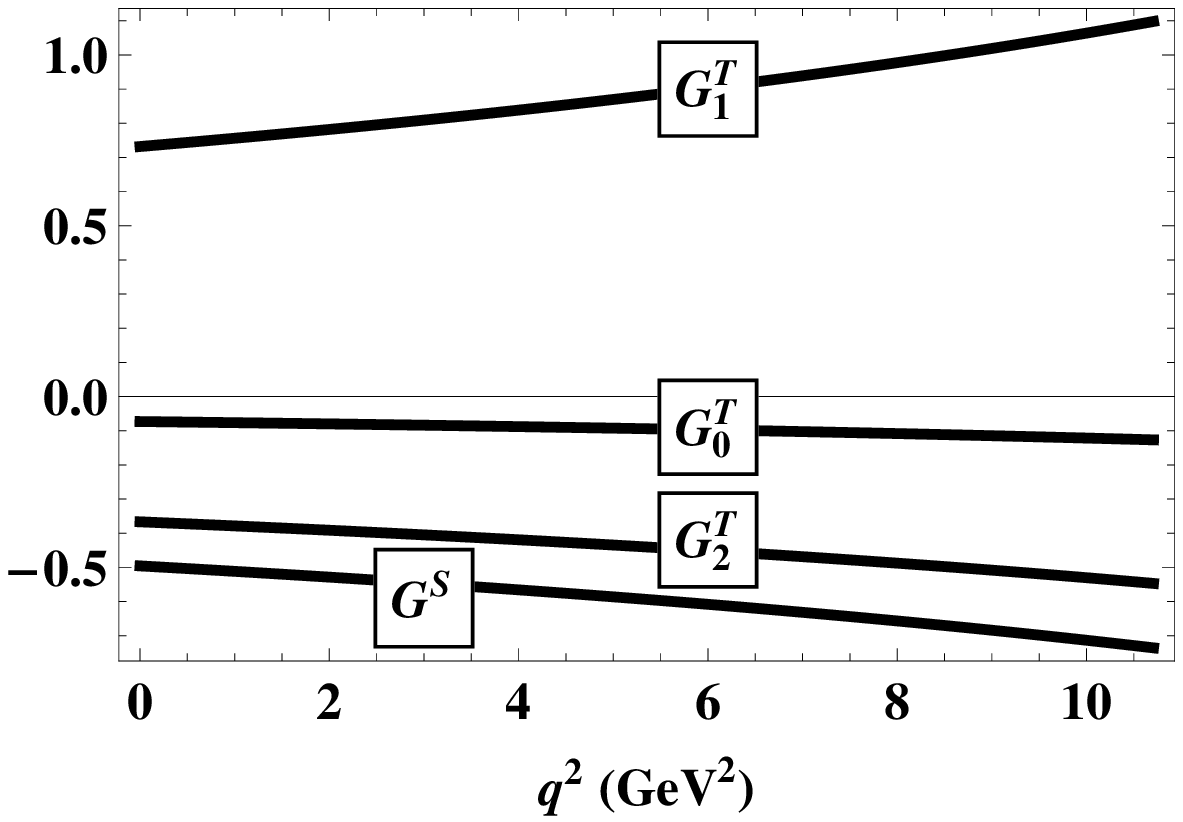}
\end{tabular}
\caption{Form factors of the transitions $\bar{B}^0 \to D$ (upper panels)
  and $\bar{B}^0 \to D^\ast$ (lower panels).}
\label{fig:formfactor}
\end{figure}

In Fig.~\ref{fig:formfactor} we present the $q^2$ dependence of 
the $\bar{B}^0 \to D^{(\ast)}$ transition form factors in the full momentum 
transfer range $0\le q^2 \le q^2_{max}=(m_{\bar{B}^0}-m_{D^{(\ast)}})^2$.

%

Finally, we briefly discuss some error estimates within our model.
We fix our model parameters (the constituent quark masses, the infrared cutoff
and the hadron size parameters) by minimizing the functional
$\chi^2 = \sum\limits_i\frac{(y_i^{\rm expt}-y_i^{\rm theor})^2}{\sigma^2_i}$
where $\sigma_i$ is the experimental  uncertainty.
If $\sigma$ is too small then we take its value of 10$\%$.
Moreover, we observed that the errors of the fitted parameters 
are of the order of  10$\%$.
Thus we estimate the model uncertainties within 10$\%$.
%
%

\section{Twofold distribution and experimental constraints}
\label{sec:constraint}
\subsection{$\bar{B}^0 \to D$ transition}
Using the helicity technique described in our recent paper~\cite{Ivanov:2015tru} one can write the helicity amplitude of the decay $\bar{B}^0 \to D\tau^-\bar\nu_{\tau}$ as
\bea
\sum\limits_{pol} |\mathcal{M}|^2 &=& \frac{G_F^2 |V_{cb}|^2}{2}\Big\lbrace |1+V_L+V_R|^2 \sum\limits_{hel} H(m) H^\dag(n) L_1(m',n')g_{mm'}g_{nn'}\nn
&&+|S_L+S_R|^2|H_P^S|^2 L_2\nn
&&+|T_L|^2\sum\limits_{hel}H_T(m,n)H_T^\dag(u,v)L_3(m',n',u',v')g_{mm'}g_{nn'}g_{uu'}g_{vv'}\nn
&&+(S_L+S_R)^\dag H_P^{\dag S}\sum\limits_{hel}H(m)L_4(m')g_{mm'}\nn
&&+(S_L+S_R)H_P^S\sum\limits_{hel}H^\dag(m)L_5(m')g_{mm'}\nn
&&+T_L^\dag \sum\limits_{hel}H_T^\dag(m,n)H(u)L_6(m',n',u')g_{mm'}g_{nn'}g_{uu'}\nn
&&+T_L\sum\limits_{hel}H_T(m,n)H^\dag(u)L_7(m',n',u')g_{mm'}g_{nn'}g_{uu'}
\Big\rbrace,
\label{eq:BDhelamp}
\ena
where the notation $\ds\sum\limits_{hel}$ means all helicity indices appearing in the expression under the symbol are summed up. The hadronic and leptonic helicity amplitudes in Eq.~(\ref{eq:BDhelamp}) are defined as follows:
\be
\begin{aligned}
H(m) &= \epsilon_\mu^\dag(m)T_1^\mu,\qquad
H_P^S = T_2,\qquad
H_T(m,n) = i\epsilon_\mu^\dag(m)\epsilon_\nu^\dag(n)T_3^{\mu\nu},\nn
L_1(m,n)&=\epsilon_\mu(m)\epsilon_\alpha^\dag(n)\Tr[(\not{k_1}+m_\tau)\gamma^\mu (1-\gamma_5) \not{k_2} (1+\gamma_5)\gamma^\alpha],\nn
L_2&=\Tr[(\not{k_1}+m_\tau)(1-\gamma_5) \not{k_2} (1+\gamma_5)],\nn
L_3(m,n,r,s)&=\epsilon_\mu(m)\epsilon_\nu(n)\epsilon_\alpha^\dag(r)\epsilon_\beta^\dag(s)\Tr[(\not{k_1}+m_\tau)\sigma^{\mu\nu} (1-\gamma_5) \not{k_2} (1+\gamma_5)\sigma^{\alpha\beta}],\nn
L_4(m)&=\epsilon_\mu(m)\Tr[(\not{k_1}+m_\tau)\gamma^\mu (1-\gamma_5) \not{k_2} (1+\gamma_5)],\nn
L_5(m)&=\epsilon_\alpha^\dag(m) \Tr[(\not{k_1}+m_\tau) (1-\gamma_5) \not{k_2} (1+\gamma_5)\gamma^\alpha],\nn
L_6(m,n,r)&=i\epsilon_\alpha^\dag(m)\epsilon_\beta^\dag(n)\epsilon_\mu(r)\Tr[(\not{k_1}+m_\tau)\gamma^\mu (1-\gamma_5) \not{k_2} (1+\gamma_5)\sigma^{\alpha\beta}],\nn
L_7(m,n,r)&=-i\epsilon_\mu(m)\epsilon_\nu(n)\epsilon_\alpha^\dag(r) \Tr[(\not{k_1}+m_\tau)\sigma^{\mu\nu} (1-\gamma_5) \not{k_2} (1+\gamma_5)\gamma^\alpha].
\end{aligned}
\label{eq:lephel}
\en
One obtains the following nonzero hadronic helicity amplitudes written in terms of the invariant form factors:
\bea
H_t   &\equiv& H(t)=\frac{1}{\sqrt{q^2}}(Pq F_+ + q^2 F_-),
\qquad
H_0 \equiv H(0)=\frac{2m_1|{\bf p_2}|}{\sqrt{q^2}}F_+,\nn
H_P^S &\equiv& (m_1+m_2)F^S,\nn
H_T &\equiv& H_T(t,0) = -H_T(0,t) = \pm H_T(\mp,\pm)=\frac{2m_1|{\bf p_2}|}{m_1+m_2}F^T.
\label{eq:hel_pp}
\ena
The differential $(q^2,\cos\theta)$ distribution is written as
\bea
\lefteqn{\frac{d^2\Gamma(\bar{B}^0\to D\tau^-\bar{\nu}_\tau)}{dq^2d\cos\theta}}\nn
&=&\frac{G_F^2|V_{cb}|^2|{\bf p_2}|q^2 v^2}{(2\pi)^3 16m_1^2}\nn
&&\times 
\Big\lbrace|1+V_L+V_R|^2\left[ |H_0|^2\sin^2\theta +2\delta_\tau |H_t-H_0\cos\theta|^2 \right]\nn
&&+|S_L+S_R|^2|H_P^S|^2
+16|T_L|^2\left[2\delta_\tau+( 1-2\delta_\tau)\cos^2\theta \right]|H_T|^2\nn
&&+2\sqrt{2\delta_\tau} {\rm Re}(S_L+S_R) H_P^S\left[ H_t-H_0\cos\theta \right]\nn
&&+8\sqrt{2\delta_\tau} {\rm Re}T_L\left[ H_0-H_t\cos\theta \right]H_T
\Big\rbrace,
\label{eq:distr2D}
\ena
where we have introduced the 
velocity-type parameter $v=1-m_\tau^2/q^2$ as well as the helicity flip factor 
$\delta_\tau = m^2_\tau/2q^2$. After integrating over $\cos\theta$ one has
\bea
\lefteqn{\frac{d\Gamma(\bar{B}^0\to D\tau^-\bar{\nu}_\tau)}{dq^2}}\nn
&=&\frac{G_F^2|V_{cb}|^2|{\bf p_2}|q^2v^2}{(2\pi)^3 12m_1^2}\nn
&&\times 
\Big\lbrace |1+V_L+V_R|^2\left[(1+ \delta_\tau) |H_0|^2+3\delta_\tau|H_t|^2 \right]\nn
&&+\frac{3}{2}|S_L+S_R|^2 |H_P^S|^2
+8|T_L|^2 ( 1+4\delta_\tau) |H_T|^2\nn
&&+ 3\sqrt{2\delta_\tau} {\rm Re} (S_L+S_R) H_P^S H_t
+12\sqrt{2\delta_\tau} {\rm Re}T_L H_0H_T
\Big\rbrace.
\label{eq:distr1D}
\ena
This $q^2$ distribution agrees with the result of Ref.~\cite{Sakaki:2013bfa}. Note that in this paper we do not consider interference terms between different NP operators.  
\subsection{$\bar{B}^0 \to D^\ast$ transition}
The helicity amplitude of the decay $\bar{B}^0 \to D^\ast\tau^-\bar\nu_{\tau}$ is written as
\bea
\sum\limits_{pol} |\mathcal{M}|^2 =&& \frac{G_F^2 |V_{cb}|^2}{2}\nn
&&\times\Big\lbrace \sum\limits_{hel}\Big[|1+V_L|^2 H_L(m,r)H_L^\dag(n,r)+|V_R|^2H_R(m,r)H_R^\dag(n,r)\nn
&&+V_R^\dag H_L(m,r)H_R^\dag(n,r)+V_RH_R(m,r)H_L^\dag(n,r)\Big] L_1(m',n')g_{mm'}g_{nn'}\nn
&&+\sum\limits_{hel}|S_R-S_L|^2 |H^S_V(r)|^2 L_2\nn
&&+\sum\limits_{hel}|T_L|^2  H_T(m,n,r)H_T^\dag(u,v,r)L_3(m',n',u',v')g_{mm'}g_{nn'}g_{uu'}g_{vv'}\nn
&&+\sum\limits_{hel}(S_R-S_L)^\dag H_L(m,r)H^{S\dag}_V(r)  L_4(m')g_{mm'}\nn
&&+\sum\limits_{hel}(S_R-S_L) H_L^\dag(m,r)H^S_V(r)  L_5(m')g_{mm'}\nn
&&+\sum\limits_{hel}T_L^\dag H_T^\dag(m,n,r)H_L(u,r)L_6(m',n',u')g_{mm'}g_{nn'}g_{uu'}\nn
&&+\sum\limits_{hel}T_L H_T(m,n,r)H_L^\dag(u,r)L_7(m',n',u')g_{mm'}g_{nn'}g_{uu'}
\Big\rbrace,
\ena
where we have defined the hadronic helicity amplitudes as follows:
\bea
H_L(m,r) &=& \epsilon_\mu^\dag(m){\epsilon_2}_\alpha^\dag(r)\mathcal{T}_{1L}^{\mu\alpha},\qquad
H_R(m,r) = \epsilon_\mu^\dag(m){\epsilon_2}_\alpha^\dag(r)\mathcal{T}_{1R}^{\mu\alpha},
\nn
H_T(m,n,r) &=& i\epsilon_\mu^\dag(m)\epsilon_\nu^\dag(n){\epsilon_2}_\alpha^\dag(r)\mathcal{T}_3^{\mu\nu\alpha},\qquad
H^S_V(r) = \epsilon^\dag_{2\alpha}(r)\mathcal{T}_2^\alpha.
\ena
The nonzero helicity amplitudes read
\bea
H_{00} &\equiv & H_L(0,0) = -H_R(0,0) = \frac{-Pq(m_1^2 - m_2^2 - q^2)A_0 + 4m_1^2|{\bf p_2}|^2 A_+}{2m_2\sqrt{q^2}(m_1+m_2)} 
,\nn
H_{t0} &\equiv & H_L(t,0) = -H_R(t,0) = \frac{m_1|{\bf p_2}|\left(Pq(-A_0+A_+)+q^2 A_-\right)}{m_2\sqrt{q^2}(m_1+m_2)},
\nn
H_{\pm\pm} &\equiv & H_L(\pm,\pm) = -H_R(\mp,\mp) = \frac{-Pq A_0\pm 2m_1|{\bf p_2}| V}{m_1+m_2},\nn
H^S_V &\equiv& H^S_V(0)=\frac{m_1}{m_2}|{\bf p_2}|G^S, \nn
H_T^\pm &\equiv & H_T(\pm,t,\pm) = \pm H_T(\pm,0,\pm)=-H_T(t,\pm,\pm)=\mp H_T(0,\pm,\pm)\nn
&=& -\frac{1}{\sqrt{q^2}}\left[
(m_1^2-m_2^2\pm\lambda^{1/2}(m_1^2,m_2^2,q^2))G_1^T+q^2G_2^T
\right],\nn
H_T^0 &\equiv &H_T(0,t,0) = H_T(+,-,0) = -H_T(t,0,0) = -H_T(-,+,0)\nn
 &=& -\frac{1}{2m_2}
\Big[
(m_1^2+3m_2^2-q^2)G_1^T +(m_1^2-m_2^2-q^2)G_2^T-\frac{\lambda(m_1^2,m_2^2,q^2)}{(m_1+m_2)^2}G_0^T
\Big].
\ena
The differential $(q^2,\cos\theta)$ distribution of the decay $\bar{B}^0\to D^\ast\tau^-\bar{\nu}_\tau$ is given in Appendix~\ref{app:twofold}. After integrating over $\cos\theta$ one has
\bea
\lefteqn{\frac{d\Gamma(\bar{B}^0\to D^\ast\tau^-\bar{\nu}_\tau)}{dq^2}}\nn
&=&\frac{G_F^2|V_{cb}|^2|{\bf p_2}|q^2v^2}{(2\pi)^3 12m_1^2}\nn
&&\times 
\Big\lbrace(|1+V_L|^2+|V_R|^2)\left[(1+\delta_\tau) (|H_{00}|^2+|H_{++}|^2+|H_{--}|^2)+3\delta_\tau |H_{t0}|^2 \right]\nn
&&-2 {\rm Re}V_R\left[(1+\delta_\tau) (|H_{00}|^2+2H_{++}H_{--})+3\delta_\tau |H_{t0}|^2 \right]\nn
&&+\frac{3}{2}|S_R-S_L|^2|H^S_V|^2
+3\sqrt{2\delta_\tau} {\rm Re}(S_R-S_L) H^S_V H_{t0}\nn
&&+8|T_L|^2 (1+4\delta_\tau)(|H_T^0|^2+|H_T^+|^2+|H_T^-|^2)\nn
&&-12\sqrt{2\delta_\tau} {\rm Re}T_L (H_{++}H_T^+ + H_{--}H_T^- + H_{00}H_T^0)
\Big\rbrace.
\label{eq:distr1Dv}
\ena
This $q^2$ distribution agrees with the result of Ref.~\cite{Sakaki:2013bfa}. Note that in this paper we do not consider interference terms between different NP operators.  
\subsection{Experimental constraints}
Assuming that NP only affects the tau modes, we integrate Eqs.~(\ref{eq:distr1D}) and~(\ref{eq:distr1Dv}) and obtain the ratios of branching fractions $R(D^{(\ast)})$ in the presence of NP operators. It is important to note that within the SM (without any NP operators) our model calculation yields
\[
R(D)=0.267,\qquad R(D^\ast)=0.238,
\]
which are consistent with other SM predictions given in Refs.~\cite{Na:2015kha, Fajfer:2012vx, Kamenik:2008tj, Lattice:2015rga} within $10\%$.
In order to acquire the allowed regions for the NP Wilson coefficients, we assume that besides the SM contribution, only one of the NP operators in Eq.~(\ref{eq:Heff}) is switched on at a time. We then compare the calculated ratios $R(D^{(\ast)})$ with the recent experimental data from the Belle, \textit{BABAR}, and LHCb collaborations~\cite{Lees:2012xj,Huschle:2015rga,Aaij:2015yra,Sato:2016svk,Abdesselam:2016xqt} given in Eq.~(\ref{eq:RD-expt}). We also take into account a theoretical error of $10\%$ for the ratios $R(D^{(\ast)})$. 

\begin{figure}[htbp]
\begin{tabular}{lr}
\includegraphics[scale=0.45]{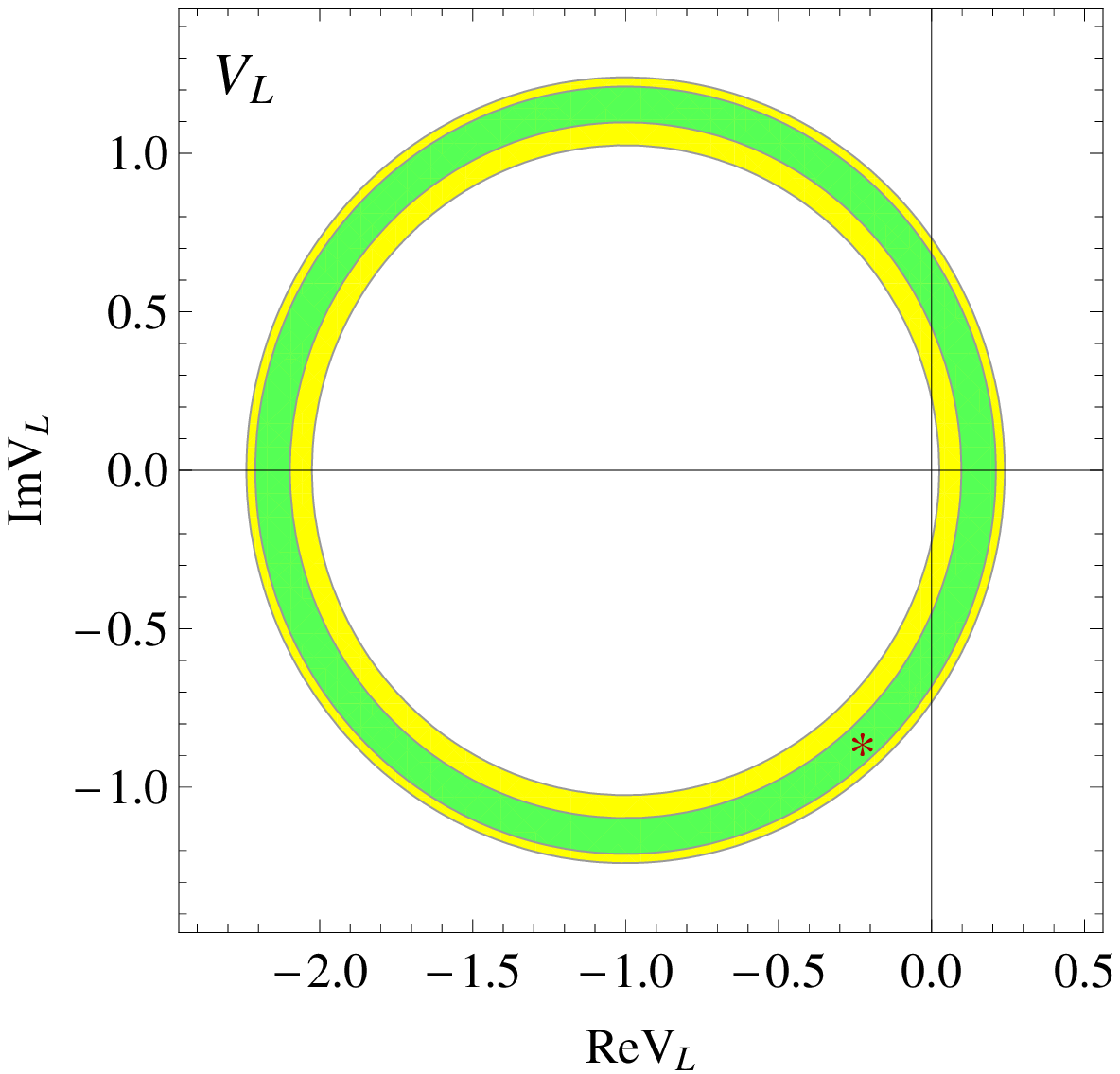}
& 
\includegraphics[scale=0.45]{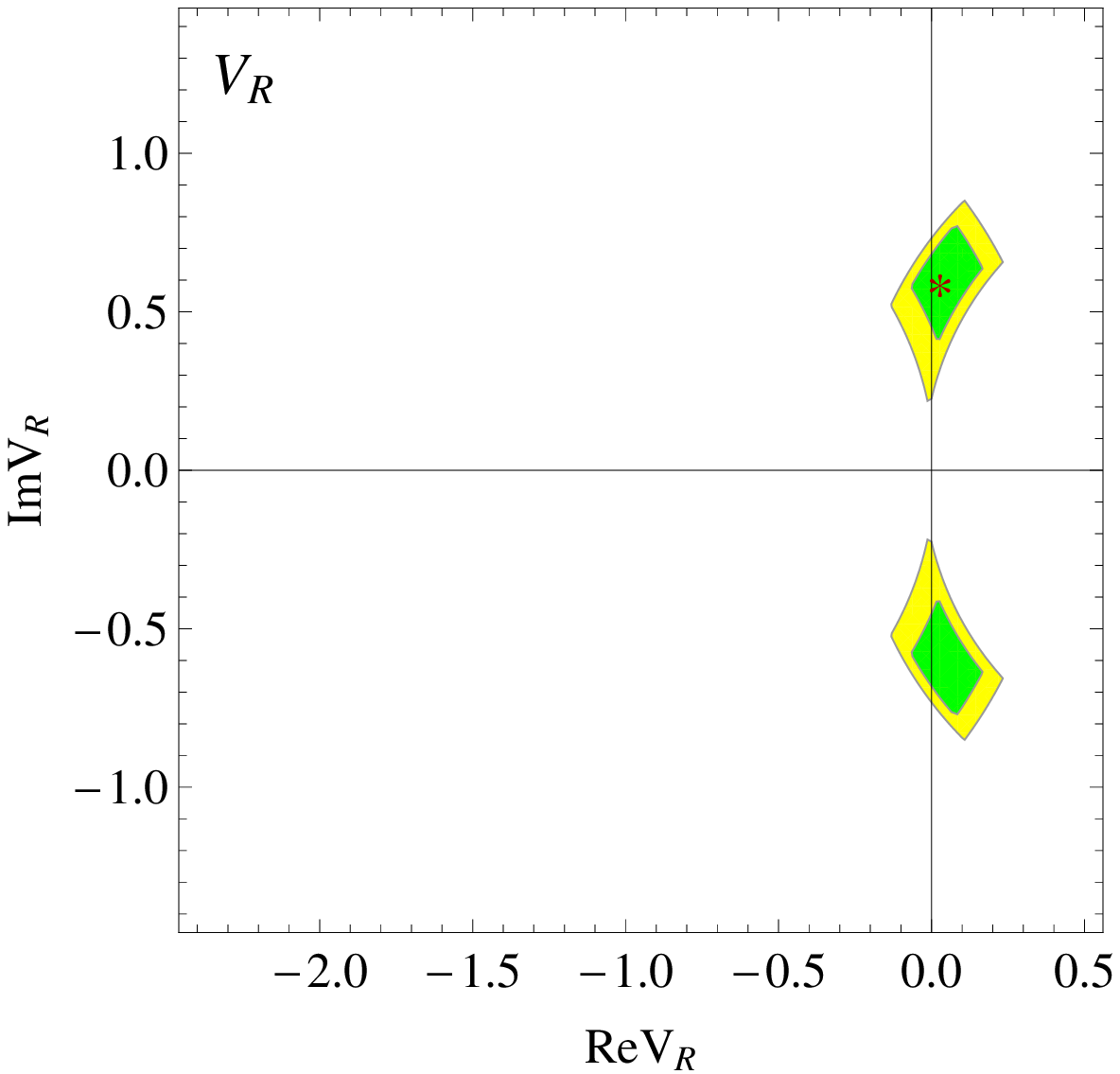}\\
\includegraphics[scale=0.45]{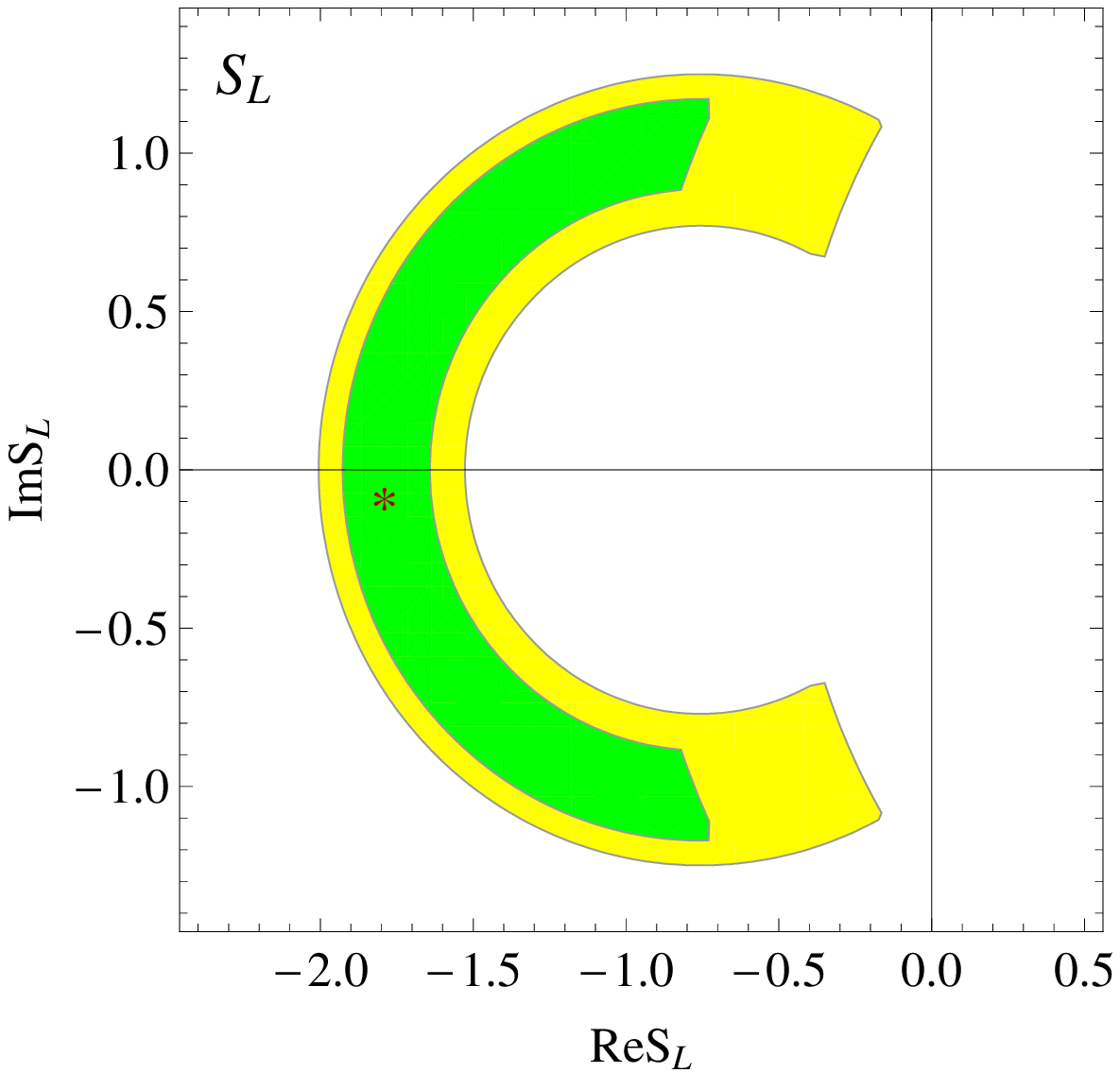}
& 
\includegraphics[scale=0.45]{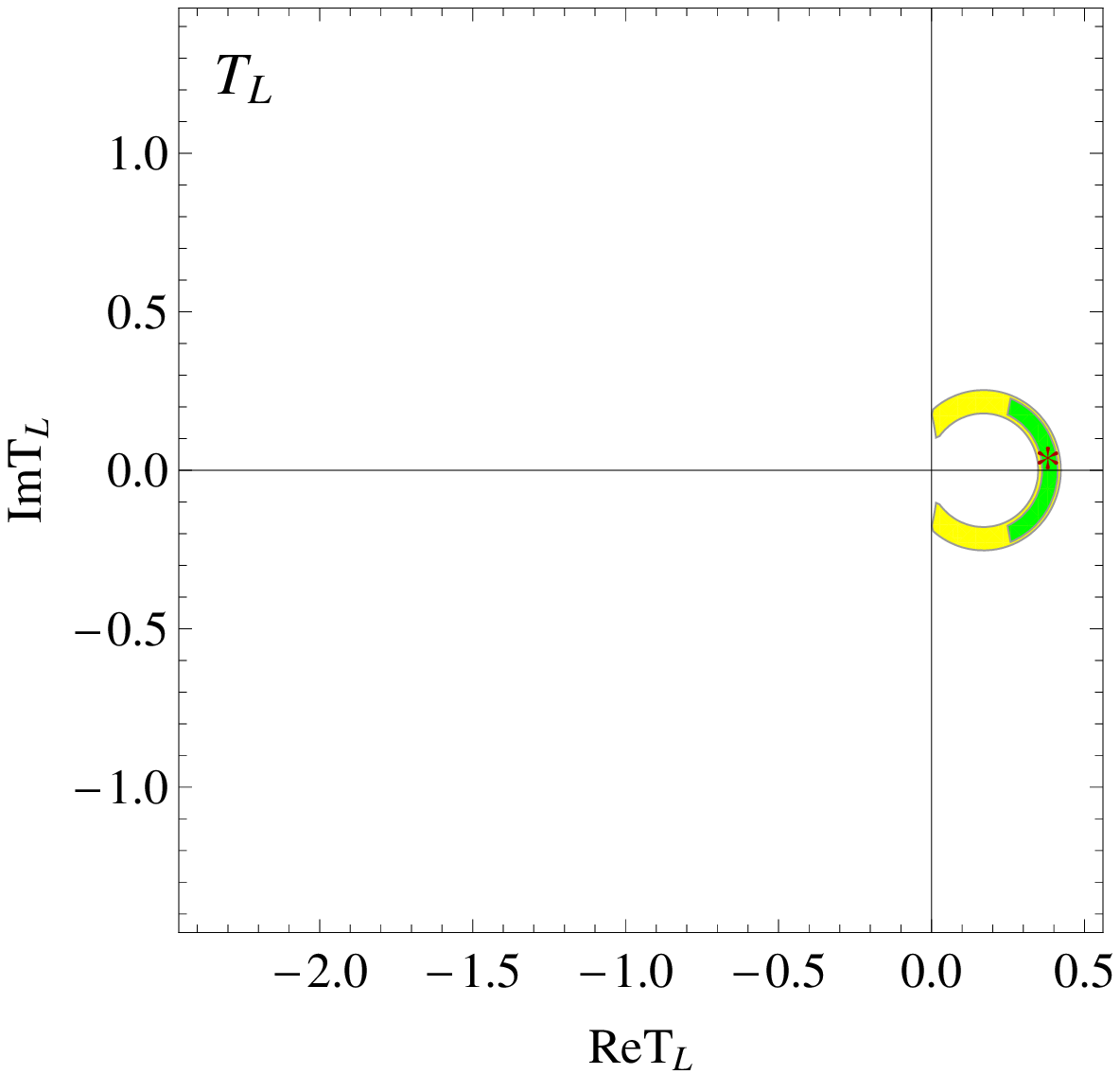}
\end{tabular}
\caption{ The allowed regions of the Wilson coefficients $V_{L,R}$, $S_L$, and 
$T_L$ within $1\sigma$ (green, dark) and $2\sigma$ (yellow, light). The best-fit value in each case
is denoted with the symbol $\ast$. The coefficient $S_R$ is disfavored at 
$2\sigma$ and therefore  is not shown here.}
\label{fig:constraint}
\end{figure}

The experimental constraints are shown in Fig.~\ref{fig:constraint}. The vector operators $\mathcal{O}_{V_{L,R}}$ and the left scalar operator $\mathcal{O}_{S_L}$ are favored while there is no allowed region for the right scalar operator $\mathcal{O}_{S_R}$ within $2\sigma$. Therefore we will not consider $\mathcal{O}_{S_R}$ in what follows. The tensor operator $\mathcal{O}_{T_L}$ is less favored, but it can still well explain the current experimental results. The stringent constraint on the tensor coupling mainly comes from the experimental data of $R(D^\ast)$. 

In each allowed region at $2\sigma$ we find the best-fit value for each NP coupling. The best-fit couplings read
\bea
\begin{aligned}
V_L &=-0.23-i0.85, \qquad & V_R &=0.03+i0.60,\\
S_L &=-1.80-i0.07, \qquad & T_L &=0.38+i0.06,
\end{aligned}
\label{eq:bestfit}
\ena
and are marked with an asterisk.
The allowed regions of the coupling coefficients are then used to analyze the effect of the NP operators on different physical observables defined in the next section. 

\section{The cascade decay 
$\bar{B}^0\to D^{\ast+}(\to D^0\pi^+)\tau^-\bar{\nu}_\tau$ and 
the angular observables}
\label{sec:NP}
 \begin{figure}[htbp]
\begin{center}
\epsfig{figure=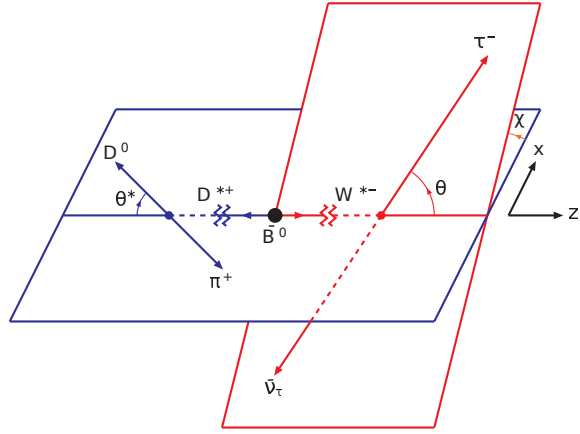,scale=.4}
\caption{Definition of the angles $\theta$, $\theta^\ast$ and $\chi$ in
the cascade decay $\bar B^0\to D^{\ast+}(\to D^0\pi^+)\tau^-\bar\nu_\tau$. See text for more details.}
\label{fig:bdangl}
\end{center}
\end{figure}
\subsection{The fourfold distribution}
In order to analyze NP effects on the polarization of the $D^\ast$ meson one uses the cascade decay $\bar{B}^0\to D^{\ast+}(\to D^0\pi^+)\tau^-\bar{\nu}_\tau$. A detailed derivation of the fourfold angular distribution (without NP) can be found in our paper~\cite{Faessler:2002ut}. The three angles $\theta$, $\theta^\ast$, and $\chi$ in the distribution are defined in Fig.~\ref{fig:bdangl}~\cite{Kang:2013jaa,Duraisamy:2014sna,Ivanov:2015tru}. Here, $\theta$ is the polar angle between the $\tau^-$ three-momentum and the direction opposite to the $D^*$ meson in the $W^-$ rest frame, and $\theta^\ast$ is the polar angle between the three-momentum of the final meson $D^0$ and the direction of the $D^*$ meson in the $(D^0\pi^+)$ rest frame. The $\bar{B}^0 \to D^{\ast+}\tau^-\bar\nu_{\tau}$ decay plane is spanned by the three-momenta of $\tau^-$ and $\bar\nu_{\tau}$ while the $D^{\ast+}\to D^0\pi^+$ decay plane is spanned by the three-momenta of the mesons $D^0$ and $\pi^+$. And finally, $\chi$ is the azimuthal angle between the two planes. We choose a right-handed $xyz$ coordinate system in the $W^-$ rest frame such that the $z$ axis is opposite to the direction of the mesons $\bar{B}^0$ and $D^\ast$, and the three-momentum of $\tau^-$ lies in the $(xz)$ plane.

One has~\cite{Feldmann:2015xsa}
\be
\frac{d^4\Gamma(\bar{B}^0\to D^{\ast+}(\to D^0\pi^+) \tau^-\bar\nu_\tau)}
     {dq^2 d\cos\theta d\chi d\cos\theta^\ast} 
=\frac{9}{8\pi}|N|^2J(\theta,\theta^\ast,\chi),
\label{eq:distr4}
\en
where
\be
|N|^2=
\frac{G_F^2 |V_{cb}|^2 |{\bf p_2}| q^2 v^2}{(2\pi)^3 12 m_1^2}\mathcal{B}(D^\ast\to D\pi).
\en
The full angular distribution $J(\theta,\theta^\ast,\chi)$ is written as
\bea
\lefteqn{J(\theta,\theta^\ast,\chi)}\nn
&=& J_{1s}\sin^2\theta^\ast + J_{1c}\cos^2\theta^\ast
+(J_{2s}\sin^2\theta^\ast + J_{2c}\cos^2\theta^\ast)\cos2\theta\nn
&&+J_3\sin^2\theta^\ast \sin^2\theta \cos2\chi
+J_4\sin2\theta^\ast \sin2\theta \cos\chi\nn
&&+J_5\sin2\theta^\ast \sin\theta \cos\chi
+(J_{6s}\sin^2\theta^\ast+J_{6c}\cos^2\theta^\ast)\cos\theta\nn
&&+J_7\sin2\theta^\ast \sin\theta \sin\chi
+J_8\sin2\theta^\ast \sin2\theta \sin\chi
+J_9\sin^2\theta^\ast \sin^2\theta \sin2\chi ,
\ena
where $J_{i(a)}$ $(i=1,\dots,9; a=s,c)$ are the angular observables. Their explicit expressions in terms of helicity amplitudes and Wilson coefficients read
\newpage
\bea
4J_{1s} &=&
\frac{3+2\delta_\tau}{4} (|1+V_L|^2+|V_R|^2)(|H_{++}|^2+|H_{--}|^2)-(3+2\delta_\tau) {\rm Re}V_RH_{++}H_{--}\nn
&&-8\sqrt{2\delta_\tau} {\rm Re}T_L (H_{++}H_T^+ +H_{--}H_T^-)+4(1+6\delta_\tau) |T_L|^2 (|H_T^+|^2+|H_T^-|^2)
,\nn
4J_{1c} &=&2|S_R-S_L|^2 |H^S_V|^2+4\sqrt{2\delta_\tau} {\rm Re} (S_R-S_L) H^S_V H_{0t}\nn
&&+(|1+V_L|^2+|V_R|^2-2{\rm Re}V_R)\Big[4\delta_\tau |H_{0t}|^2+\Big(1+2\delta_\tau\Big)|H_{00}|^2\Big]\nn
&&-16\sqrt{2\delta_\tau} {\rm Re}T_L H_{00}H_T^0+16(1+2\delta_\tau)|T_L|^2|H_T^0|^2,\nn
4J_{2s} &=&\frac{1}{4}(1-2\delta_\tau)\Big[
 (|1+V_L|^2+|V_R|^2)(|H_{++}|^2+|H_{--}|^2)\nn
&&-4{\rm Re}V_RH_{++}H_{--} -16|T_L|^2(|H_T^+|^2 + |H_T^-|^2)
\Big],\nn
4J_{2c} &=&(1-2\delta_\tau)\Big[-
(|1+V_L|^2+|V_R|^2-2{\rm Re}V_R)|H_{00}|^2+16|T_L|^2|H_T^0|^2\Big],\nn
4J_3 &=&(1-2\delta_\tau)\Big[-(|1+V_L|^2+|V_R|^2)H_{++}H_{--}\nn
&&+{\rm Re}V_R(|H_{++}|^2+|H_{--}|^2)+16|T_L|^2H_T^+H_T^-
\Big],\nn
4J_4 &=&\frac{1}{2}(1-2\delta_\tau)\Big[
(|1+V_L|^2+|V_R|^2-2{\rm Re}V_R)H_{00}(H_{++}+H_{--})\nn
&&-16|T_L|^2H_T^0(H_T^++H_T^-)
\Big],\nn
4J_5 &=&
(|1+V_L|^2-|V_R|^2)H_{00}(H_{--}-H_{++})\nn
&&+2\delta_\tau(|1+V_L|^2+|V_R|^2-2{\rm Re}V_R)H_{t0}(H_{++}+H_{--})\nn
&&+\sqrt{2\delta_\tau}
{\rm Re}(S_R-S_L)H^S_V(H_{++}+H_{--})\nn
&&+4\sqrt{2\delta_\tau}{\rm Re}T_L[H_T^0(H_{++}-H_{--})-H_T^-(H_{00}+H_{t0})-H_T^+(H_{t0}-H_{00})]\nn
&&-32\delta_\tau |T_L|^2H_T^0(H_T^+ -H_T^-),\nn
4J_{6s} &=&
(|1+V_L|^2-|V_R|^2)(|H_{--}|^2-|H_{++}|^2)\nn
&&+8\sqrt{2\delta_\tau} {\rm Re}T_L(H_{++}H_T^+ -H_{--}H_T^-)-32\delta_\tau|T_L|^2(|H_T^+|^2-|H_T^-|^2),\nn
4J_{6c} &=&
-8\delta_\tau(|1+V_L|^2+|V_R|^2-2{\rm Re}V_R)H_{t0}H_{00}\nn
&&-4\sqrt{2\delta_\tau}{\rm Re}(S_R-S_L)H^S_V H_{00}+16\sqrt{2\delta_\tau}{\rm Re}T_L H_{t0}H_T^0,\nn
4J_7 &=& 
\sqrt{2\delta_\tau}{\rm Im}(S_R-S_L)H^S_V(H_{++}-H_{--})
-4\delta_\tau {\rm Im}V_R H_{t0}(H_{++}-H_{--})\nn
&&+4\sqrt{2\delta_\tau} {\rm Im}T_L [H_T^0(H_{++}+H_{--})-H_T^-(H_{t0}+H_{00})+H_T^+(H_{t0}-H_{00})],\nn
4J_8 &=& (1-2\delta_\tau) {\rm Im}V_R  H_{00}(H_{--}-H_{++}),\nn
4J_9 &=&(1-2\delta_\tau) {\rm Im}V_R  (|H_{++}|^2-|H_{--}|^2).
\label{eq:angular}
\ena
\newpage
The results for the angular functions $J_i$ in Eq.~(\ref{eq:angular}) agree with those of Ref.~\cite{Duraisamy:2014sna} except for a difference in the sign of $J_8$, $J_9$, and the first two terms of $J_7$. However, we find agreement with the results of our previous paper~\cite{Ivanov:2015tru} in the case of $J_7$, $J_8$, and $J_9$. Note again that in this paper we do not consider interference terms between different NP operators. In our quark model all helicity amplitudes are real, which implies the
vanishing of all terms proportional to $\sin\chi$ and $\sin2\chi$, namely $J_{7,8,9}$, within the SM. This does not necessarily hold when considering complex 
NP Wilson coefficients, as can be seen in Eq.~(\ref{eq:angular}).   
\subsection{The $q^2$ distribution and the ratios of branching fractions $R(D^{(\ast)})$}
\begin{figure}[htbp]
\begin{tabular}{lr}
\includegraphics[scale=0.4]{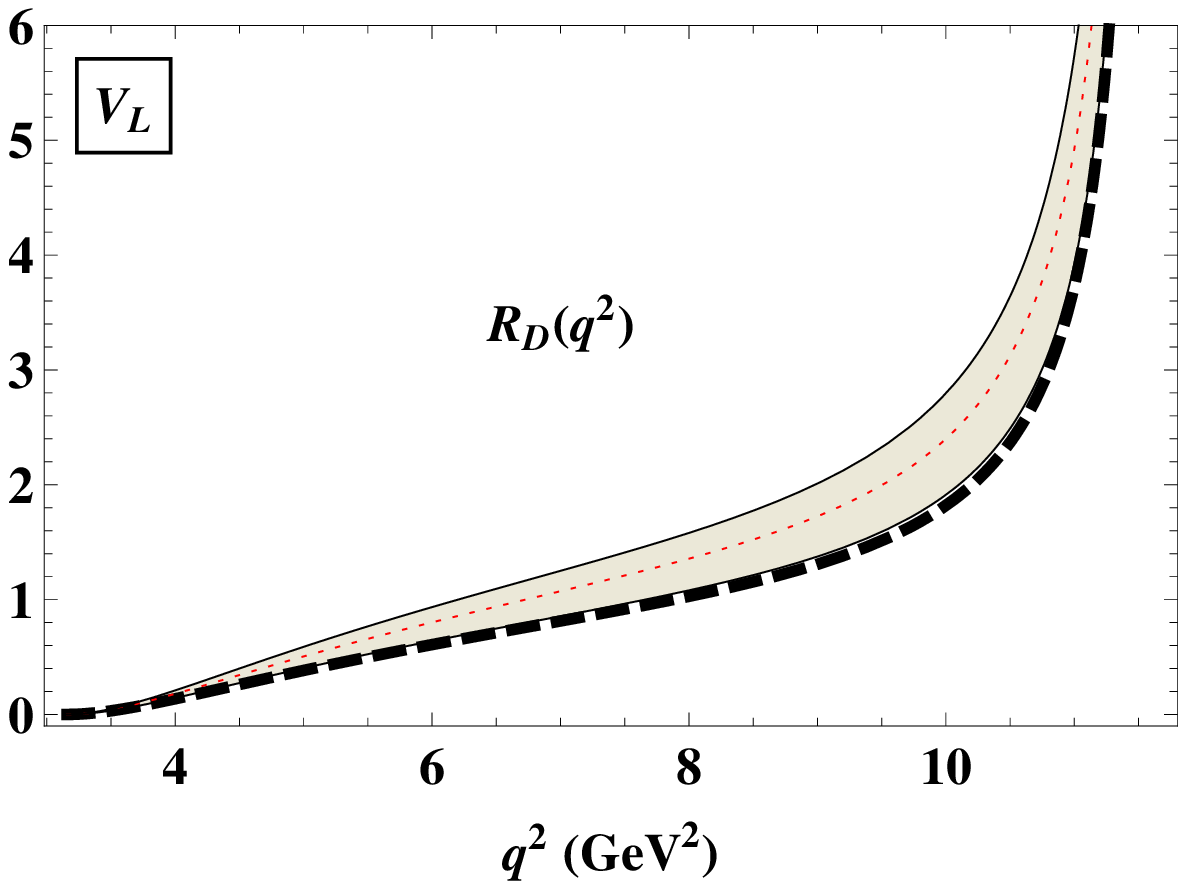}&
\includegraphics[scale=0.4]{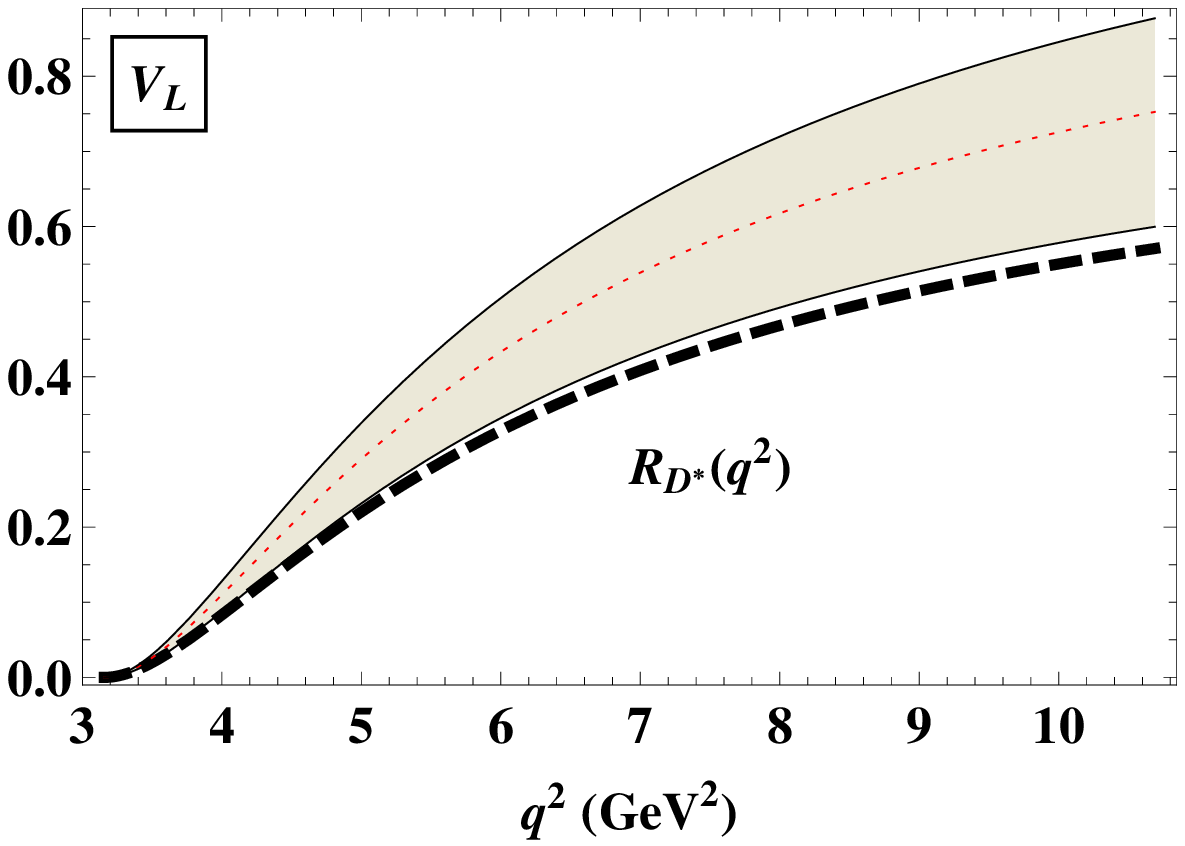}\\
\includegraphics[scale=0.4]{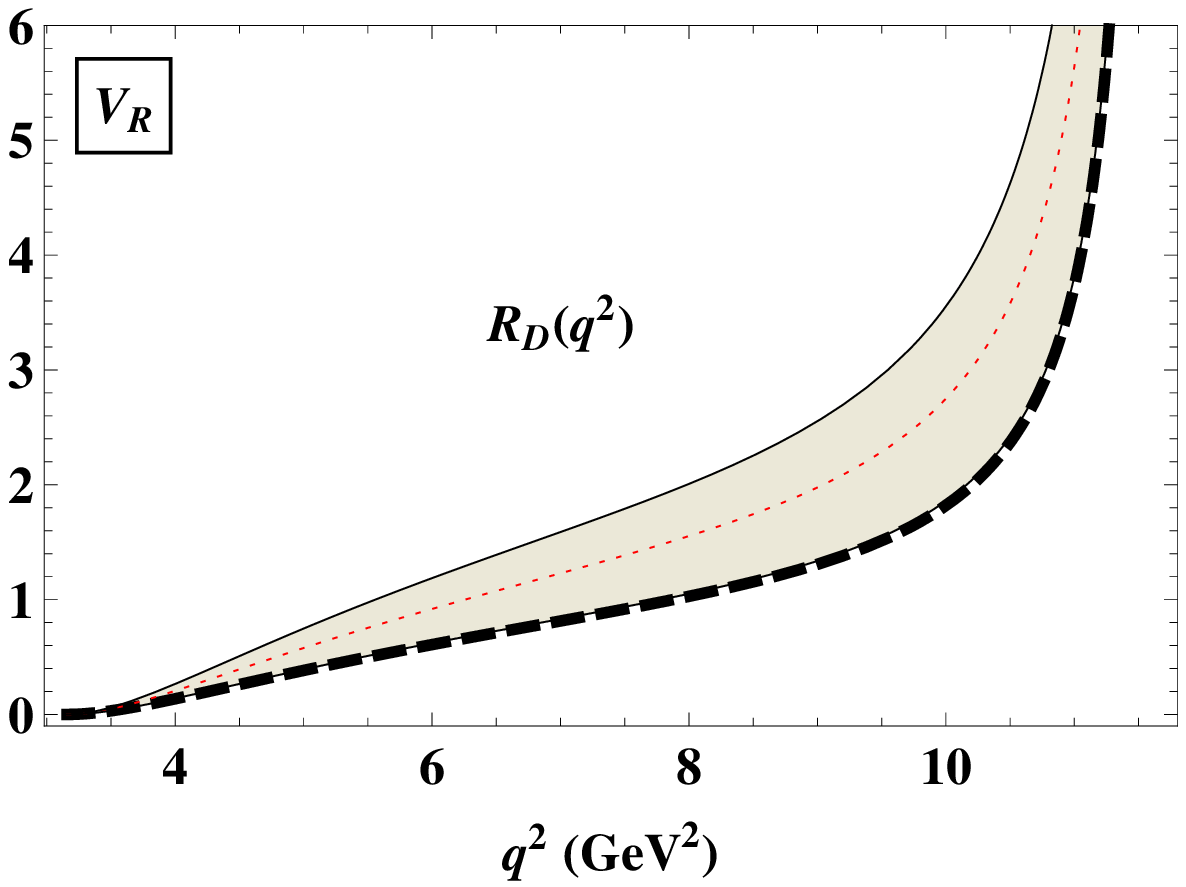}&
\includegraphics[scale=0.4]{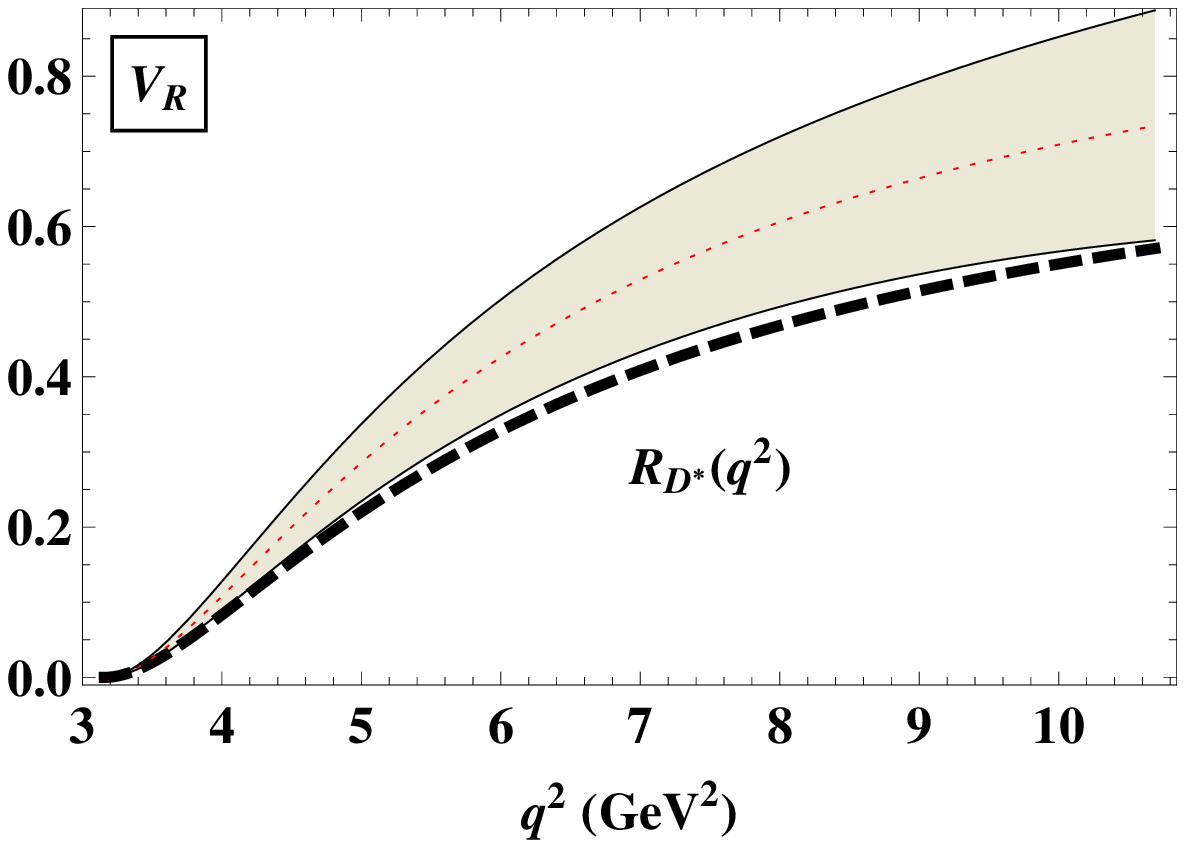}\\
\includegraphics[scale=0.4]{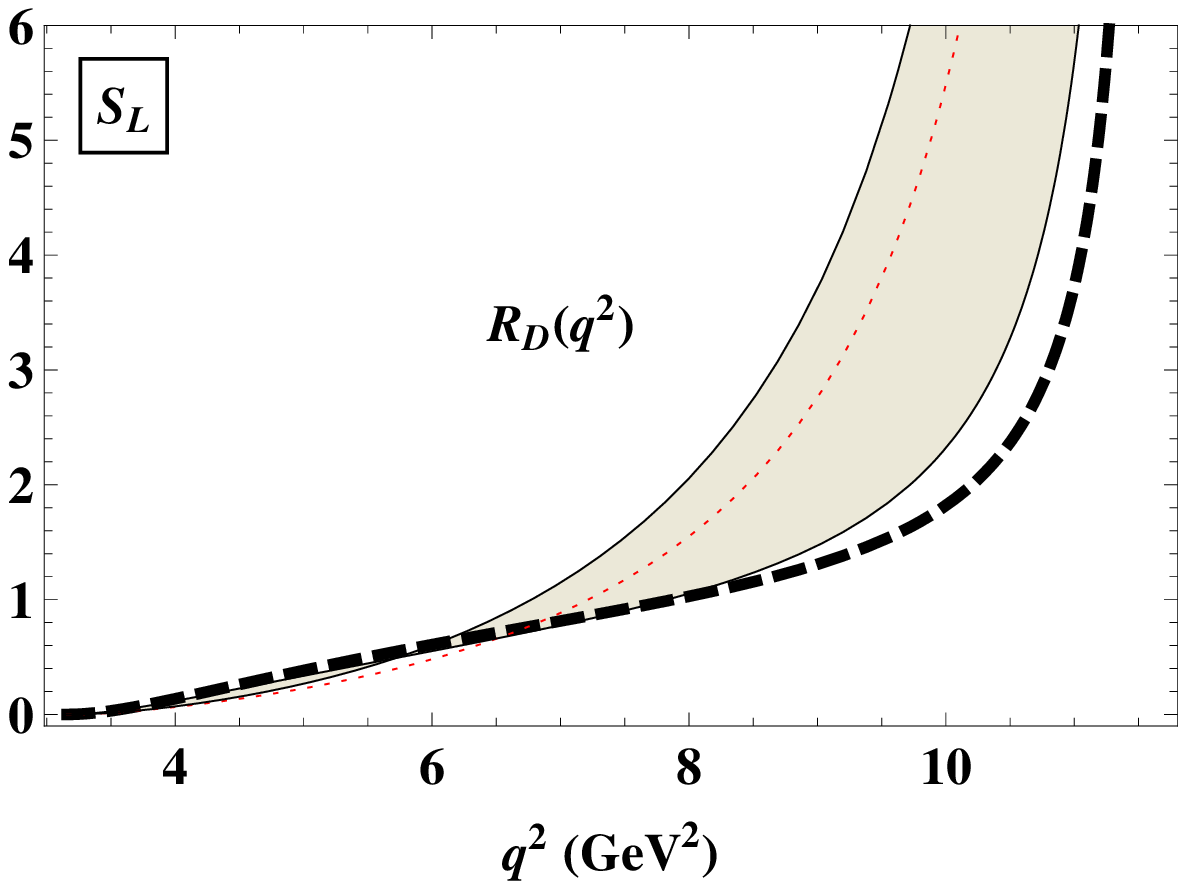}&
\includegraphics[scale=0.4]{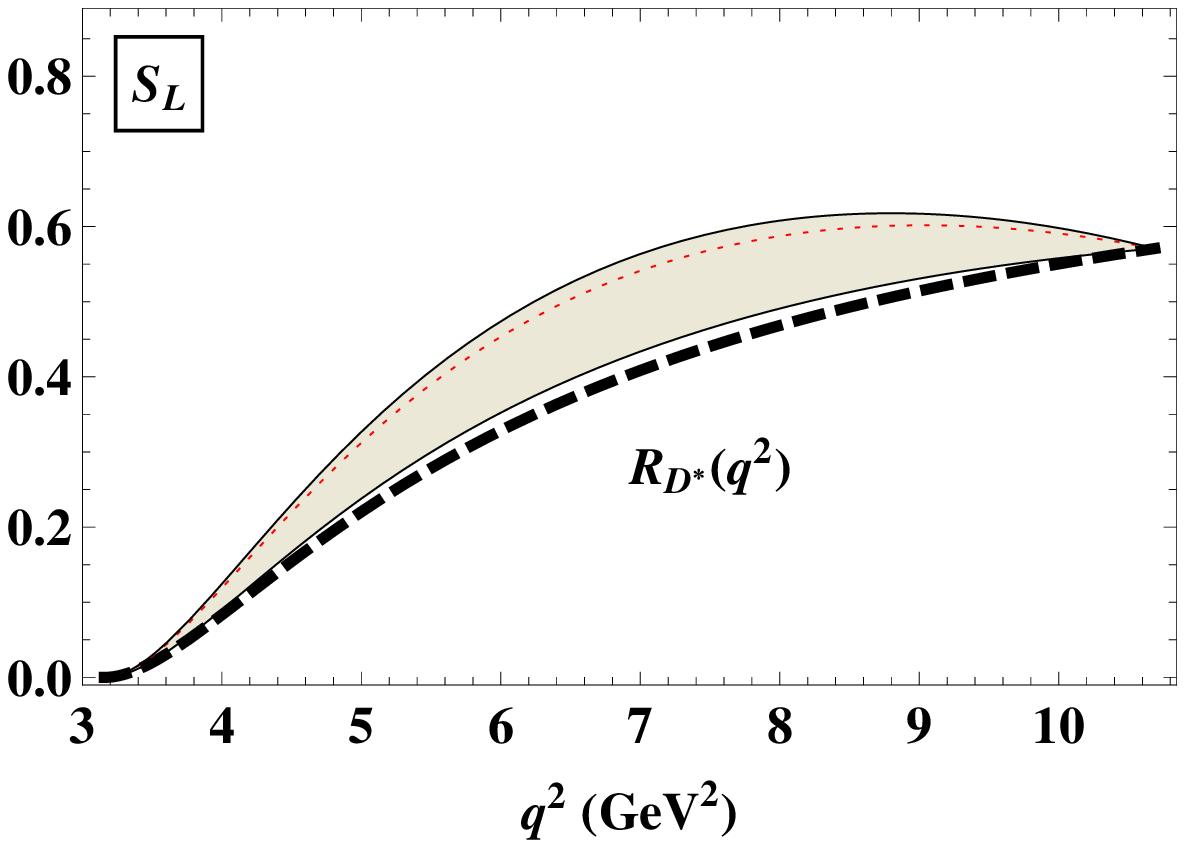}\\
\includegraphics[scale=0.4]{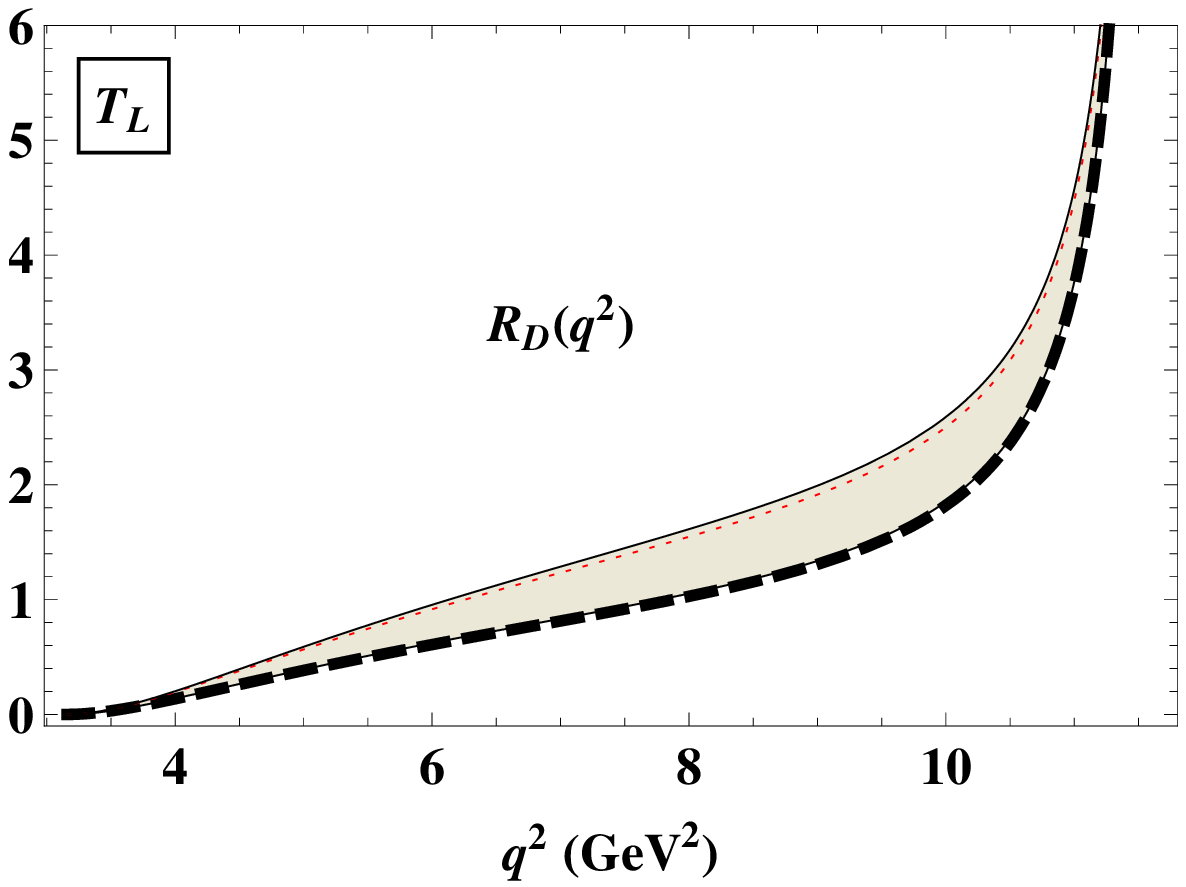}&
\includegraphics[scale=0.4]{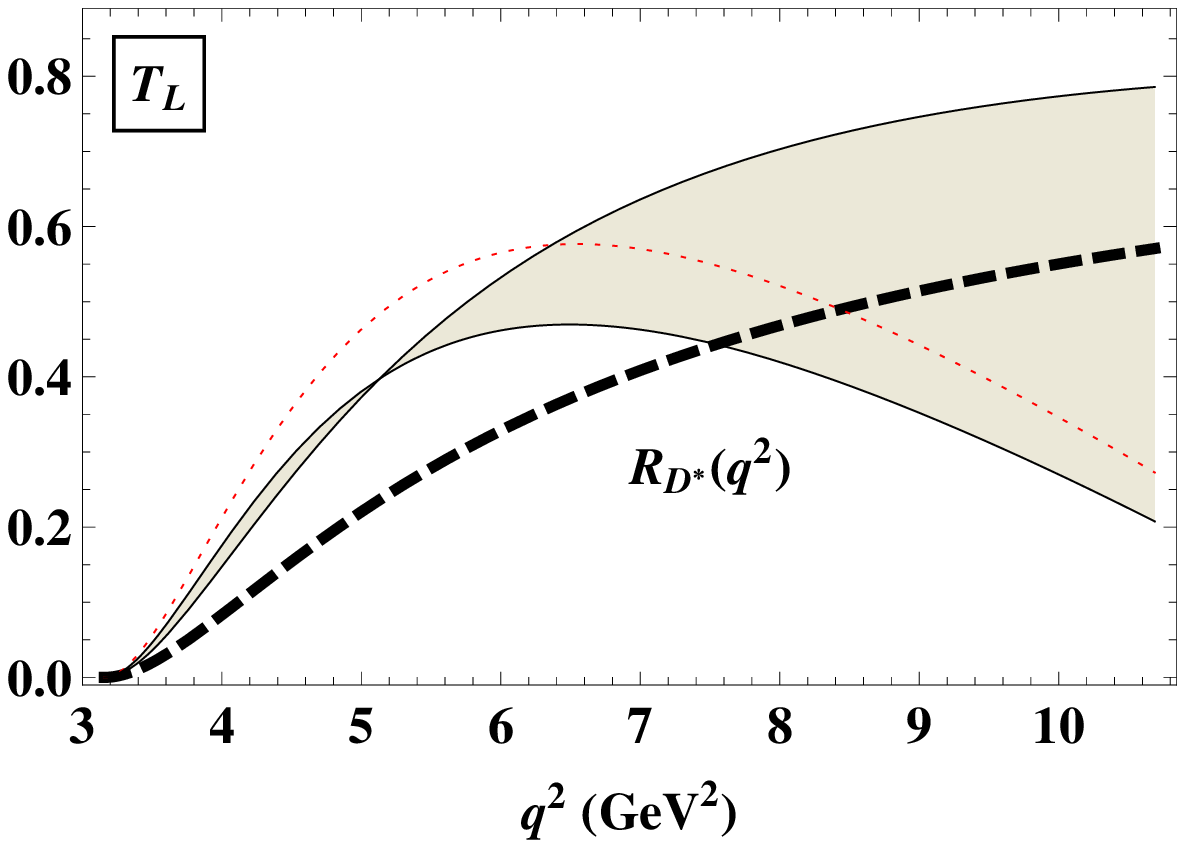}
\end{tabular}
\caption{Differential rate ratios $R_{D}(q^2)$ (left) and $R_{D^\ast}(q^2)$ (right). The thick black dashed lines are the SM prediction; the gray bands include NP effects corresponding to the $2\sigma$ allowed regions in Fig.~\ref{fig:constraint}; the red dotted lines represent the best-fit values of the NP couplings given in Eq.~(\ref{eq:bestfit}).}
\label{fig:RD}
\end{figure}

 The fourfold distribution allows one to define a large set of observables which can help probe NP in the decay. 
First, by integrating Eq.~(\ref{eq:distr4}) over all angles one obtains 
\be
\frac{d\Gamma(\bar{B}^0\to D^{\ast} \tau^-\bar\nu_\tau)}{dq^2} =
|N|^2 J_{\rm tot} = |N|^2 (J_L+J_T),
\label{eq:distr1}
\en
where $J_L$ and $J_T$ are the longitudinal and transverse polarization amplitudes of the $D^\ast$ meson, given by
\be
J_L=3J_{1c}-J_{2c},\qquad J_T=2(3J_{1s}-J_{2s}).
\en
The decay rate in Eq.~(\ref{eq:distr1}) is often normalized over the corresponding muon mode in order to dismiss the poorly known $V_{cb}$ and to partially cancel uncertainties from the hadronic form factors. In Fig.~\ref{fig:RD} we present the 
$q^{2}$ dependence
of the rate ratios 
\be
R_{D^{(\ast)}}(q^{2})=\frac{d\Gamma(\bar{B}^0\to D^{(\ast)}\tau^{-}\bar \nu_{\tau})}
{dq^{2}}
\bigg/\frac{d\Gamma(\bar{B}^0\to D^{(\ast)}\mu^{-}\bar \nu_{\mu})}{dq^{2}}
\en
in different NP scenarios. It is interesting to note that unlike the vector and scalar operators, which tend to increase both ratios, the tensor operator can lead to a decrease of the ratio $R(D^\ast)$ for $q^2 \gtrsim 8~\text{GeV}^2$. Moreover, while the ratio $R(D^\ast)$ is minimally sensitive to the scalar coupling $S_L$ (in comparison with other couplings, i.e. $V_{L,R}$, $T_L$), the ratio $R(D)$ shows maximal sensitivity to $S_L$. These behaviors can help discriminate between different NP operators.
%
%
\subsection{The $\cos\theta$ distribution, the forward-backward asymmetry,  and the lepton-side convexity parameter}
\begin{figure}[htbp]
\begin{tabular}{lr}
\includegraphics[scale=0.4]{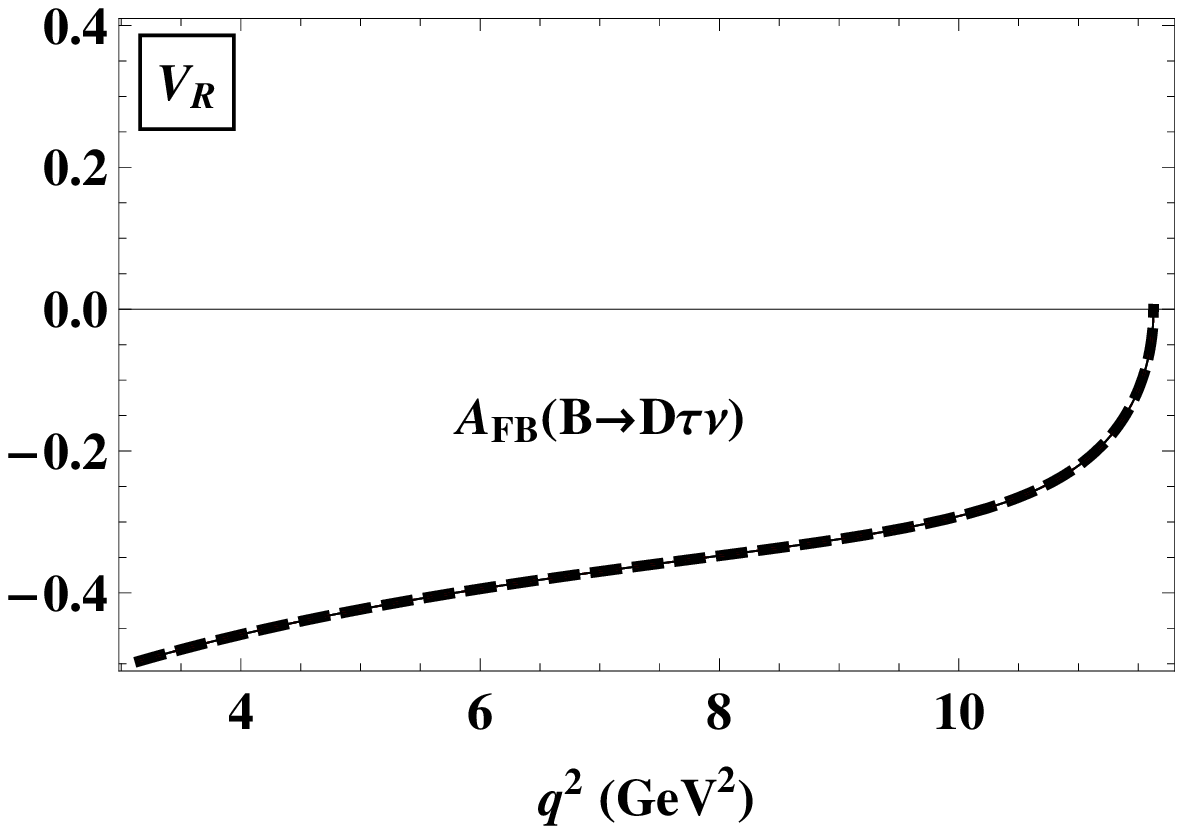}&
\includegraphics[scale=0.4]{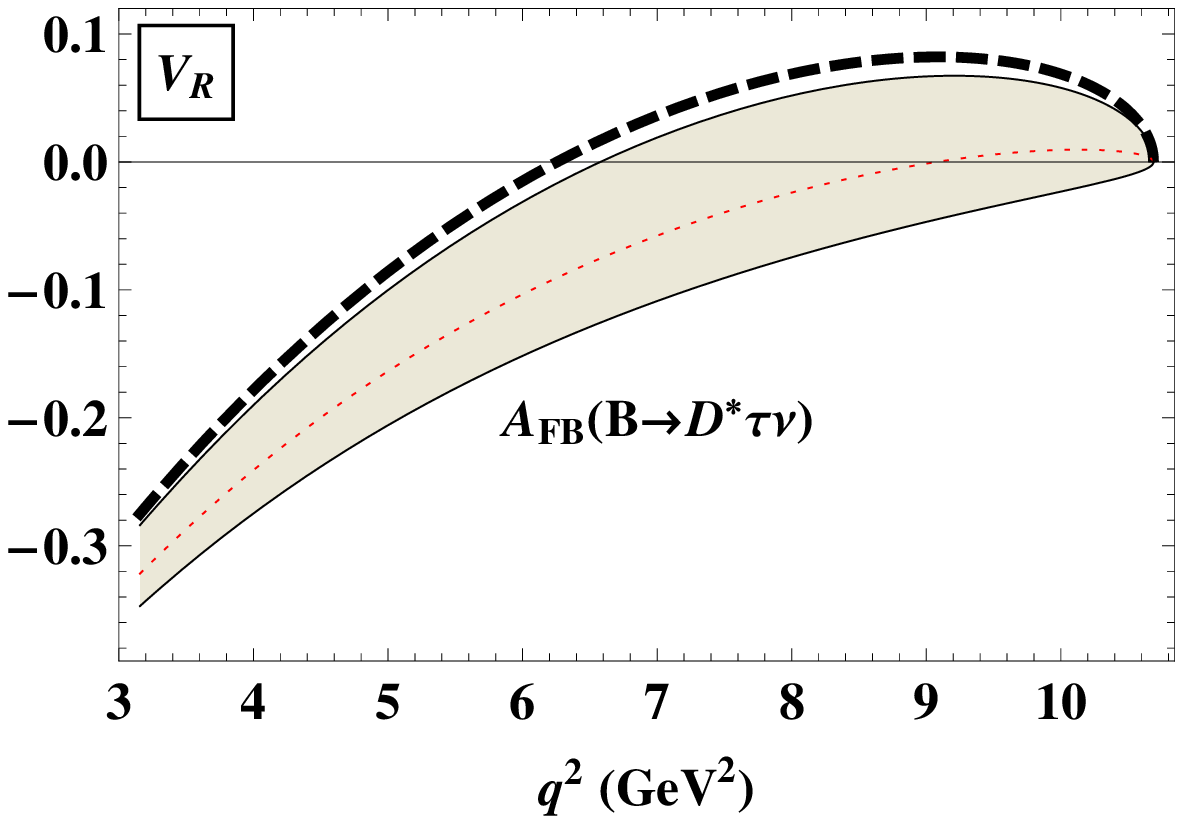}\\
\includegraphics[scale=0.4]{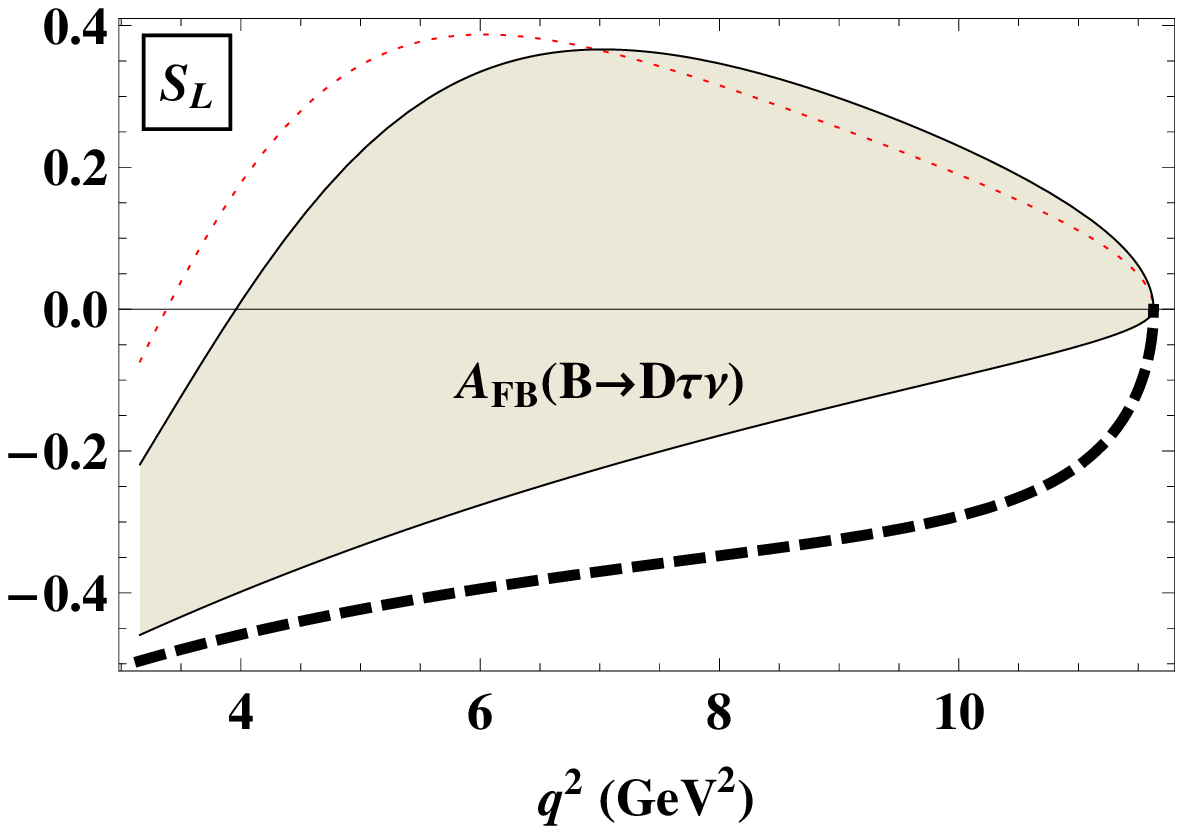}&
\includegraphics[scale=0.4]{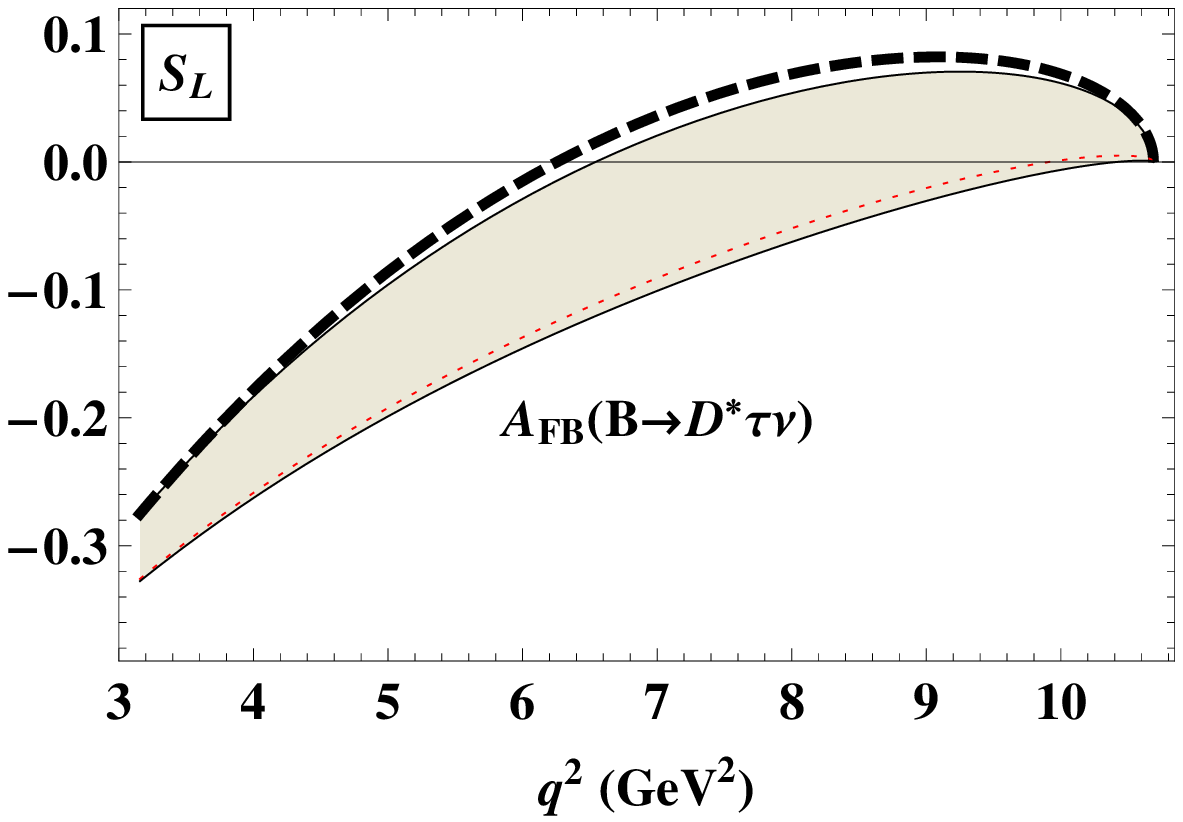}\\
\includegraphics[scale=0.4]{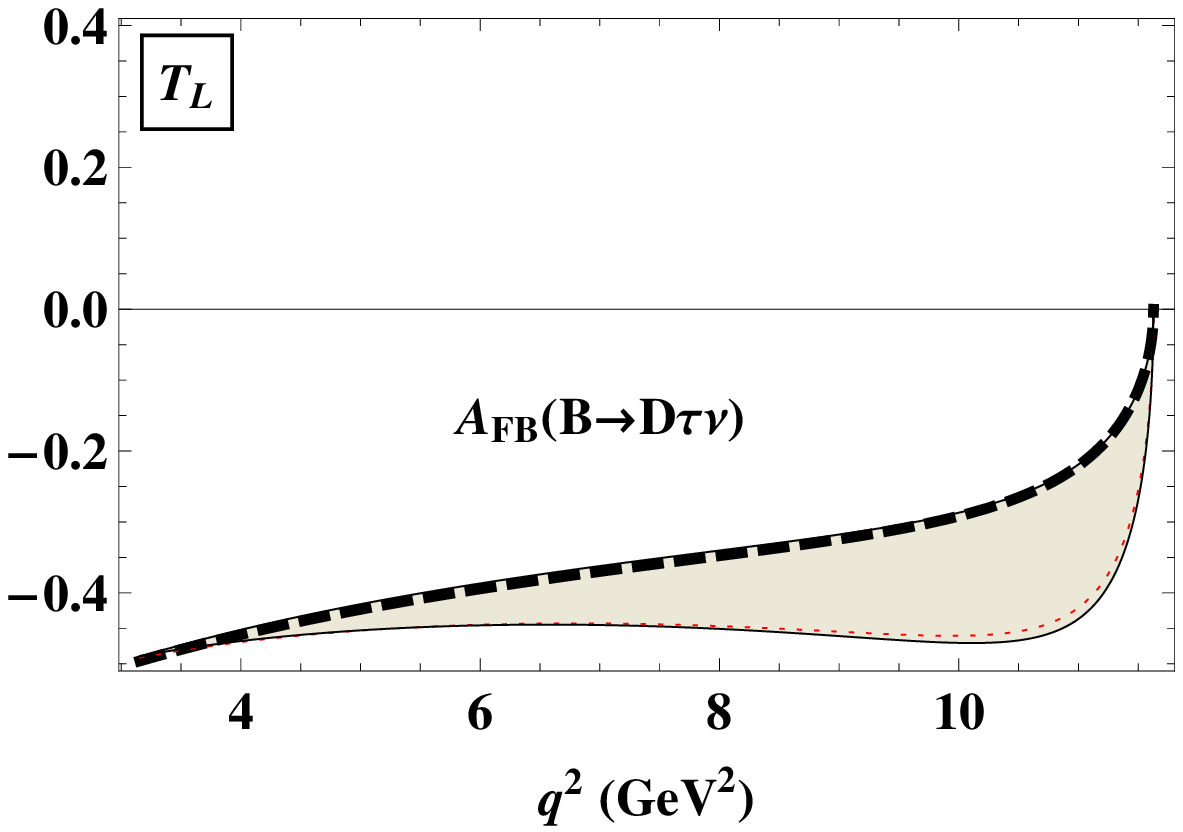}&
\includegraphics[scale=0.4]{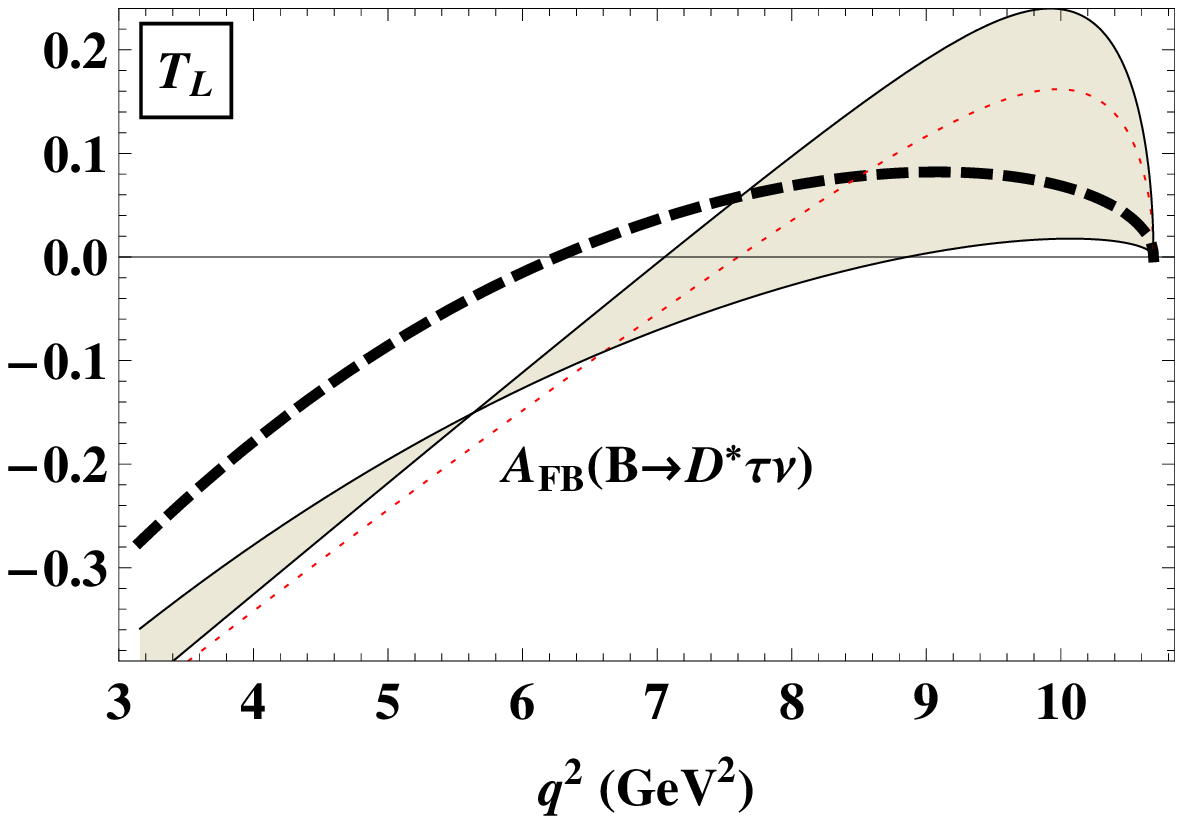}
\end{tabular}
\caption{Forward-backward asymmetry $\mathcal{A}_{FB}$ for $\bar{B}^0 \to D\tau^-\bar\nu_{\tau}$ (left) and $\bar{B}^0 \to D^\ast\tau^-\bar\nu_{\tau}$ (right). Notations are the same as in Fig.~\ref{fig:RD}.}
\label{fig:AFB}
\end{figure}

We define a normalized angular decay distribution 
$\widetilde J(\theta^\ast,\theta,\chi)$ through
\be
\widetilde J(\theta^\ast,\theta,\chi)=\frac{9}{8\pi}\frac{J(\theta^\ast,\theta,\chi)}
{J_{\rm tot}},
\label{eq:normdis}
\en
where $J_{\rm tot}=3J_{1c}+6J_{1s}-J_{2c}-2J_{2s}$.
The normalized angular decay distribution 
$\widetilde J(\theta^\ast,\theta,\chi)$ obviously integrates to $1$ after
$\cos\theta^\ast,\,\cos\theta$, and $\chi$ integration. By integrating 
Eq.~(\ref{eq:distr4}) over $\cos\theta^\ast$ and 
$\chi$ one obtains the differential $\cos\theta$ distribution which is described by a 
tilted parabola. The normalized form of the parabola reads
\be
\widetilde J(\theta)=\frac{a+b\cos\theta+c\cos^{2}\theta}{2(a+c/3)}.
\en
The linear
coefficient $b/2(a+c/3)$ can be projected out by defining a 
forward-backward asymmetry given by 
\bea
\mathcal{A}_{FB}(q^2) = 
\frac{ \int_{0}^{1} d\cos\theta\, d\Gamma/d\cos\theta
      -\int_{-1}^{0} d\cos\theta\, d\Gamma/d\cos\theta }
     { \int_{0}^{1} d\cos\theta\, d\Gamma/d\cos\theta
      +\int_{-1}^{0} d\cos\theta\, d\Gamma/d\cos\theta} 
= \frac{b}{2(a+c/3)}=\frac32 \frac{J_{6c}+2J_{6s}}{J_{\rm tot}}.
\label{fbAsym}
\ena
The coefficient $c/2(a+c/3)$ of the quadratic contribution is obtained
by taking the second derivative of $\widetilde J(\theta)$. Accordingly, we
define a convexity parameter by writing 
\be
C_F^\tau(q^2) = \frac{d^{2}\widetilde J(\theta)}{d(\cos\theta)^{2}}
= \frac{c}{a+c/3} 
= \frac{6(J_{2c}+2J_{2s})}{J_{\rm tot}}.
\label{eq:convex_lep}
\en 
\noindent
When calculating the $q^{2}$ averages of the forward-backward asymmetry
and the convexity parameter, one 
has to multiply the numerator and denominator of Eqs.~(\ref{fbAsym}) and~(\ref{eq:convex_lep}) by the 
$q^{2}$-dependent piece of the phase-space factor 
$
C(q^2) = |\mathbf{p_2}| q^2 v^2.
$
For example, the mean forward-backward asymmetry can then be calculated 
according to
\be
\langle \mathcal{A}_{FB}\rangle = \frac32 
\frac{\int dq^{2} C(q^{2})\big(J_{6c}+2J_{6s}\big)}
{\int dq^{2} C(q^{2})J_{\rm tot}}.
\label{eq:FBint}
\en
The $q^2$ dependence of the forward-backward asymmetry is shown in Fig.~\ref{fig:AFB}. The coupling $V_L$ does not effect $\mathcal{A}_{FB}$ in both decays since it stands before the SM operator and drops out in the definition of the observable. In the case of the $\bar{B}^0\to D^{\ast}$ transition, the operators $\mathcal{O}_{V_R}$, $\mathcal{O}_{S_L}$, and $\mathcal{O}_{T_L}$ behave mostly similarly: they tend to decrease the forward-backward asymmetry and shift the zero-crossing point to greater values than the SM one. However, the tensor operator can also increase the asymmetry in the high-$q^2$ region. In the case of the $\bar{B}^0\to D$ transition, the operator $\mathcal{O}_{V_R}$ does not affect $\mathcal{A}_{FB}$, the tensor operator $\mathcal{O}_{T_L}$ tends to lower $\mathcal{A}_{FB}$, and the scalar operator $\mathcal{O}_{S_L}$ thoroughly changes $\mathcal{A}_{FB}$: it can increase the forward-backward asymmetry by up to $200\%$ and implies a zero-crossing point, which is impossible in the SM. This unique effect of $\mathcal{O}_{S_L}$ would clearly distinguish it from the other NP operators.

\begin{figure}[htbp]
\begin{tabular}{lr}
\includegraphics[scale=0.4]{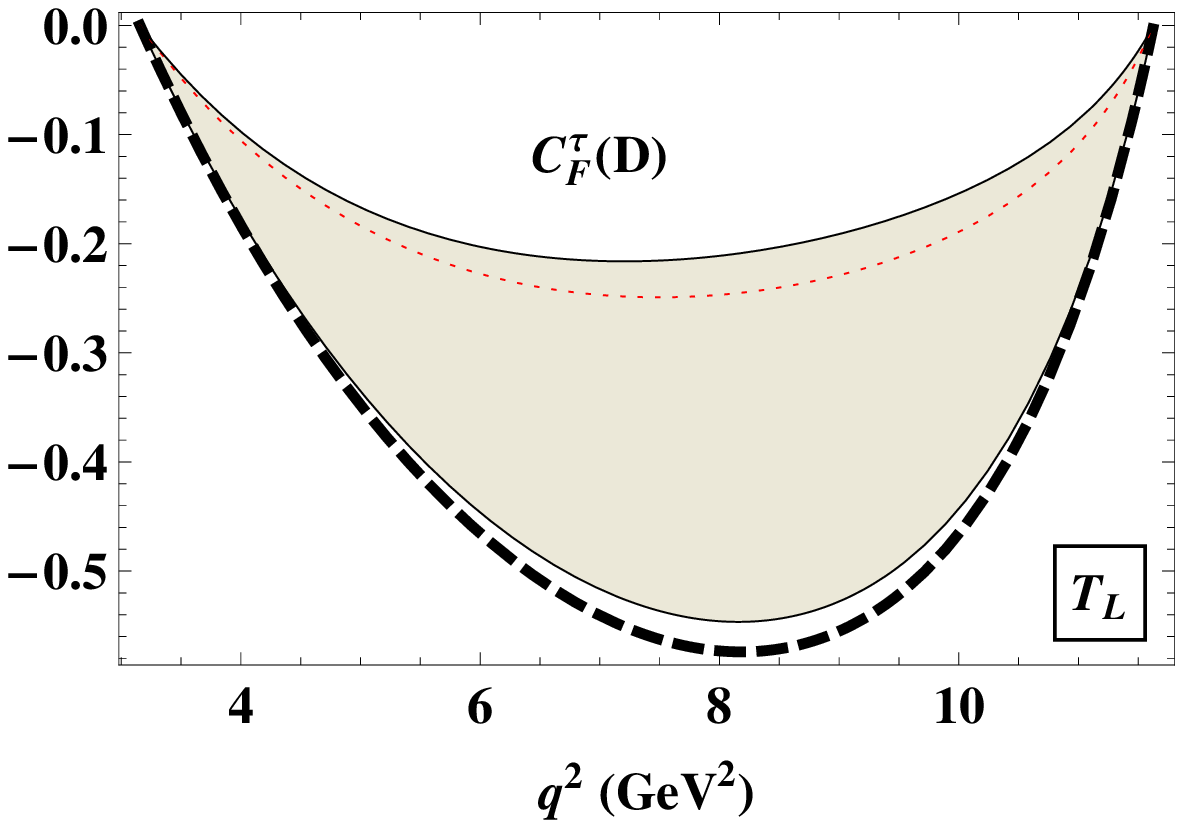}&
\includegraphics[scale=0.4]{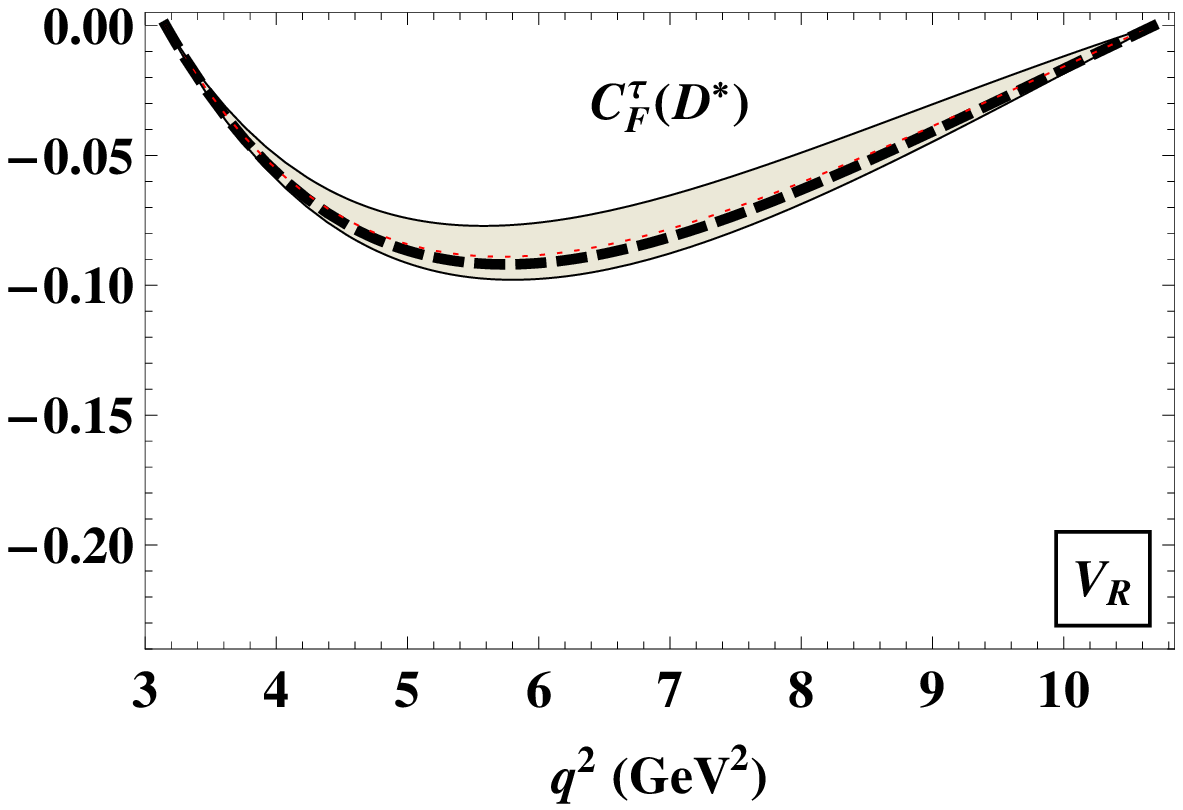}\\
\includegraphics[scale=0.4]{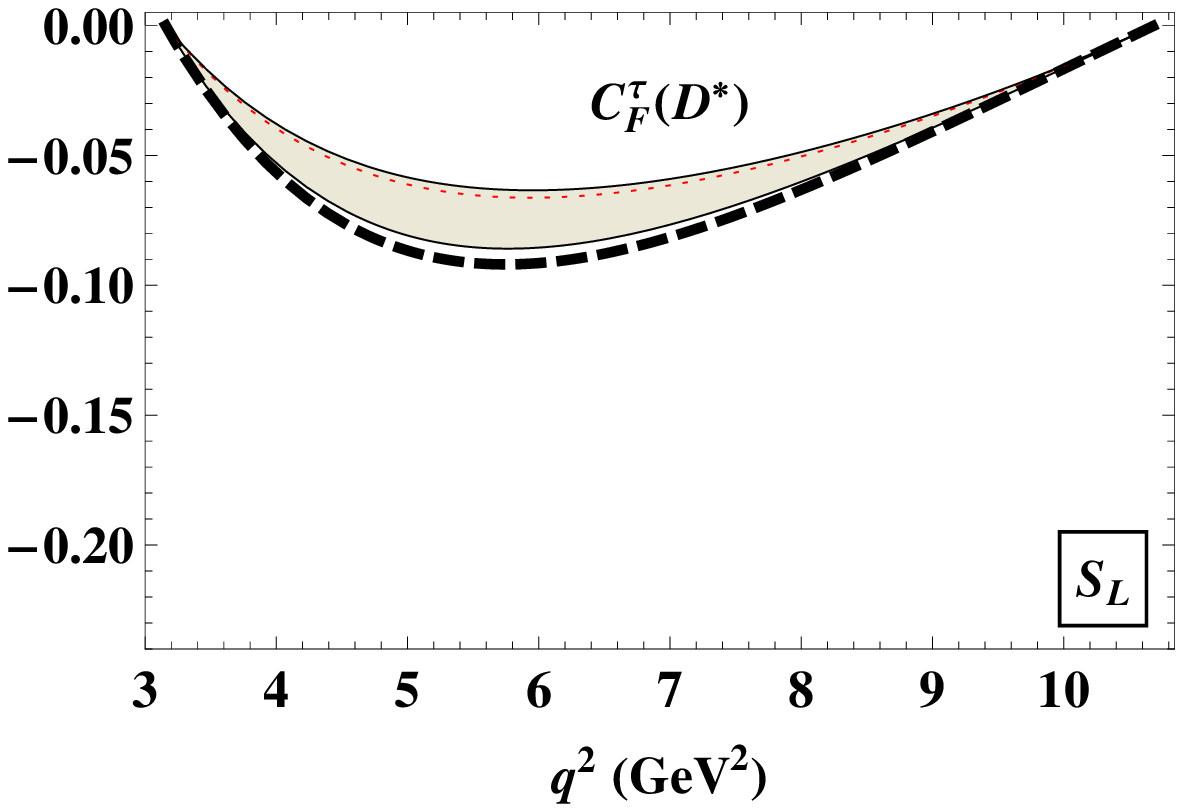}&
\includegraphics[scale=0.4]{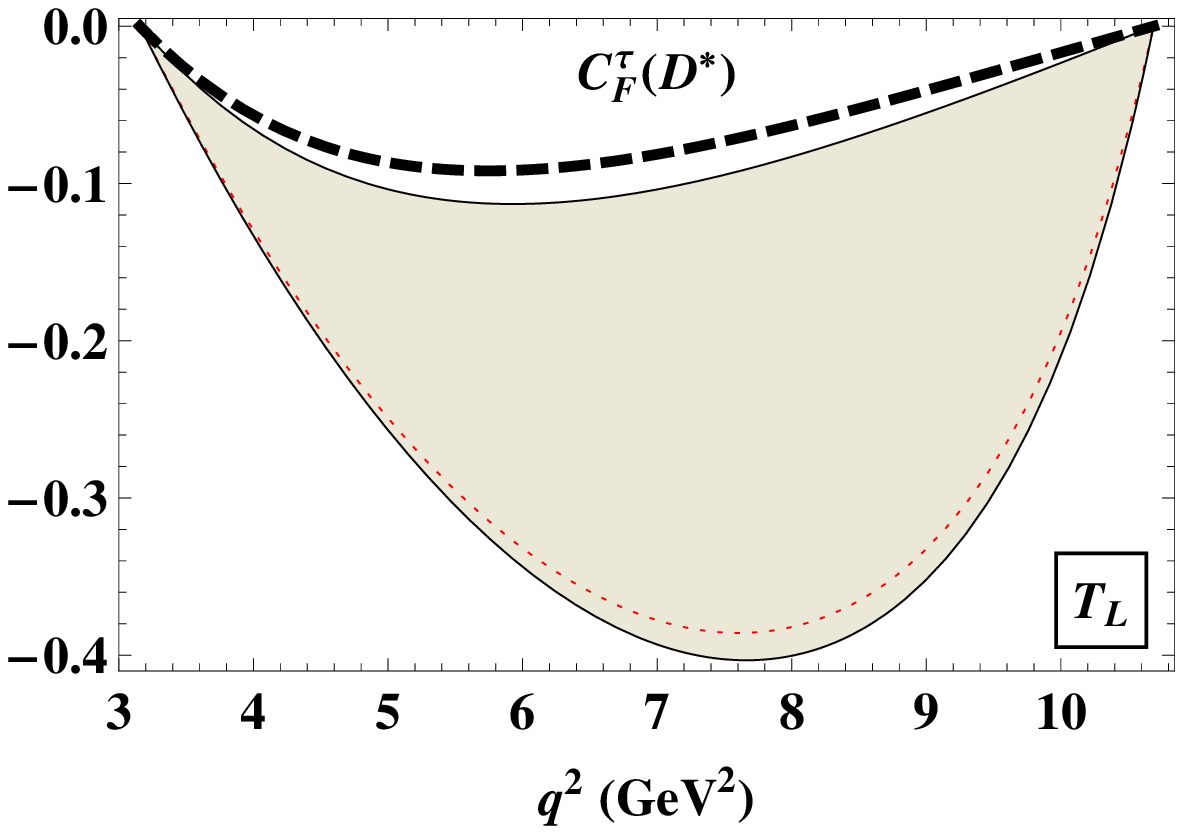}
\end{tabular}
\caption{Lepton-side convexity parameter $C_F^\tau(q^2)$. Notations are the same as in Fig.~\ref{fig:RD}. In the case of the $D$ meson, NP effects come from the tensor operator only.}
\label{fig:CFL}
\end{figure}

In Fig.~\ref{fig:CFL} we present the lepton-side convexity parameter $C_F^\tau(q^2)$. While $C_F^\tau(D)$ is only sensitive to $\mathcal{O}_{T_L}$, $C_F^\tau(D^{\ast})$ is sensitive to $\mathcal{O}_{S_L}$, $\mathcal{O}_{V_R}$, and $\mathcal{O}_{T_L}$. Unlike $\mathcal{O}_{S_L}$, which can only increase $C_F^\tau(D^{\ast})$, the operator $\mathcal{O}_{T_L}$ can only lower the parameter. It is worth mentioning that $C_F^\tau(D)$ and $C_F^\tau(D^{\ast})$ are extremely sensitive to $\mathcal{O}_{T_L}$: it can change $C_F^\tau(D^{(\ast)})$ by a factor of 4 at $q^2\approx 7~\text{GeV}^2$.
\subsection{The $\cos\theta^\ast$ distribution and the hadron-side convexity parameter}
By integrating 
Eq.~(\ref{eq:distr4}) over $\cos\theta$ and 
$\chi$ one obtains the hadron-side $\cos\theta^\ast$ distribution described 
by an untilted parabola (without a linear term). The normalized form of the  
$\cos\theta^\ast$ distribution
reads $\widetilde {J} (\theta^\ast)=(a'+c'\cos^{2}\theta^\ast)/2(a'+c'/3)$,
which can again be characterized by its convexity 
parameter given by
\be
C_F^h(q^2) = \frac{d^{2}\widetilde J(\theta^\ast)}{d(\cos\theta^{\ast})^{2}}
=\frac{c'}{a'+c'/3}=
\frac{3J_{1c}-J_{2c}-3J_{1s}+J_{2s}}{J_{\rm tot}/3}.
\label{eq:convex_had}
\en
The $\cos\theta^\ast$ distribution can be written as
\be
\widetilde J(\theta^\ast)=\frac34\left(2F_L(q^2)\cos^2\theta^\ast+F_T(q^2)\sin^2\theta^\ast\right),
\en
where $F_L(q^2)$ and $F_T(q^2)$ are the polarization fractions of the $D^\ast$ meson and are defined as
\be
F_L(q^2)=\frac{J_L}{J_L+J_T},\qquad F_T(q^2)=\frac{J_T}{J_L+J_T},\qquad F_L(q^2)+F_T(q^2)=1.
\en
\noindent
The hadron-side convexity parameter and the polarization fractions of the $D^\ast$ meson are related by
\be 
C_F^h(q^2)=\frac32 \left( 2F_L(q^2)-F_T(q^2) \right)=\frac32 \left(3F_L(q^2)-1\right).
\en
\begin{figure}[htbp]
\begin{tabular}{ccc}
\includegraphics[scale=0.40]{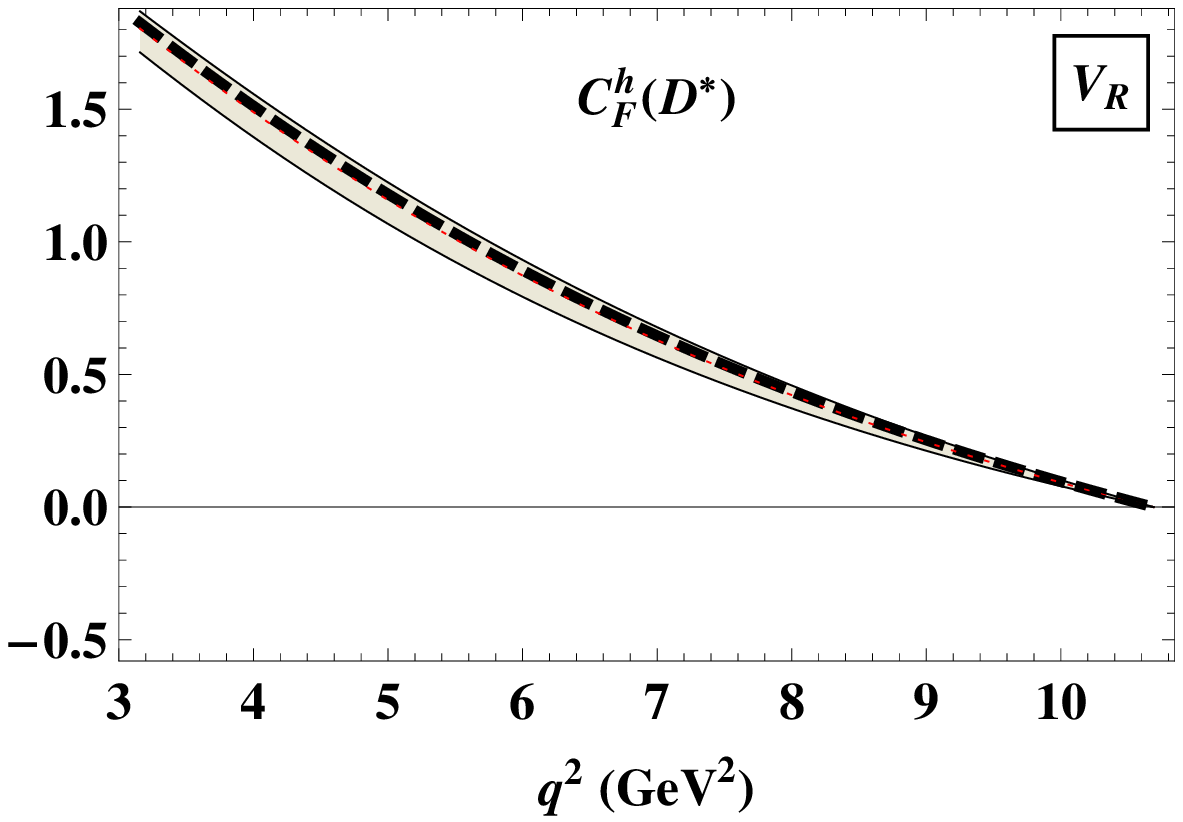}&
\includegraphics[scale=0.40]{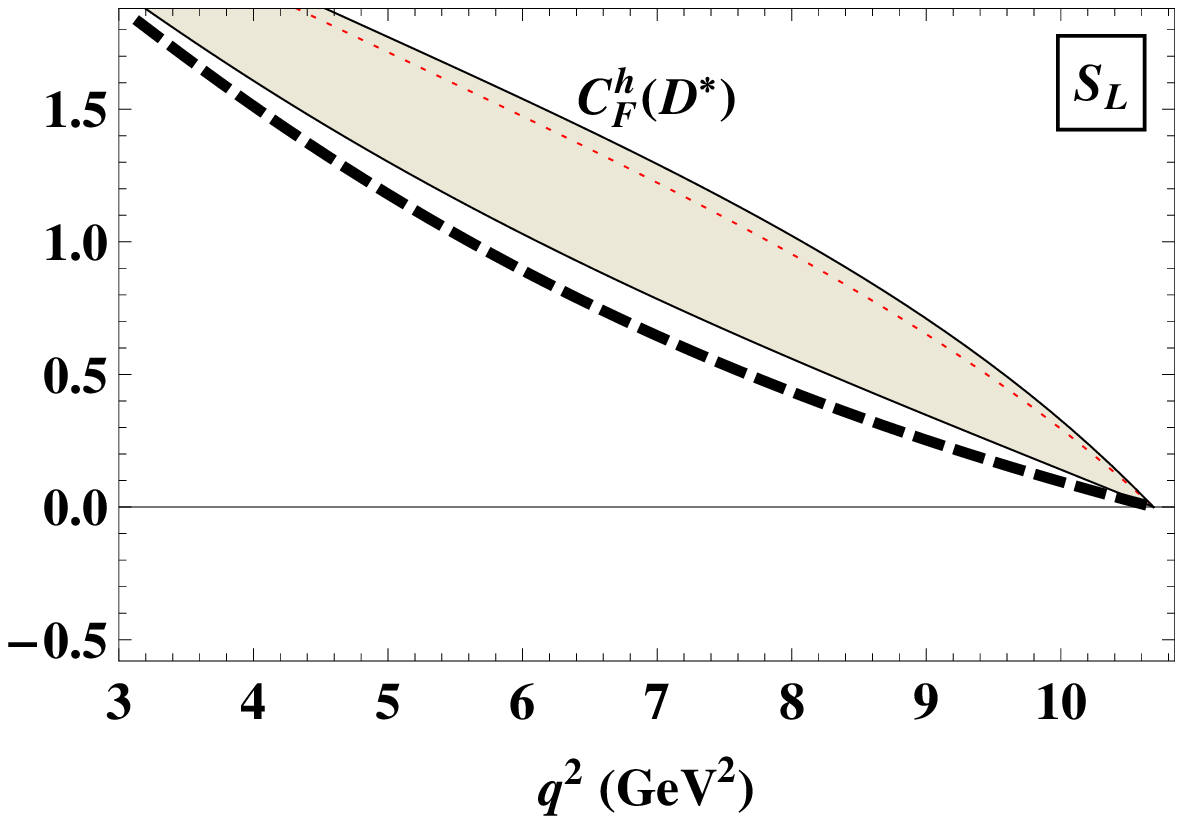}&
\includegraphics[scale=0.40]{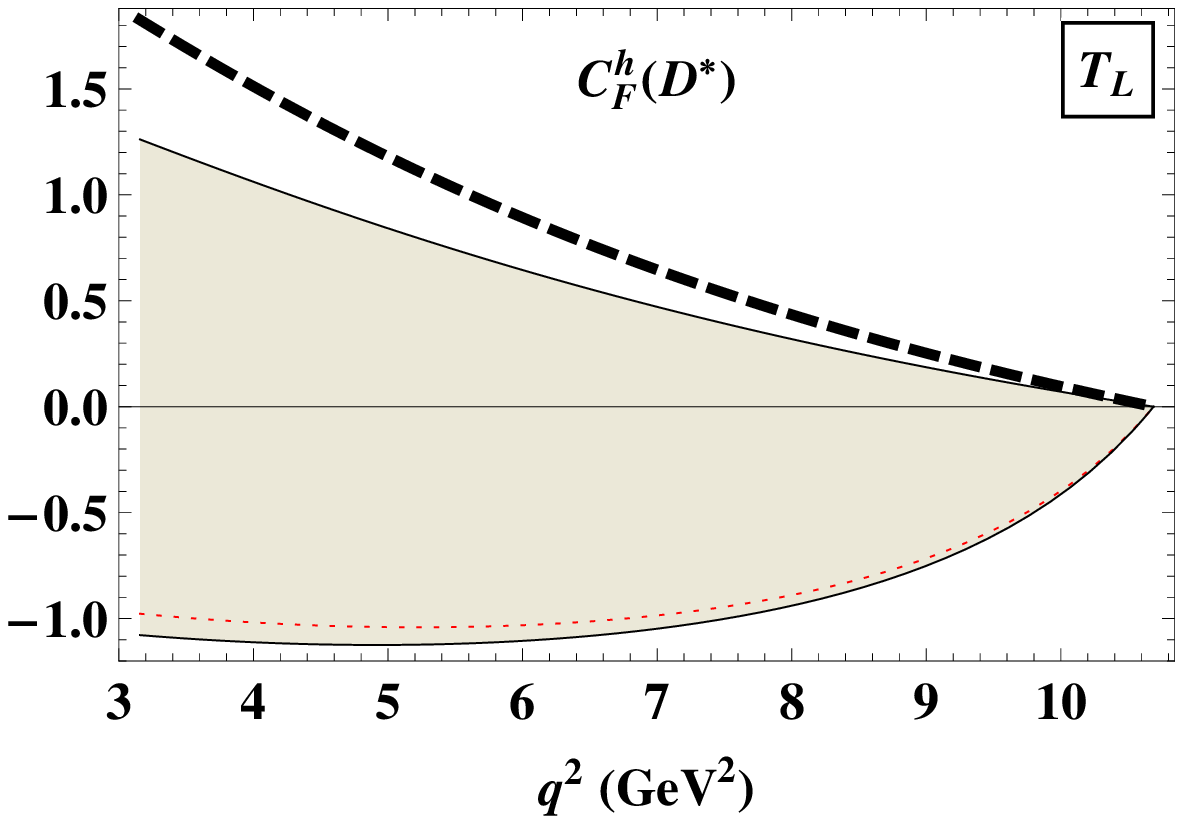}
\end{tabular}
\caption{Hadron-side convexity parameter $C_F^h(q^2)$. Notations are the same as in Fig.~\ref{fig:RD}.}
\label{fig:CFH}
\end{figure}
The effects of NP operators on the hadron-side convexity parameter $C_F^h(q^2)$ are shown in Fig.~\ref{fig:CFH}. Each NP operator can change $C_F^h(q^2)$ in a unique way: the vector operator $\mathcal{O}_{V_R}$ almost does nothing to the parameter; the scalar operator $\mathcal{O}_{S_L}$ increases the parameter by about $50\%$ nearly in the whole range of $q^2$; the tensor operator $\mathcal{O}_{T_L}$ lowers the parameter (by up to $200\%$ at low $q^2$), and it also allows negative values of $C_F^h(q^2)$, which are impossible in the SM.
\subsection{The $\chi$ distribution and the trigonometric moments}
By integrating 
Eq.~(\ref{eq:distr4}) over $\cos\theta$ and $\cos\theta^\ast$
one obtains the $\chi$ distribution whose normalized form reads
\be
\widetilde J^{(I)}(\chi)=\frac{1}{2\pi}\Big[1+A_C^{(1)}(q^2)\cos 2\chi+A_T^{(1)}(q^2)\sin 2\chi\Big],
\en
where $A_C^{(1)}(q^2)=4J_3/J_{\rm tot}$ and $A_T^{(1)}(q^2)=4J_9/J_{\rm tot}$. Besides, one can also define other angular distributions in the angular variable $\chi$ as follows~\cite{Duraisamy:2013kcw}:
\be
J^{(II)}(\chi)=\Big[\int_0^1-\int_{-1}^0\Big]d\cos \theta^\ast\int_{-1}^1d\cos \theta\frac{d^4\Gamma}
     {dq^2 d\cos\theta d\chi d\cos\theta^\ast}, 
\en
\be
J^{(III)}(\chi)=\Big[\int_0^1-\int_{-1}^0\Big]d\cos \theta^\ast\Big[\int_0^1-\int_{-1}^0\Big]d\cos \theta\frac{d^4\Gamma}
     {dq^2 d\cos\theta d\chi d\cos\theta^\ast}.
\en
The normalized forms of these distributions read
\be
\widetilde J^{(II)}(\chi)=\frac14\Big[A_C^{(2)}(q^2)\cos\chi+A_T^{(2)}(q^2)\sin\chi\Big],
\en
\be
\widetilde J^{(III)}(\chi)=\frac{2}{3\pi}\Big[A_C^{(3)}(q^2)\cos\chi+A_T^{(3)}(q^2)\sin\chi\Big],
\en
where 
\be
A_C^{(2)}(q^2)=\frac{3J_5}{J_{\rm tot}},\qquad A_T^{(2)}(q^2)=\frac{3J_7}{J_{\rm tot}},\qquad A_C^{(3)}(q^2)=\frac{3J_4}{J_{\rm tot}},\qquad A_T^{(3)}(q^2)=\frac{3J_8}{J_{\rm tot}}.
\en
Another method to project the coefficient functions $J_i$ $(i=3,4,5,7,8,9)$ out from the
threefold angular decay distribution in Eq.~(\ref{eq:distr4}) is to take the
appropriate trigonometric moments of the normalized decay distribution
$\widetilde J(\theta^\ast,\theta,\chi)$~\cite{Ivanov:2015tru}. The trigonometric moments are
defined by
\be
W_{i} = \int d\cos\theta d\cos\theta^\ast d\chi
M_{i}(\theta^\ast,\theta,\chi)\widetilde J(\theta^\ast,\theta,\chi) 
\equiv  \left\langle M_{i}(\theta^\ast,\theta,\chi) \right\rangle,
\en
where $M_{i}(\theta^\ast,\theta,\chi)$ defines the trigonometric moment that 
is being taken.
One finds 
\bea
W_T(q^2) &\equiv& \left\langle \cos 2\chi \right\rangle  
= \frac{2J_3}{J_{\rm tot}} = \frac12 A_C^{(1)}(q^2),
\nn
W_{IT}(q^2) &\equiv& \left\langle \sin 2\chi \right\rangle  
= \frac{2J_9}{J_{\rm tot}} = \frac12 A_T^{(1)}(q^2),
\nn
W_A(q^2) &\equiv& \left\langle \sin\theta\cos\theta^{\ast}\cos \chi \right\rangle 
= \frac{3\pi}{8} \frac{J_5}{J_{\rm tot}} = \frac{\pi}{8} A_C^{(2)}(q^2),
\nn
W_{IA}(q^2) &\equiv& \left\langle \sin\theta\cos\theta^{\ast}\sin \chi \right\rangle 
= \frac{3\pi}{8} \frac{J_7}{J_{\rm tot}} = \frac{\pi}{8} A_T^{(2)}(q^2),
\nn
W_I(q^2) &\equiv& \left\langle   
\cos\theta\cos\theta^{\ast}\cos \chi \right\rangle  
= \frac{9\pi^2}{128}\frac{J_4}{J_{\rm tot}} = \frac{3\pi^2}{128} A_C^{(3)}(q^2),
\nn
W_{II}(q^2) &\equiv& \left\langle   
\cos\theta\cos\theta^{\ast}\sin \chi \right\rangle  
= \frac{9\pi^2}{128}\frac{J_8}{J_{\rm tot}} = \frac{3\pi^2}{128} A_T^{(3)}(q^2).
\label{eq:W}
\ena
These coefficient functions can also be projected out by
taking piecewise sums and differences of different sectors of the angular 
phase space~\cite{Korner:1989ve,Korner:1989qb,Gratrex:2015hna,Becirevic:2016hea}. 

\begin{figure}[htbp]
\begin{tabular}{ccc}
\includegraphics[scale=0.40]{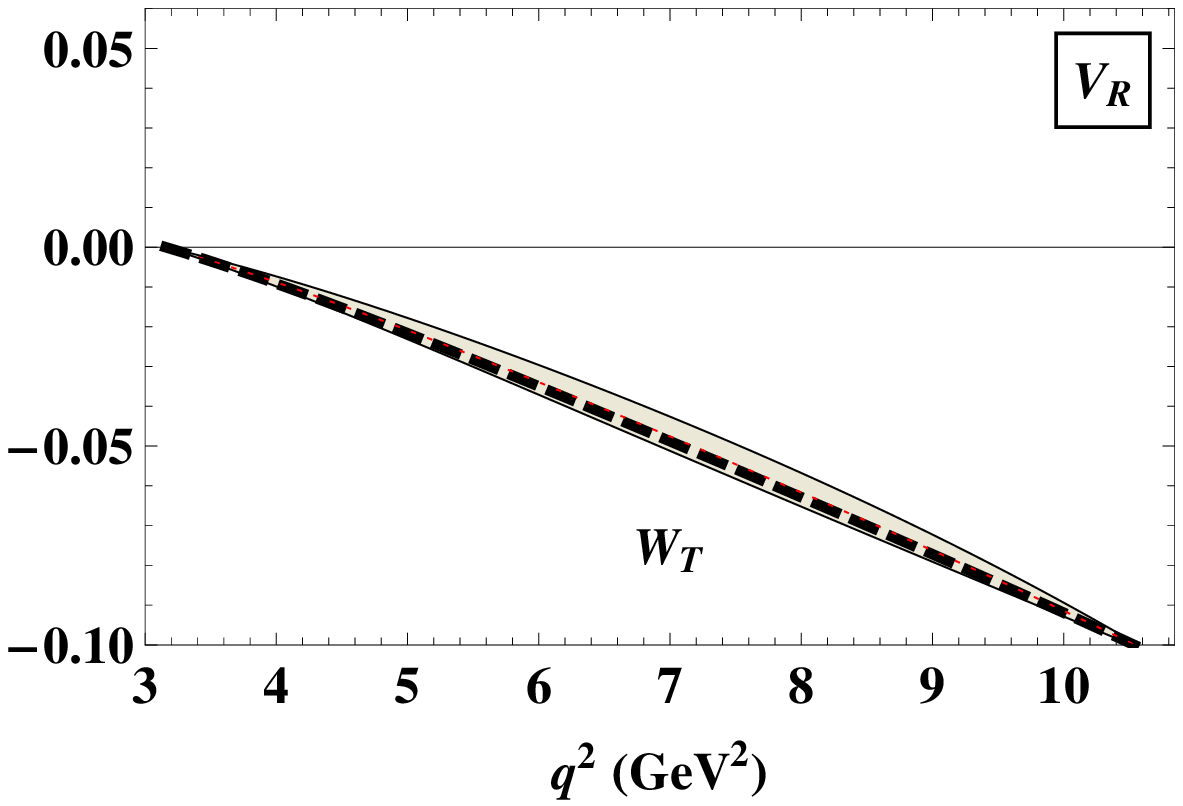}&
\includegraphics[scale=0.40]{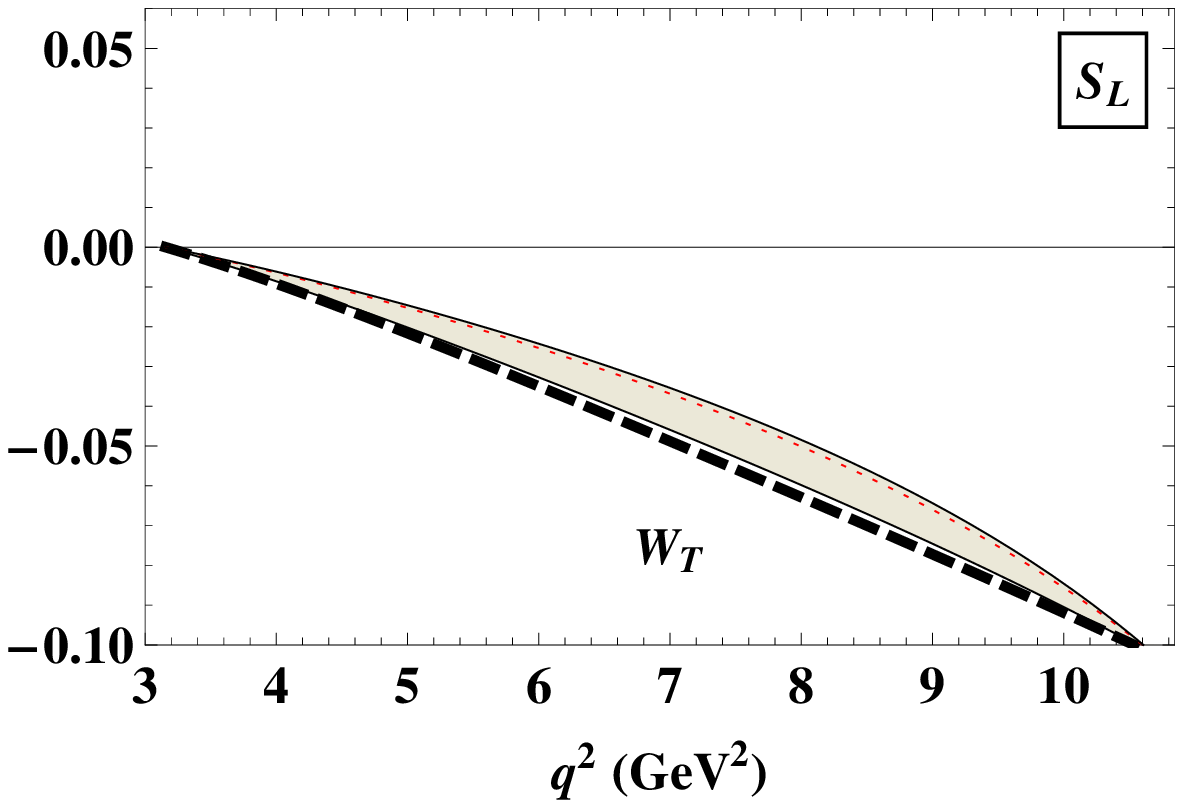}&
\includegraphics[scale=0.40]{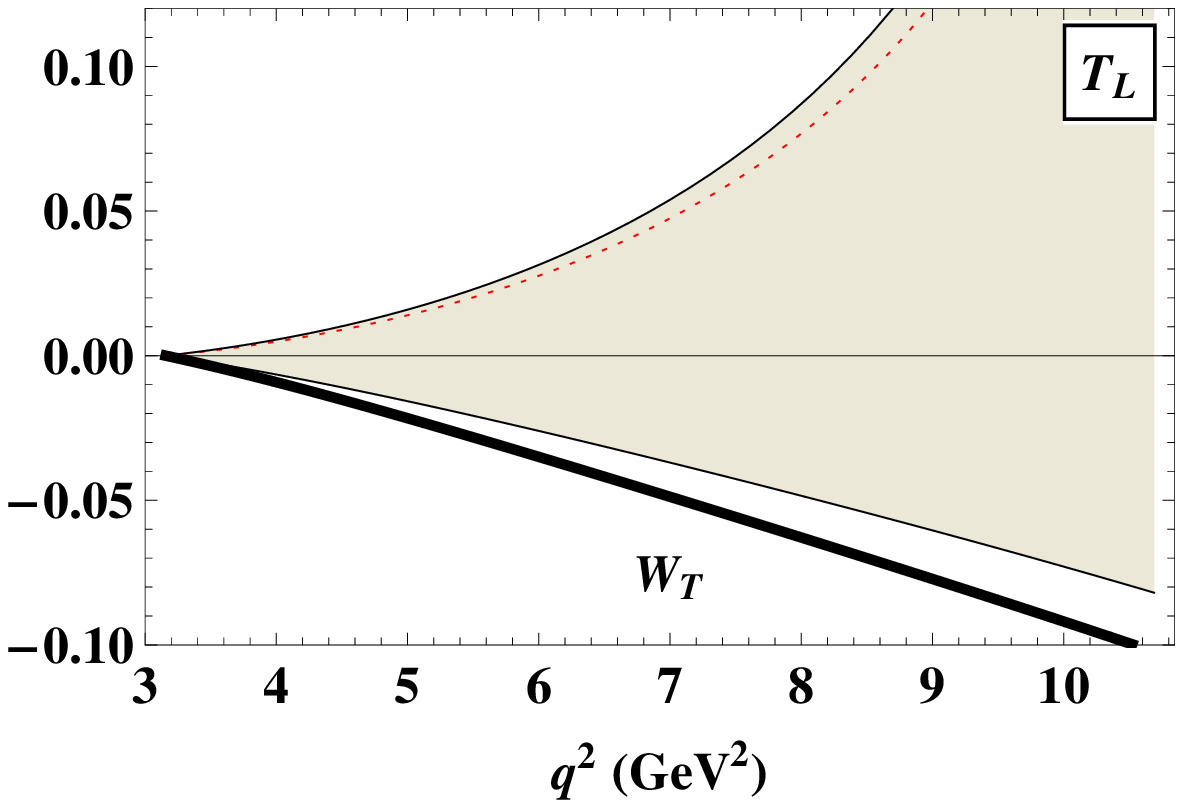}\\
\includegraphics[scale=0.40]{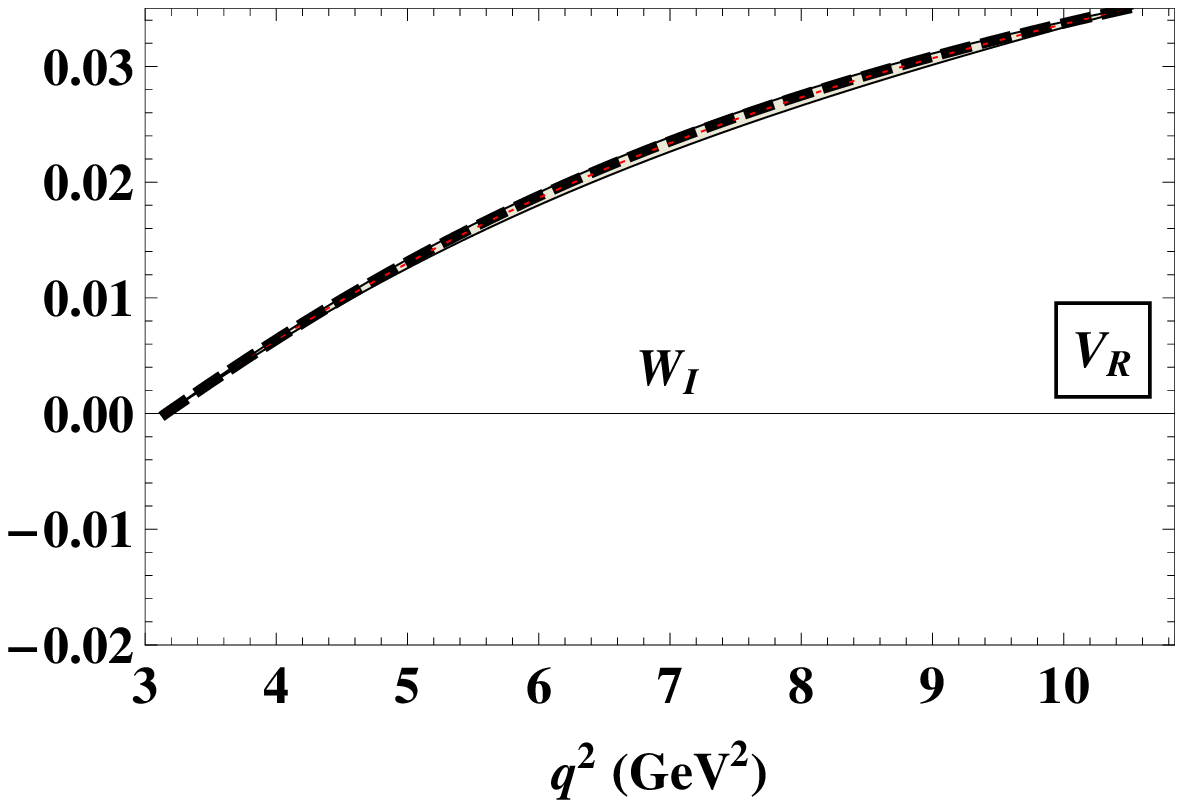}&
\includegraphics[scale=0.40]{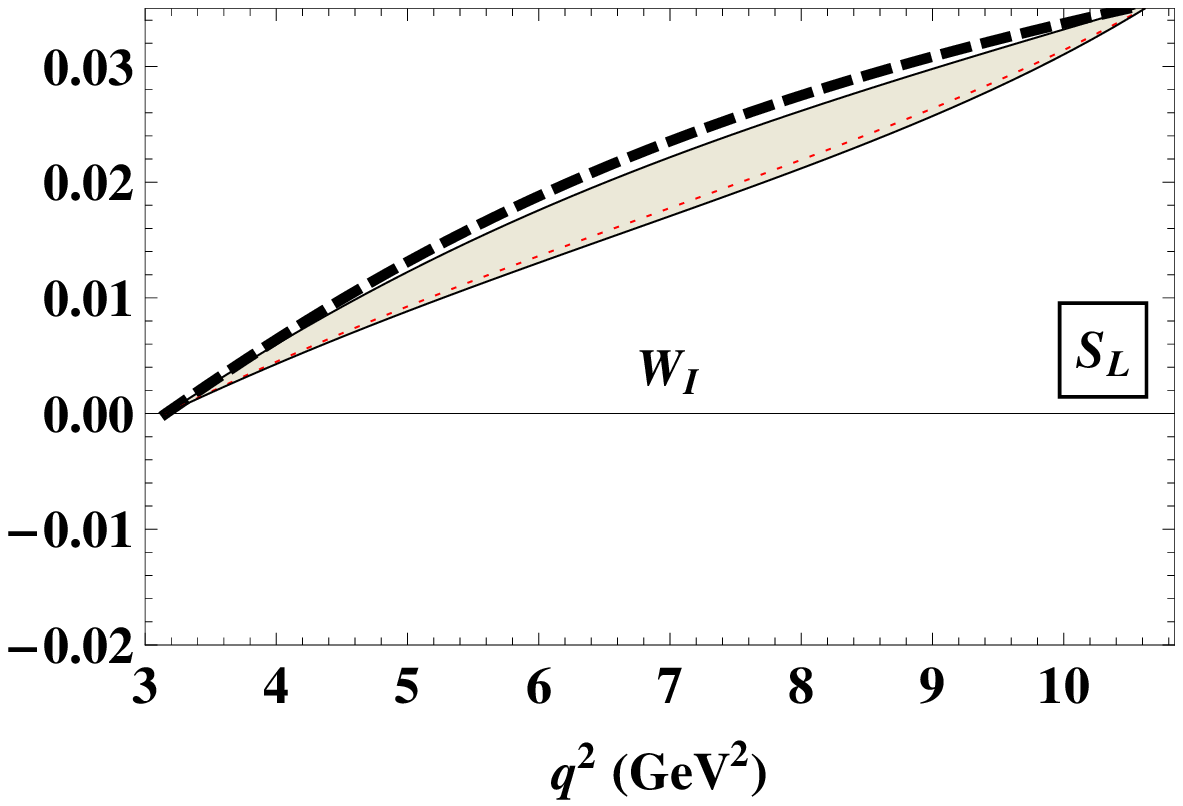}&
\includegraphics[scale=0.40]{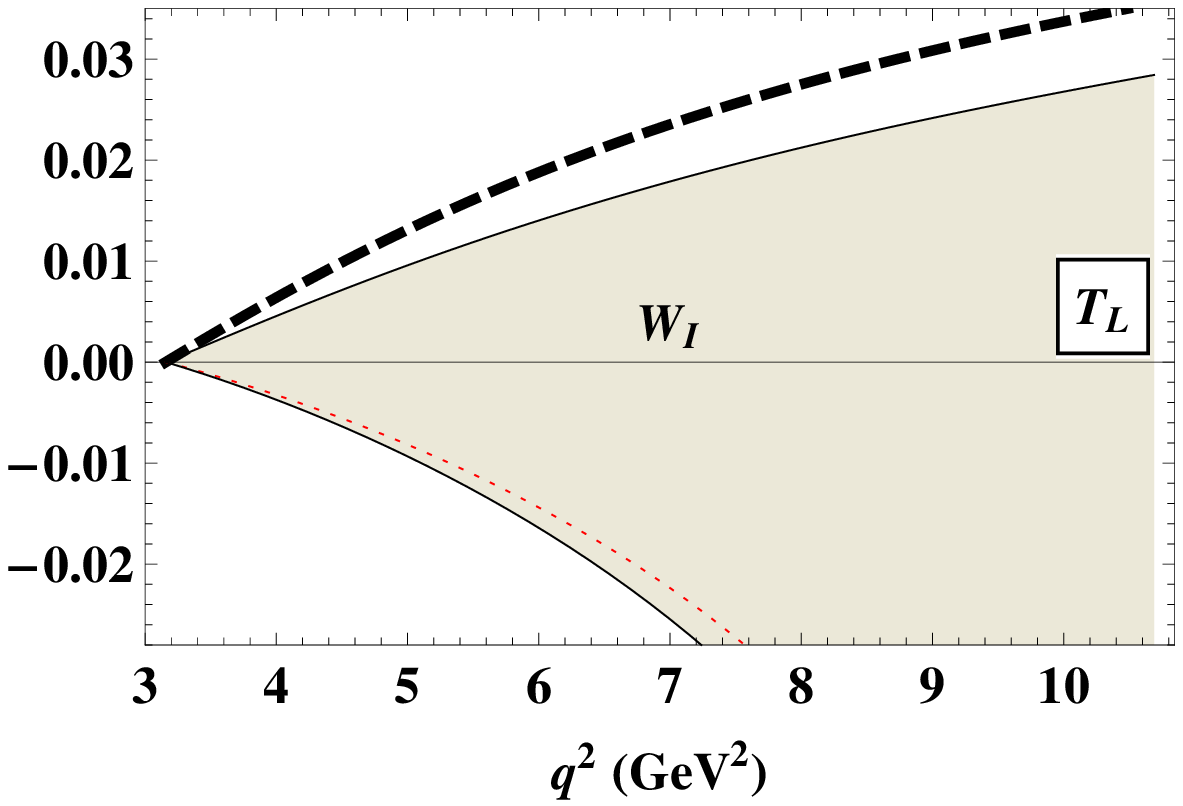}\\
\includegraphics[scale=0.40]{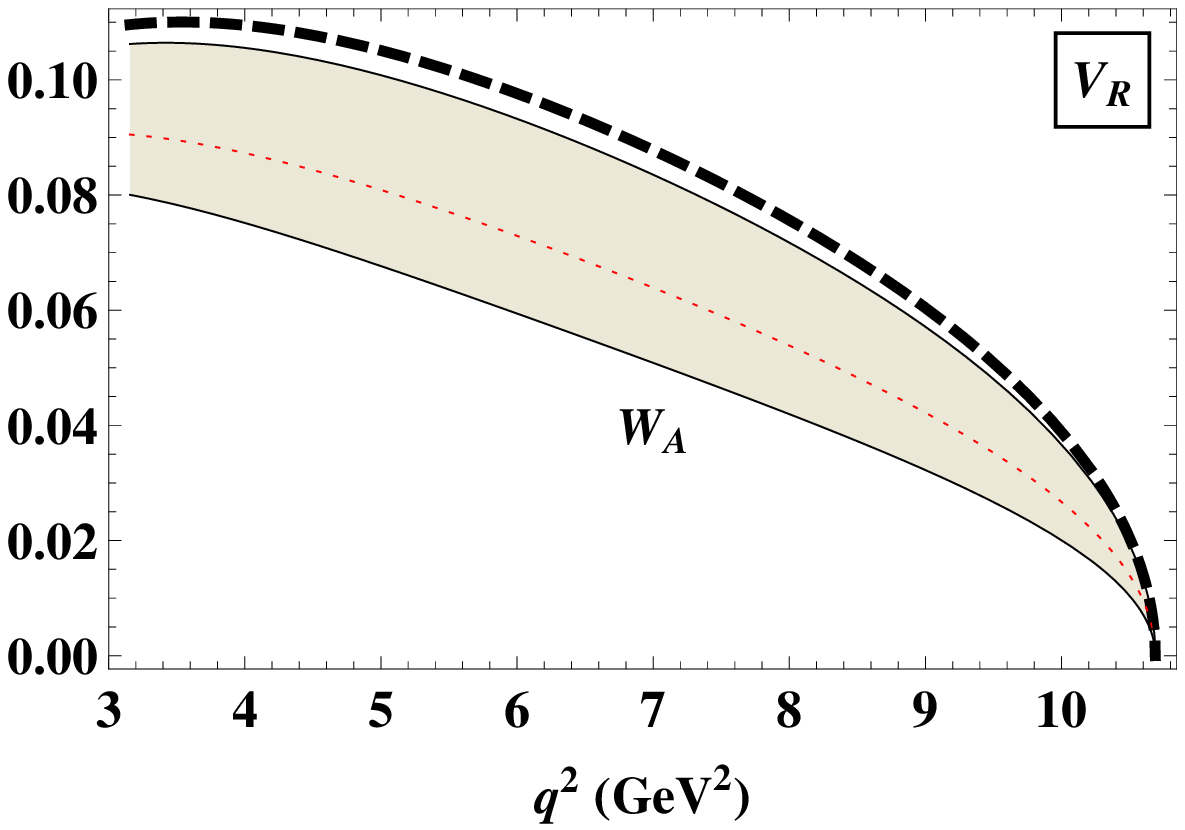}&
\includegraphics[scale=0.40]{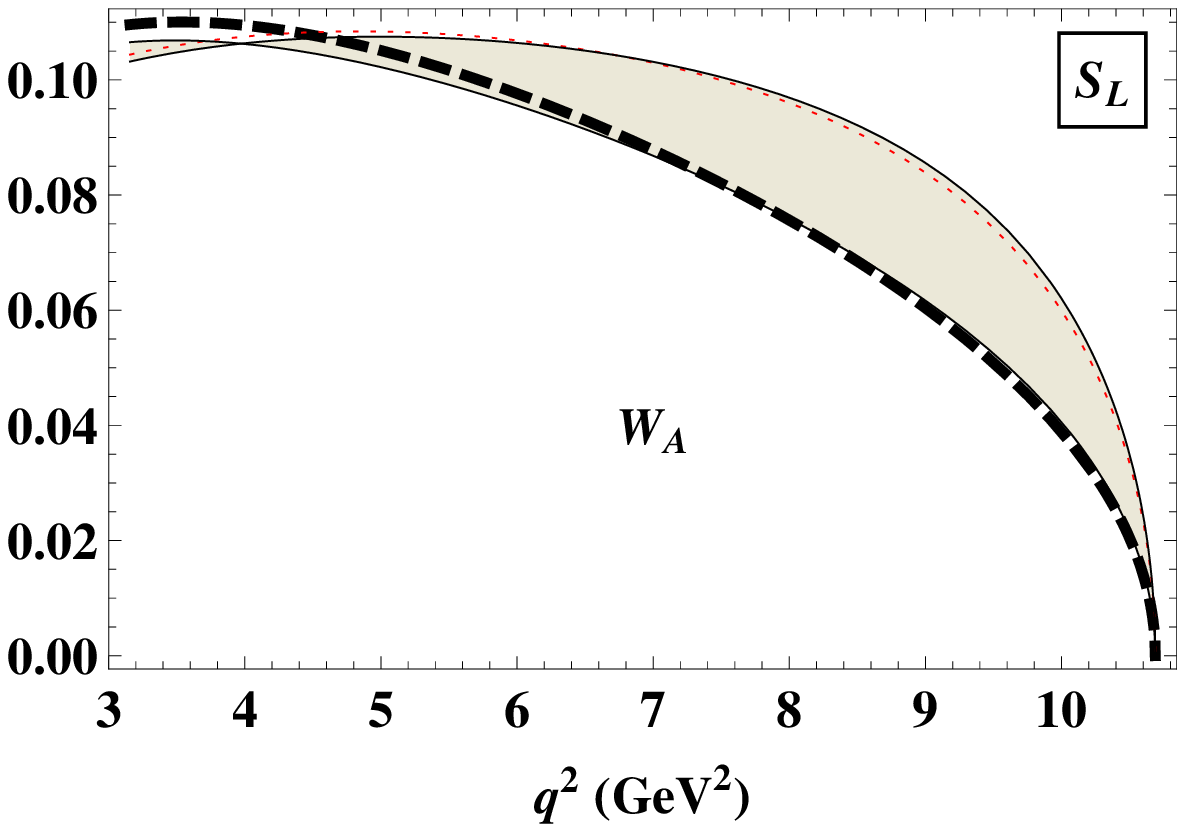}&
\includegraphics[scale=0.40]{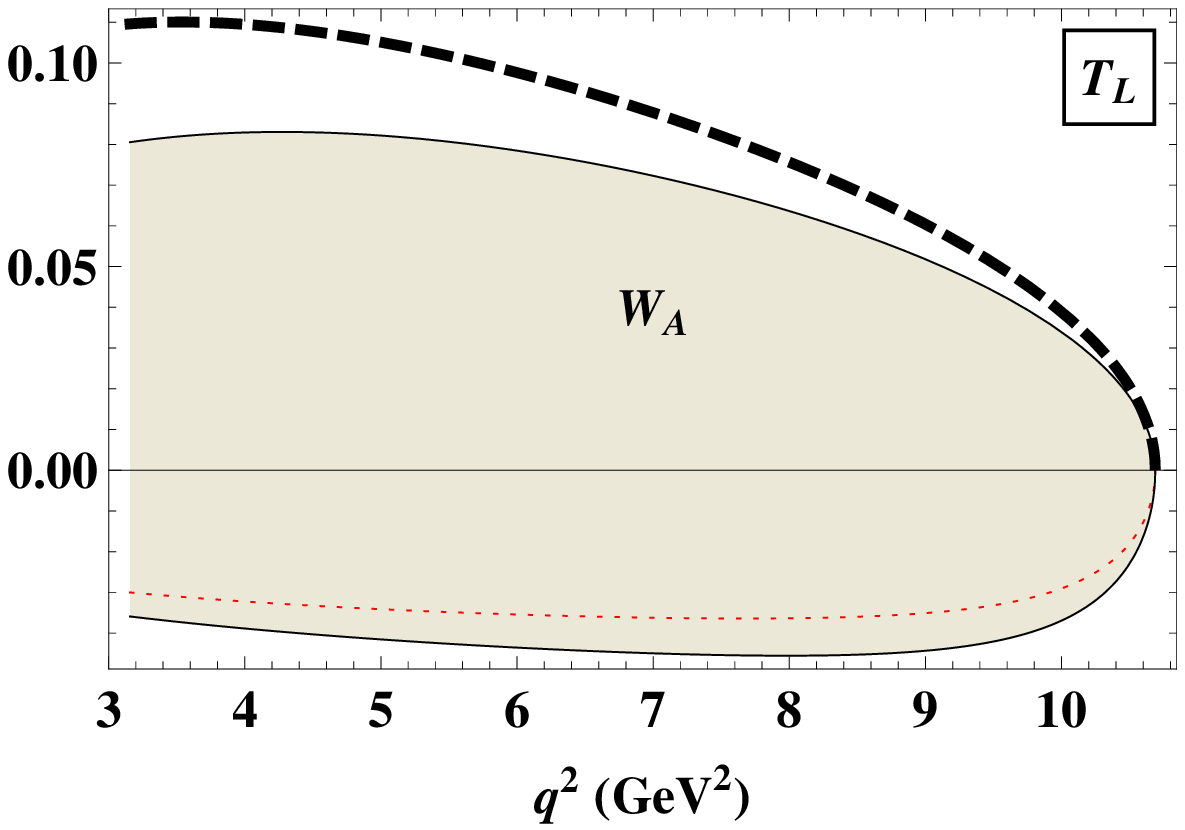}\\
\includegraphics[scale=0.40]{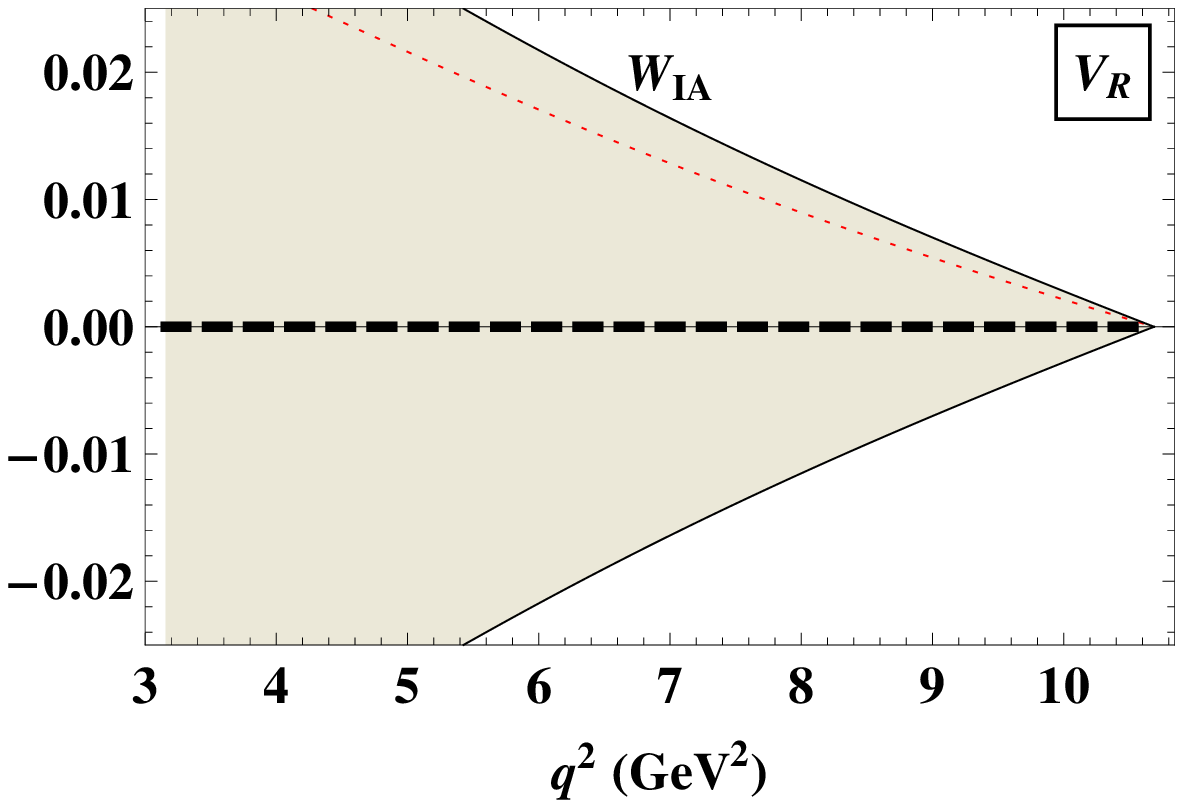}&
\includegraphics[scale=0.40]{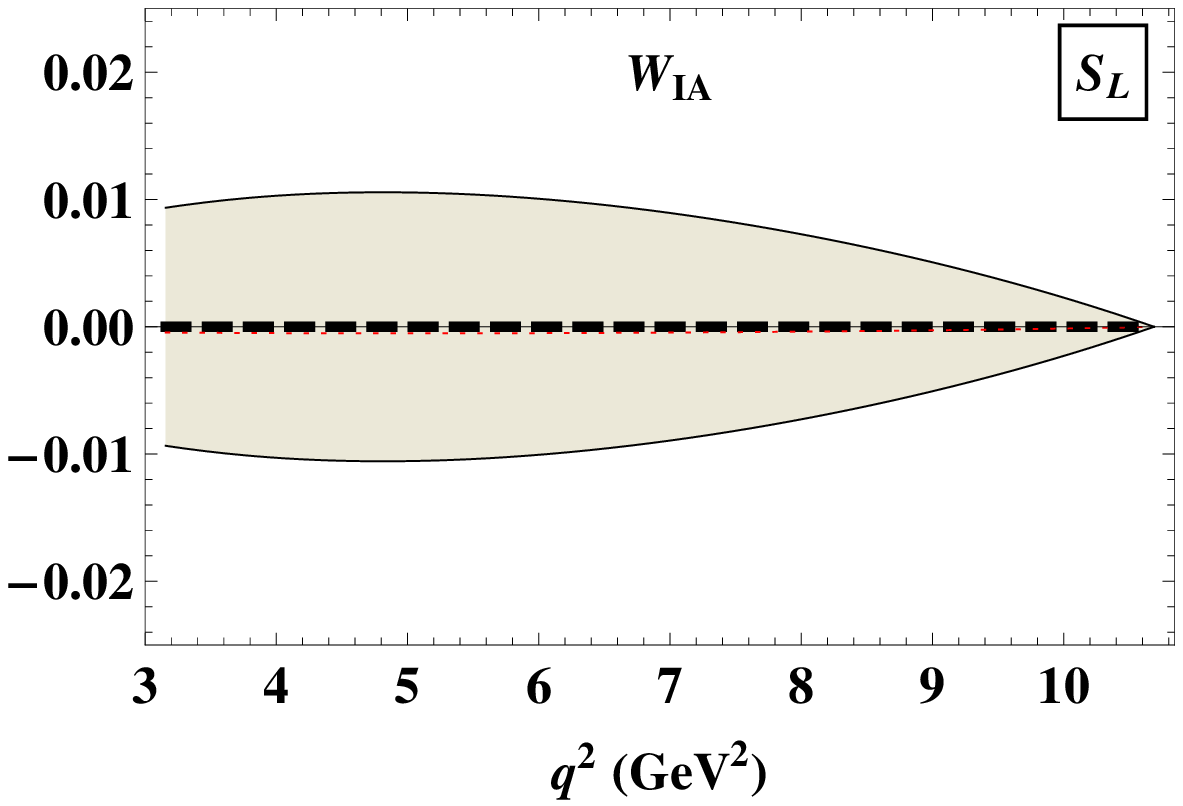}&
\includegraphics[scale=0.40]{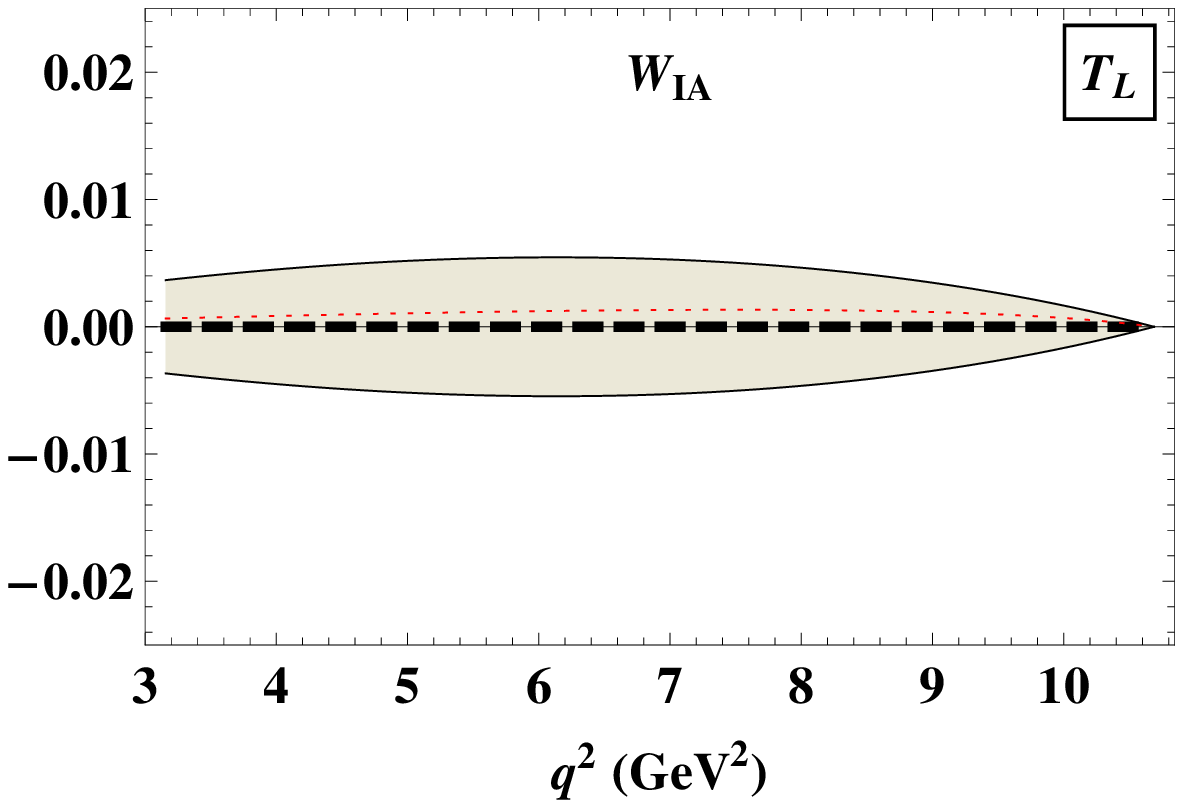}
\end{tabular}
\caption{Trigonometric moments $W_T(q^2)$, $W_I(q^2)$, $W_A(q^2)$, and $W_{IA}(q^2)$. Notations are the same as in Fig.~\ref{fig:RD}.}
\label{fig:WTAI}
\end{figure}
In Fig.~\ref{fig:WTAI} we show the $q^2$ dependence of the trigonometric moments $W_T(q^2)$, $W_I(q^2)$, $W_A(q^2)$, and $W_{IA}(q^2)$. The moments $W_T(q^2)$ and  $W_I(q^2)$ are almost insensitive to $\mathcal{O}_{V_R}$ but highly sensitive to $\mathcal{O}_{T_L}$. The scalar and tensor operators  are likely to raise $W_T(q^2)$ and to lower $W_I(q^2)$ in general. The moment $W_A(q^2)$ shows great sensitivity to $\mathcal{O}_{V_R}$, $\mathcal{O}_{S_L}$, and $\mathcal{O}_{T_L}$. Both $\mathcal{O}_{V_R}$ and $\mathcal{O}_{T_L}$ tend to decrease $W_A(q^2)$ while $\mathcal{O}_{S_L}$ tries to do the opposite. It is worth noting that all three moments $W_T(q^2)$, $W_I(q^2)$, and $W_A(q^2)$ are extremely sensitive to the tensor operator and their sign can change in the presence of $\mathcal{O}_{T_L}$. Regarding the moment $W_{IA}(q^2)$, the three operators act in the same manner: they can change the moment in both directions and the sensitivity is maximal in the case of $\mathcal{O}_{V_R}$.
\begin{figure}[htbp]
\begin{tabular}{lr}
\includegraphics[scale=0.40]{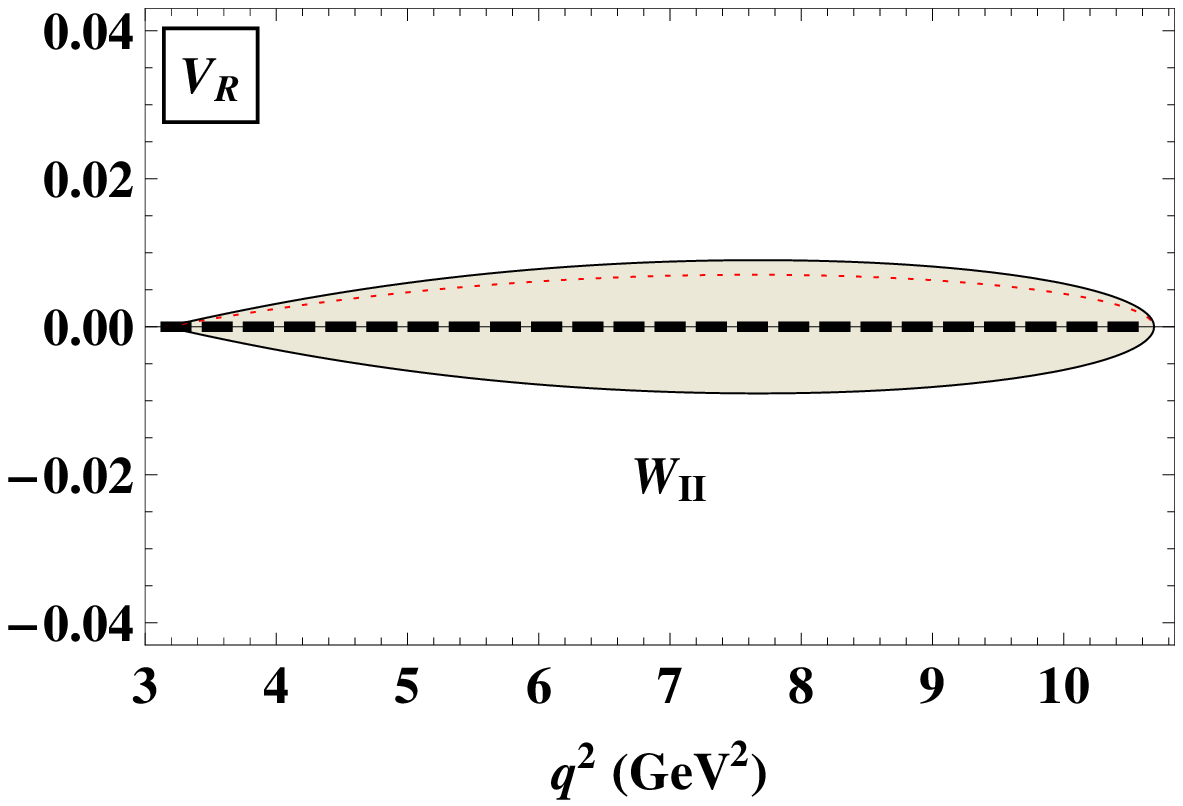}&
\includegraphics[scale=0.40]{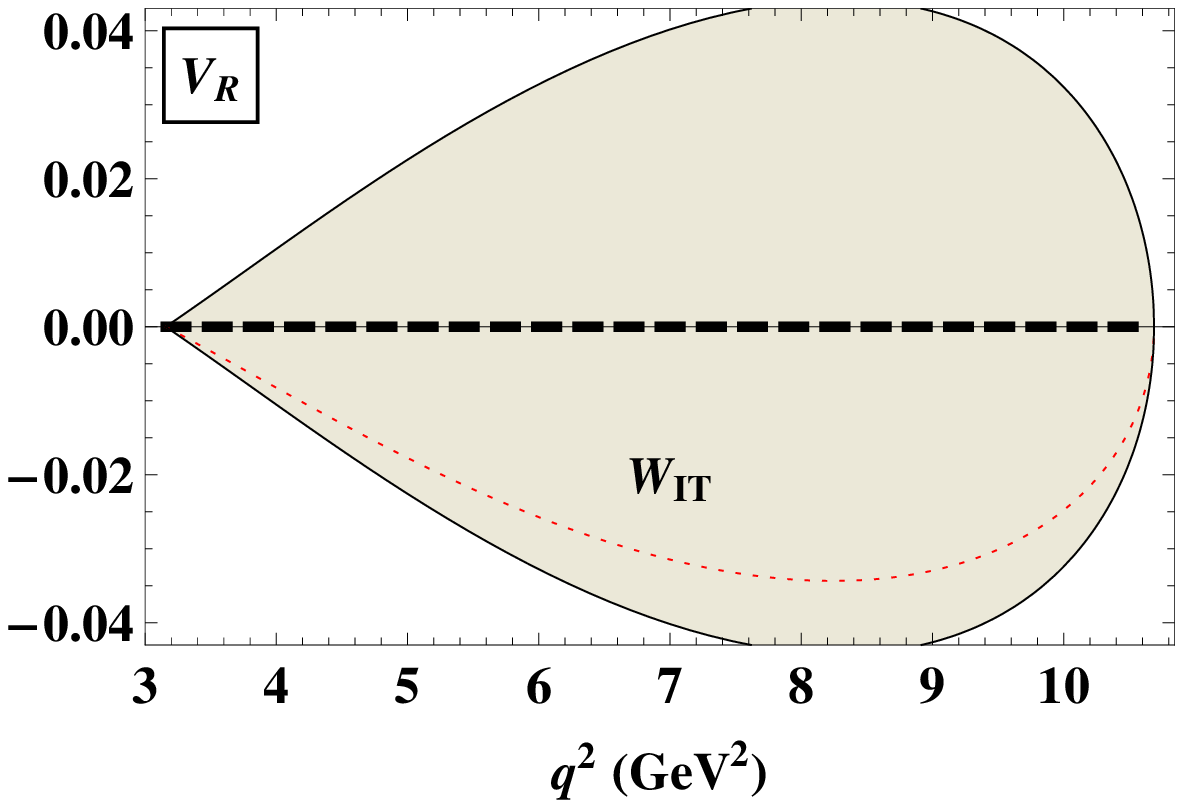}
\end{tabular}
\caption{Trigonometric moments $W_{II}(q^2)$ (left), and $W_{IT}(q^2)$ (right). Notations are the same as in Fig.~\ref{fig:RD}.}
\label{fig:WIIT}
\end{figure}

The trigonometric moments $W_{II}(q^2)$ and $W_{IT}(q^2)$ are equal to zero in the SM and obtain a nonzero contribution only from the right-chiral vector operator
$\mathcal{O}_{V_R}$, as depicted in Fig.~\ref{fig:WIIT}. Both moments are proportional to the imaginary part of $V_R$ and the effect of $\mathcal{O}_{V_R}$ cancels in their ratio.

\begin{figure}[htbp]
\includegraphics[scale=0.6]{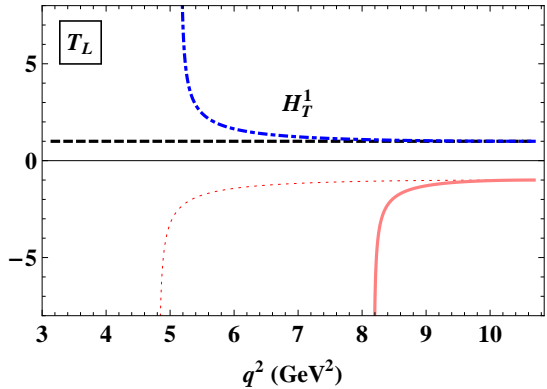}
\caption{Optimized angular observable $H_T^{(1)}(q^2)$ defined in Eq.~(\ref{eq:HT1}). The black dashed line is the SM prediction. The red dotted line represents the best-fit value of $T_L$. The blue dot-dashed line and the pink solid line are the predictions for $T_L=0.04-0.17i$ and $T_L=0.18+0.23i$, respectively.}
\label{fig:HT1}
\end{figure}

One can also consider certain combinations of angular observables where the form factor dependence drops out (at least in most NP scenarios), as described in Ref.~\cite{Feldmann:2015xsa}. As a demonstration, we consider the optimized observable 
\be
H_T^{(1)} = \frac{\sqrt{2}J_4}{\sqrt{-J_{2c}(2J_{2s}-J_3)}},
\label{eq:HT1}
\en
which is equal to one not only in the SM but also in all NP scenarios except the tensor one, as shown in Fig.~\ref{fig:HT1}. Therefore $H_T^{(1)}(q^2)$ plays a prominent role in confirming the appearance of the tensor operator $\mathcal{O}_{T_L}$ in the decay $\bar{B}^0 \to D^{\ast} \tau^- \bar{\nu}_{\tau}$.
%

\section{Summary and discussion}
\label{sec:summary}
We have provided a thorough analysis of possible NP in the semileptonic decays $\bar{B}^0 \to D^{(\ast)}\tau^-\bar\nu_{\tau}$ using the form factors obtained from our covariant quark model. Starting with a general effective Hamiltonian including NP operators, we have derived the full angular distribution and defined a large set of physical observables which helps discriminate between NP scenarios. Assuming NP only affects leptons of the third generation and only one NP operator appears at a time, we have gained the allowed regions of NP couplings based on recent measurements at $B$ factories and studied the effects of each operator on the observables. It has turned out that the current experimental data of $R(D)$ and $R(D^\ast)$ prefer the operators $\mathcal{O}_{S_L}$ and $\mathcal{O}_{V_{L,R}}$, the operator $\mathcal{O}_{T_L}$ is less favored, and the operator $\mathcal{O}_{S_R}$ is disfavored at $2\sigma$. 

Our analysis has been done under the assumption of one-operator dominance. However, the large observable set has revealed unique behaviors of several observables and provided many correlations between them, which allows one to distinguish between NP operators. Our analysis can serve as a map for setting up various strategies to identify the origins of NP, one of which is as follows: first, one uses the null tests $W_{IT}(q^2) = 0$ and $H_T^{(1)}(q^2) -1 = 0$ to probe the operators $\mathcal{O}_{V_R}$ and $\mathcal{O}_{T_L}$, respectively. Second, one measures the forward-backward asymmetry in the decay $\bar{B}^0 \to D\tau^-\bar\nu_{\tau}$. If $\mathcal{A}_{FB}^D(q^2)$ has a zero-crossing point, then it is a clear sign of the operator $\mathcal{O}_{S_L}$. The coupling $V_L$ is more difficult to test because it is just a multiplier of the SM operator. However, if the tests above disconfirm $\mathcal{O}_{V_R}$, $\mathcal{O}_{T_L}$, and $\mathcal{O}_{S_L}$ at the same time, then the modification of $V_L$ to $R(D)$ and $R(D^\ast)$ is a must. In the future when more precise data will be collected, one can adopt the strategies described in this paper as a useful tool to discover NP in these decays if the deviation from the SM still remains. 


\begin{acknowledgments}
The authors thank the Heisenberg-Landau Grant for providing support 
for their collaboration.  
M.A.I.\ acknowledges the support from  PRISMA cluster of excellence 
(Mainz Uni.).
M.A.I. and C.T.T.  greatly appreciate warm hospitality 
of the Mainz Institute for Theoretical Physics (MITP) 
throughout their visit.
\end{acknowledgments}

\appendix

\section{Relations for the form factors in the spectator quark model}
\label{app:HQET}

The HQET relations for the form factors can be calculated using the
formulas (see e.g., Ref.~\cite{Korner:1987kd})
\bea
\langle D(p_2)\,|\,\bar c \Gamma b\,|\,\bar B(p_1)\rangle
&=&-\frac{1}{4\sqrt{m_1 m_2}}
  \Tr[(\slp_2-m_2)\gamma_5\Gamma (\slp_1+m_1)\gamma_5]\cdot\xi(w),
\nn
  \langle D^\ast(p_2)\,|\,\bar c \Gamma b\,|\,\bar B(p_1)\rangle
&=&\,-\frac{1}{4\sqrt{m_1 m_2}}
   \Tr[(\slp_2-m_2)\epsilon_2^\ast\Gamma(\slp_1 +m_1)\gamma_5]\cdot\xi(w),
\label{eq:HQET}
\ena
where $\Gamma$ denotes the Dirac operator that induces the transition,
 \[
w=\frac{2p_1p_2}{2m_1m_2}=\frac{m_1^2+m_2^2-q^2}{2m_1m_2},\qquad 
q=p_1-p_2 \,,
\]
and $\xi(w)$ is the universal Isgur-Wise function normalized to $1$ at
zero recoil $w=1$. These relations allow one to express all form
factors defined in Eqs.~(\ref{eq:BD-ff}) and  (\ref{eq:BDv-ff})
in terms of the Isgur-Wise function as
\bea
  F_\pm(q^2)&=& \pm\frac{m_1 \pm m_2}{2\sqrt{m_1 m_2}} \cdot\xi(w),\qquad
  F^S(q^2) = \frac{\sqrt{m_1 m_2}}{m_1+m_2}(w+1)\cdot\xi(w),
\nn
  F^T(q^2) &=& \frac{m_1 +m_2}{2\sqrt{m_1 m_2}}\cdot\xi(w),
\nn[2ex]
    A_0(q^2)&=&
    \frac{\sqrt{m_1m_2}}{(m_1-m_2)}(w +1)  \cdot\xi(w),
\nn
    A_+(q^2)&=&-A_-(q^2)= V(q^2)=\frac{m_1+m_2}{2\sqrt{m_1 m_2}}\cdot\xi(w),
\nn
 G_S(q^2)&=&- \frac{m_2}{\sqrt{m_1 m_2}}\cdot\xi(w), \qquad G_0^T(q^2)=0,
\nn
  G_1^T(q^2) &=& \frac{m_1+m_2}{2\sqrt{m_1 m_2}}\cdot\xi(w),\qquad
  G_2^T(q^2)=-\frac{m_1-m_2}{\sqrt{m_1 m_2}}\cdot\xi(w).
\label{eq:HQET-relations}
\ena

One can use the equations of motion (EOMs) to obtain relations between
the various form factors. The EOMs for the charm and the bottom fields read
\be
\label{eom}
\bar \psi_c \slp_c =m_c \bar \psi_c,
\qquad \qquad \slp_b\psi_b =m_b\psi_b.
\en
Using $p_1-p_2=p_b-p_c$ and the EOMs (\ref{eom}) one can then relate
the vector (axial) form factors to the scalar (pseudoscalar) form
factors, and the tensor form factors to the vector (axial) form factors.
One obtains
\be
\label{bd2}
  F_+(q^2)(m_1^2-m_2^2)  + F_-(q^2) q^2=
 (m_b-m_c)(m_1+m_2)F^S(q^2),
  \en
    \bea
\label{bd3}
-\frac{q^2}{(m_1+m_2)}F^T&=&-(m_b +m_c)F_+ + (m_1+m_2)F^S, \\
\label{bd4}
  \frac{(m_1^2-m_2^2)}{(m_1+m_2)}F^T&=&-(m_b +m_c)F_-,
  \ena
  \be
\label{bds5}
(m_b+m_c) G^S(q^2)=-(m_1-m_2) A_0(q^2)+(m_1-m_2) A_+(q^2)+
\frac{q^2}{(m_1+m_2)} A_-(q^2),
\en
\bea
\label{bds6}
(m_1^2 -m_2^2)G^T_{1} +q^2 G^T_{2} &=& +(m_b-m_c)\frac{(m_1^2 -m_2^2)}{m_1+m_2}
A_0, \\
\label{bds7}
-G^T_{1} +\frac{q^2}{(m_1 +m_2)^2} G^T_{0}
&=& -\frac{(m_b -m_c)}{m_1+m_2}A_+ +G^S, \\
\label{bds8}
-G^T_{2} - \frac{m_1^2 -m_2^2}{(m_1 +m_2)^2} G^T_{0}&=&
+ \frac{m_b -m_c}{(m_1 +m_2)}A_-.
\ena
One can check that the HQET form factors satisfy the three relations
Eqs.~(\ref{bds6}), (\ref{bds7}), and~(\ref{bds8}) for $m_b-m_c=m_1-m_2$.

\section{Heavy quark limit in the covariant quark model}

In our approach the HQL was explored in great detail
in Ref.~\cite{Ivanov:2015tru} for the  heavy-to-heavy transition 
$B\to D(D^\ast)$. It was explicitly shown that in the limit
$m_B=m_b + E,\,\,\,m_b\to\infty $ and
          $m_D=m_{D^\ast}=m_c + E,\,\,\,m_c\to\infty $ 
we reproduce all relations given by Eqs.~(\ref{bd3}), (\ref{bd4}), 
(\ref{bds6}), (\ref{bds7}), and~(\ref{bds8}). In addition the Isgur-Wise
function $\xi(w)$ has been calculated as follows:
\be
\xi(w) = \frac{J_3(E,w)}{J_3(E,1)}, \qquad
J_3(E,w) = \int\limits_0^1 \frac{d\tau}{W}
\int\limits_0^\infty\!\! du\, \widetilde\Phi^2(z) 
\left( \sigma_S(z) + \sqrt{\frac{u}{W}} \sigma_V(z) \right),
\label{eq:IW}
\en
where $W=1+2\tau(1-\tau)(w-1)$, $z=u - 2 E \sqrt{u/W}$, and
\[
\widetilde\Phi(z) = \exp(-z/\Lambda^2), \qquad
 \sigma_S(z) = \frac{m_u}{m_u^2 +z},  \qquad
 \sigma_V(z) = \frac{1}{m_u^2 +z}.
\]
It is readily seen that this function has the correct normalization
at zero recoil, i.e. $\xi(w=1)=1$.

The subleading corrections to the heavy quark limit arising from
finite quark masses have been investigated within our framework
in Ref.~\cite{Ivanov:1992wx}. In particular, it was found that the matrix
element of the semileptonic decay $B\to D \ell\bar\nu$ 
calculated in the heavy quark limit up to the next-to-leading $1/m_Q$ order
is written as
\bea
<D(p_2)\,|\,\bar c \gamma^\mu b\, |\, B(p_1)>
&=&
\sqrt{m_bm_c}
\Big\{
  (v_1+v_2)^\mu\Big[\xi(w) + (\frac{1}{m_c}+\frac{1}{m_b})\,\xi^{(1)}_+(w)\Big]
\nn
&+& (v_1-v_2)^\mu  \Big(\frac{1}{m_c}-\frac{1}{m_b}\Big)\,\xi^{(1)}_-(w)
\Big\},
\label{eq:HQL-NLO}
\ena
where the subleading function $\xi^{(1)}_+(w)$ is equal to zero at $w=1$.
Therefore, the subleading  $1/m_Q$ corrections vanish at zero
recoil $v_1=v_2$ ($w=1$) in accordance with  Luke's theorem  
\cite{Luke:1990eg} (for review, see  Ref.~\cite{Neubert:1993mb}).

One has to emphasize that HQL is just a great simplification of reality. 
In our approach hadrons are treated as bound states of quarks which are bound  
and confined.  The matrix elements describing the transitions between hadrons 
are defined by the Feynman diagrams with virtual off-shell quarks. 
It is for this reason that the EOM relations are only approximately fulfilled 
in our model in the finite heavy quark mass case. The situation is rather 
different from the naive spectator quark model where quarks are supposed to be 
free and on shell. 
The calculation of the matrix elements in the CCQM  is straightforward and 
does not need to rely on either the EOM or the HQL. 
We mention that a discussion on decay constants and form factors in the 
$B\to D(D^\ast)$ transitions in the HQL can also be found in 
Ref.~\cite{DeVito:2004zs}. 

\section{Twofold distribution of 
${\bm \bar{B}^0 \to D^\ast \tau^-\bar{\nu}_\tau}$}
\label{app:twofold}

In this appendix we provide the explicit differential $(q^2,\cos\theta)$ 
distribution of the decay $\bar{B}^0 \to D^\ast \tau^-\bar{\nu}_\tau$ for easy 
comparison with other studies. The distribution reads
\bea
\lefteqn{\frac{d^2\Gamma(\bar{B}^0\to D^\ast\tau^-\bar{\nu}_\tau)}{dq^2d\cos\theta}}\nn
&=&\frac{G_F^2|V_{cb}|^2|{\bf p_2}|q^2v^2}{32 (2\pi)^3 m_1^2}\nn
&&\times 
\Big\lbrace
|1+V_L|^2
\Big[
(1-\cos\theta)^2|H_{++}|^2+(1+\cos\theta)^2|H_{--}|^2+2\sin^2\theta|H_{00}|^2\nn
&&+2\delta_\tau
\Big(
\sin^2\theta(|H_{++}|^2+|H_{--}|^2)+2|H_{t0}-H_{00}\cos\theta|^2
\Big)
\Big]\nn
&&+|V_R|^2
\Big[
(1-\cos\theta)^2|H_{--}|^2+(1+\cos\theta)^2|H_{++}|^2+2\sin^2\theta|H_{00}|^2\nn
&&+2\delta_\tau
\Big(
\sin^2\theta(|H_{++}|^2+|H_{--}|^2)+2|H_{t0}-H_{00}\cos\theta|^2
\Big)
\Big]\nn
&&-4 {\rm Re}V_R
\Big[
(1+\cos^2\theta)H_{++}H_{--}+\sin^2\theta|H_{00}|^2\nn
&&+2\delta_\tau
\Big(
\sin^2\theta H_{++}H_{--}+|H_{t0}-H_{00}\cos\theta|^2
\Big)
\Big]\nn
&&+2|S_R-S_L|^2|H^S_V|^2\nn
&&+4 \sqrt{2\delta_\tau} {\rm Re}(S_R-S_L)H^S_V(H_{t0}-H_{00}\cos\theta)\nn
&&+16|T_L|^2 \Big[
|H_T^0|^2\Big( 1+2\delta_\tau +(1-2\delta_\tau)\cos2\theta \Big)\nn
&&+2|H_T^+|^2\sin^2\frac{\theta}{2}\Big( 1+2\delta_\tau+(1-2\delta_\tau)\cos\theta \Big)\nn
&&+2|H_T^-|^2\cos^2\frac{\theta}{2}\Big( 1+2\delta_\tau-(1-2\delta_\tau)\cos\theta \Big)
\Big]\nn
&&-16\sqrt{2\delta_\tau} {\rm Re}T_L\Big[ H_{++}H_T^+ +H_{--}H_T^- +H_{00}H_T^0 \nn
&&-\Big(H_{++}H_T^+ -H_{--}H_T^- +H_{t0}H_T^0\Big)\cos\theta \Big]
\Big\rbrace.
\label{Ap:distr2Dv}
\ena

 \ed